\documentclass[4pt,aps,pre,reprint,superscriptaddress
]{revtex4-2} %

\usepackage[utf8]{inputenc}
\usepackage[T1]{fontenc}
\usepackage{graphicx}
\usepackage{mathrsfs}
\usepackage{bm}
\usepackage[dvipsnames]{xcolor}
\usepackage{mathtools,amsmath,amssymb}
\usepackage{hyperref}
\usepackage{siunitx}
\usepackage{fmtcount}
\usepackage{textcomp}
\usepackage{flafter}
\usepackage{float}
\usepackage{amsthm}

\usepackage{listings}
\usepackage{graphicx}
\usepackage{dcolumn}

\begin{document}
\title{
The Geometric Foundations of Microcanonical Thermodynamics: Entropy Flow Equation and Thermodynamic Equivalence}
\author{Loris Di Cairano}
\email{l.di.cairano.92@gmail.com, loris.dicairano@uni.lu}

\affiliation{Department of Physics and Materials Science, University of Luxembourg, L-1511 Luxembourg City, Luxembourg}

\date{\today}% It is always \today, today,
%%% DATE: 09/09/2025 (Bianca) 

\begin{abstract}
We develop a geometric foundation of microcanonical thermodynamics in which entropy and its derivatives are determined from the geometry of phase space, rather than being introduced through an \emph{a priori} ensemble postulate. Once the minimal structure needed to measure constant--energy manifolds is made explicit, the microcanonical measure emerges as the natural hypersurface measure on each energy shell. Thermodynamics becomes the study of how these shells deform with energy: the entropy is the logarithm of a geometric area, and its derivatives satisfy a deterministic hierarchy of \emph{entropy flow equations} driven by microcanonical averages of curvature invariants (built from the shape/Weingarten operator and related geometric data). Within this framework, phase transitions correspond to qualitative reorganizations of the geometry of energy manifolds, leaving systematic signatures in the derivatives of the entropy.

Two general structural consequences follow. First, we reveal a \emph{thermodynamic covariance}: the reconstructed thermodynamics is invariant under arbitrary descriptive choices such as reparametrizations and equivalent representations of the same conserved dynamics. Second, a \emph{geometric microcanonical equivalence} is found: microscopic realizations that share the same geometric content of their energy manifolds (in the sense of entering the curvature sources of the flow) necessarily yield the same microcanonical thermodynamics. We demonstrate the full practical power of the formalism by reconstructing microcanonical response and identifying criticality across paradigmatic systems, from exactly solvable mean-field models to genuinely nontrivial short-range lattice field theories and the 1D long-range XY model with $1/r^\alpha$ interactions.
\end{abstract}

\maketitle

\section{Introduction and Motivation}

Given a complete description of microscopic dynamics through the symplectic Hamiltonian formalism, one may ask a sharp foundational question:

\begin{quote}
\emph{Can one deduce where a phase transition occurs without introducing external statistical prescriptions?}
\end{quote}

Following conventional statistical mechanics, this question is nonsense: thermodynamics is usually founded on the introduction of a weight function $\rho(H)$---a statistical ensemble---on top of Hamiltonian mechanics. This scheme is extraordinarily effective and has generated remarkable predictions. Yet, it retains a conceptual asymmetry: Hamiltonian mechanics is \emph{intrinsically generative}---the symplectic system $(\Lambda,\omega)$ contains the rules of evolution within itself---whereas thermodynamics remains \emph{descriptive}: counting accessible states, and hence macroscopic properties, depends on a choice of measure that does not uniquely follow from the symplectic structure.

This asymmetry can be stated as an obstruction. The symplectic form $\omega$ identifies the energy level sets
\[
\Sigma_E=\{\bm x\in\Lambda:\,H(\bm x)=E\},
\]
through Hamilton's equations $\iota_{\bm X_H}\omega=dH$, and generates the Hamiltonian flow tangent to $\Sigma_E$. However, it provides no intrinsic notion of \emph{size} for these hypersurfaces: it does not select a canonical surface measure on $\Sigma_E$, and therefore cannot, by itself, determine the density of states $\Omega(E)$ on which the microcanonical entropy is built. Historically, this gap is bridged by introducing the microcanonical ensemble as an external prescription: postulate that all states at fixed energy are equally weighted. This is physically well-motivated for isolated systems; yet, it is still an addition to the Hamiltonian system, not a consequence of it. The question thus persists: can the relevant measure be derived from the Hamiltonian structure itself?

This work proposes that the answer is yes, once one makes explicit what is mathematically missing. To speak of ``volume'' and ``area'', one needs orthogonality and an induced surface element, which are metric notions. Introducing a Riemannian metric $\eta$ is therefore not an arbitrary statistical assumption; it is the geometric minimum required to complete what the symplectic structure leaves under-determined. The metric provides a distinguished normal direction to $\Sigma_E$ through the gradient $\nabla_{\eta}H$, and induces a canonical hypersurface measure $d\mu^{\eta}_E$ by the geometry of $\Sigma_E$ itself. In this way, the notion of ``accessibility'' becomes a rigorous geometric quantity.

Once phase space is equipped with the complete structure $(\Lambda,\omega,H,\eta)$, without assuming a statistical ensemble, the metric induces a natural transverse generator
\[
\bm\xi_\eta=\frac{\nabla_{\eta}H}{\|\nabla_{\eta}H\|_{\eta}^{2}}, \qquad dH(\bm\xi_\eta)=1,
\]
and a corresponding decomposition of the tangent space,
\[
T_{\bm x}\Lambda=\mathrm{span}\{\bm\xi_\eta\}\oplus T_{\bm x}\Sigma_E.
\]
This decomposition cleanly separates two evolutions acting along orthogonal directions: the Hamiltonian flow $\Phi_t^{\mathrm{sym}}$, tangent to $\Sigma_E$, governs microscopic dynamics within a given energy hypersurface; the geometric flow $\Phi_{\bm\xi}^{\mathrm{diff}}$, transverse to the foliation, describes how the hypersurface itself deforms as $E$ varies. Thermodynamics, in this framework, is not introduced as a static equilibrium scheme: it is derived from how the geometry of the energy surface evolves along $E$.

The crucial consequence is that the natural measure $d\mu^\eta_E$ on $\Sigma_E$ is induced by $\bm\xi_\eta$ through the presence of $\eta$, namely, $\iota_{\bm\xi_\eta}d\mu_{\Lambda}=d\mu_{E}^\eta$. With this construction, this induced measure is shown to exactly equate the microcanonical density of states as described by Boltzmann:
\[
\Omega_\eta(E)=\int_{\Sigma_E} d\mu^\eta_E\equiv \int_{\Lambda}\delta(H(\bm x)-E)\,d\mu_{\Lambda}=\Omega_{\rm Boltz}(E),
\]
so that the microcanonical ensemble is \emph{not} postulated as an external prescription; it emerges as the measure induced on every energy hypersurface by the metric structure. The ensemble becomes not an addition to the theory but an unavoidable geometric consequence.

We then reveal a fundamental \emph{thermodynamic covariance}: the Hamiltonian flow preserves the induced measure $d\mu^\eta_E$ but does not rigidly determine the metric tensor itself. There exists an entire equivalence class $[\eta]$ of metrics that induce the same microcanonical measure on every $\Sigma_E$, and therefore yield identical thermodynamics---same entropy $S(E)$, same temperature $\beta(E)$, same response functions. Thermodynamic content is invariant under geometric representation, provided the natural measure on energy shells is preserved. Within this equivalence class, we construct a particularly natural representative---a \emph{unit-norm gauge} metric $g\in[\eta]$ obtained through a controlled conformal rescaling---in which the induced measure $d\mu^g_E$ coincides exactly with the intrinsic area element of the hypersurface $\Sigma_E$. In this gauge, we establish a \emph{microcanonical geometric equivalence}: counting accessible states becomes literally measuring occupied geometric area, $\Omega_g(E)\equiv\mathrm{area}_g(\Sigma_E)$, and entropy follows as $S_g(E)=\ln\mathrm{area}_g(\Sigma_E)$. Thermodynamics is now manifestly and purely geometric.

Once entropy is recognized as a geometric area, we derive deterministic transport laws for its energy derivatives. Differentiating $\mathrm{area}_g(\Sigma_E)$ along the transverse flow $\Phi^\mathrm{diff}_{\bm\xi_g}$ generated by the unit normal $\bm\xi_g\equiv\nabla_g H$ naturally produces the \emph{Entropy Flow Equation} (EFE): 
\[
    \partial_E^2S_g(E)+[\partial_ES_g(E)]^2=\Upsilon^{(2)}_g(E),
\] 
where the source term
\begin{equation*}
    \Upsilon^{(2)}_g(E)=\frac12\int_{\Sigma_{E}}\!\!\left[\text{Tr}[W_{\bm\xi_g}]^{2}-\text{Tr}[W_{\bm\xi_g}^2]+R^g_{\Sigma_{E}}-R^g_{\Lambda}\right]d\rho^g_E,
\end{equation*}
is a purely geometric term involving curvature invariants such as traces of the Weingarten operator and Ricci scalars of $\Sigma_E$ averaged with respect to the area-measure $d\rho^g_E:=d\mu^g_E/\text{area}^g(E)$.

This is not an imposed dynamical hypothesis but a non-perturbative deterministic equation derived from first-principles. Phase transitions correspond to qualitative reorganizations of the geometry of energy manifolds, leaving systematic signatures in the curvature sources that drive the entropy flow.

The crucial predictive power of this framework is that knowledge of the geometry alone permits the prediction of where phase transitions emerge. To demonstrate this, we compute the geometric observables entering the EFE for three paradigmatic systems, solve the EFE numerically, and compare the reconstructed entropy derivatives with independent thermodynamic estimates obtained from microcanonical simulations via the Pearson-Halicioglu-Tiller (PHT) method.

For the 1D mean-field XY model, we first derive the exact analytical solution for the first and second derivatives of the microcanonical entropy in terms of modified Bessel functions through purely statistical-mechanical approaches. This serves as a rigorous benchmark to validate both the geometric approach and the microcanonical inflection-point analysis (MIPA). We then perform an analytical expansion of the mean curvature, namely, the trace of the Weingarten operator, in powers of the magnetization $M$ near the disordered branch. We show that 
\[
    \mathrm{Tr}W_{\bm\xi
    }(M)=C^{\infty}_0(\epsilon)+C^{\infty}_2(\varepsilon)\,M^2+O(M^4)\,,
\]
where $\epsilon=E/N$ is the specific energy and the coefficient that controls the curvature response
\[
    C^{\infty}_2(\varepsilon)=\frac{J}{(2\varepsilon-J)^2}\Big(\varepsilon-\frac{3J}{4}\Big),
\]
vanishes precisely at $\varepsilon_c=3J/4$---the known critical energy---without any statistical input. This geometric criterion directly predicts the transition. Moreover, we explicitly demonstrate the geometric mechanism underlying the transition: at low energy, the system exhibits an elliptic cylindrical geometry (ordered phase), which transforms through a hyperbolic neck (one-sheet hyperboloid at criticality) and finally into a two-sheet hyperboloid geometry in the disordered phase. The transition is encoded in how the eigenvalue spectrum of the Hessian of $H$ reorganizes, shifting from positive (convex) to mixed-sign (saddle) curvatures as collective modes soften.

For the 2D $\phi^4$ model with nearest-neighbor interactions, we compute $\Upsilon^{(2)}_g(E)$ from microcanonical simulations, solve the EFE, and reconstruct $S'(E)$ and $S''(E)$. The comparison with independent PHT estimates shows quantitative agreement across the entire energy range and for multiple system sizes. The geometric source $\Upsilon^{(2)}_g$ exhibits a pronounced peak at $\varepsilon_c\simeq 11.1$, exactly where $S''(E)$ develops a negative-valued maximum, signaling the finite-size precursor of the second-order transition. The agreement confirms that thermodynamic response is faithfully encoded in---and reconstructable from---the geometry of energy hypersurfaces.

Finally, for the 1D XY model with genuine long-range interactions $\propto 1/r^{\alpha}$ with $\alpha:=1+\sigma$, we investigate the weak long-range regime ($1<\alpha<1.5$, namely, $0<\sigma <0.5$) where the system interpolates between mean-field and short-range behavior. We demonstrate three critical points: (i) the geometric reconstruction remains quantitatively accurate even for true long-range interactions, with the EFE-reconstructed $S'(E)$ and $S''(E)$ tracking the PHT estimates across all $\sigma$ values and system sizes; (ii) MIPA consistently identifies finite-size transition precursors, with the characteristic energy $\varepsilon_c(\sigma)$ drifting systematically as the interaction range is tuned; (iii) the method does not fabricate spurious transitions---the geometric response $\Upsilon^{(2)}_g(E)$ faithfully mirrors the thermodynamic channel $\partial^2_\varepsilon S(E)$.

This reverses a common causal hierarchy. It is not the case that probability weights generate thermodynamics and that geometry comes afterward; rather, geometry generates thermodynamics, and probability emerges as a consequence of the induced measure on $\Sigma_E$. We do not seek to replace traditional statistical mechanics. We address a deeper question: \emph{where do probabilistic notions reside within a deterministic framework?} Here, they exist in the Riemannian geometry that completes the symplectic structure: fluctuations, correlations, and transition-like anomalies are encoded in how the measure of energy hypersurfaces deforms.

A concise version of this geometric viewpoint has already been put forward in a recent Letter \cite{di2025phase}, where the central idea---that thermodynamic information is carried by the geometry of the energy foliation---was presented in its most essential form.
The aim of the present work is to unfold that message into a complete construction.

\section{State of the art}\label{sec:sota}
A long-standing motivation for geometric and topological approaches to phase transitions is that standard equilibrium narratives (canonical ensemble, thermodynamic limit, scale separation) can be either unavailable or conceptually non-fundamental for many systems of current interest (finite platforms, constrained dynamics, long-range nonadditivity). A parallel line of thought, developed over several decades, asks whether the ``mechanism'' of a phase transition can be formulated in terms of intrinsic structures that already exist at the microscopic level: the geometry of energy shells in phase space, the geometry and topology of configuration-space submanifolds, or geometric structures induced on spaces of macrostates. In this view, singular thermodynamic behavior is not postulated but must emerge from well-defined geometric quantities (curvatures, second fundamental forms, Morse indices, Euler characteristics) and their ensemble averages.

Historically, one of the earliest systematic bridges between microscopic dynamics and thermodynamics was built through chaos. Numerical and theoretical work in the 1990s showed that dynamical indicators such as Lyapunov exponents often display pronounced changes across energies where thermodynamic anomalies appear. This observation was crystallized into a geometric program in which Hamiltonian flows can be reinterpreted as geodesic flows on manifolds endowed with Jacobi/Maupertuis-type metrics, and dynamical instability can be related to curvature fluctuations and parametric instability mechanisms \cite{CASETTI2000237,pettini2007geometry,Firpo1998,pettini1993geometrical,casetti1996riemannian,cerruti1996geometric}. In mean-field models undergoing a continuous transition, this framework yields quantitative estimates for the largest Lyapunov exponent and connects its scaling to thermodynamic criticality \cite{Firpo1998,campa2009statistical}. The lasting legacy of this line is methodological: it puts geometry at the level of microscopic dynamics and demonstrates that curvature-related objects can carry thermodynamic information, even when one does not start from a canonical potential.

A conceptually complementary milestone is due to Rugh, who proved that microcanonical thermodynamic derivatives can be written as phase-space averages of divergences of suitable vector fields restricted to the constant-energy hypersurface \cite{Rugh1997,Rugh1998,Rugh2001}. This result is important for two reasons. First, it makes temperature and response functions \emph{intrinsically microcanonical} objects, without using canonical identities. Second, it gives them a manifest geometric/differential form, thereby legitimizing the idea that macroscopic observables may be recovered from geometric data of energy shells. Subsequent work clarified subtleties that become central in finite systems: microcanonical entropy (and its derivatives) may exhibit non-analyticities already at finite $N$, and the inclusion of kinetic energy can modify the differentiability class of the entropy, which matters when one interprets sharp features in derivatives as transition-like signals \cite{CasettiKastner2006,CasettiKastnerNerattini2009}. This is precisely the regime where a purely geometric reconstruction is nontrivial: it must be predictive but also conservative, i.e., it must not manufacture spurious singularities when only finite-size structure is present.

In parallel, a second major geometric direction focuses on the \emph{topology} of configuration space and energy landscapes. The ``topological hypothesis'' posits that phase transitions are accompanied by topology changes of suitable families of configuration-space submanifolds, often formulated in terms of sublevel sets $M_v=\{q:V(q)\le v\}$ and level sets $V^{-1}(v)$, with a Morse-theoretic organization in terms of critical points of the potential \cite{FranzosiPettiniSpinelli2000,FranzosiPettini2007,pettini2019origin,PhysRevLett.79.4361}. A large body of work then explored how invariants such as the Euler characteristic, Betti numbers, and critical-point indices behave across transition energies, including explicit demonstrations in mean-field-like and glassy-inspired models \cite{AngelaniEtAl2003,angelani2005topology}. More recently, this perspective has been sharpened by constructing minimal or analytically tractable models in which symmetry breaking transitions can be traced back to global geometric features of equipotential manifolds (e.g., ``dumbbell-shaped'' level sets), by deriving criteria that are framed as necessary and sufficient conditions for $\mathbb{Z}_2$ symmetry breaking, and by carrying out explicit Morse-theoretic analyzes of the mean-field $\phi^4$ model and related topological toy models \cite{Baroni2011,Baroni2019,BaroniPRE2020,BaroniEPJB2020,BaroniJSTAT2020}. At the same time, rigorous and computational counterexamples sharpened the scope of the hypothesis: there exist phase transitions whose transition energy is not associated with any stationary point of the potential energy landscape in a straightforward sense, showing that the identification ``phase transition $\Leftrightarrow$ stationary point at $v_c$'' may not be universal \cite{Kastner2008,KastnerMehta2011,MehtaHauensteinKastner2012}. The modern reading of this literature is therefore nuanced: topology changes can be necessary in broad classes and can provide powerful diagnostics, but topology alone can typically be insufficient unless it is coupled with measure concentration and the way the microcanonical weight distributes over the manifolds as $N$ grows.

A third family of geometric approaches works at the level of macrostates rather than microscopic manifolds: thermodynamic geometry and information geometry introduce Riemannian metrics on the space of equilibrium states (or probability models), with curvature often interpreted as a measure of correlations and sometimes diverging at criticality. Classical constructions include the Weinhold and Ruppeiner metrics, and covariant reformulations such as geometrothermodynamics \cite{Weinhold1975,Ruppeiner1995,Quevedo2007}. These frameworks are valuable because they reveal universal geometric patterns of criticality and can be applied across disparate systems (including black holes), but they are logically downstream: they assume that an entropy or thermodynamic potential is already available. For a program that aims to \emph{derive} the entropy from microscopic geometry, thermodynamic geometry plays a complementary role, serving as a target structure for consistency checks rather than as a starting point.

For finite systems, an additional methodological pillar is the development of systematic microcanonical diagnostics that remain meaningful without invoking the thermodynamic limit. This work was initiated by D.H.E. Gross et al. \cite{gross2000phase,gross2005microcanonical,gross2001microcanonical,chomaz2002phase,pathria1983phase,gulminelli1999critical,chomaz1999energy,matty2017comparison,hilbert2014thermodynamic} who investigated physical systems that often lack a well-defined thermodynamic limit but exhibit phase transition-like behavior or ensemble inequivalence. In particular, Gross proposed the microcanonical analysis, where phase transitions are identified with the presence of convex regions of the microcanonical entropy. Only recently, this line of research has been generalized by Bachmann et al. reaching a final formulation known as the microcanonical inflection-point analysis (MIPA). This analysis classifies phase transitions via well-defined patterns in the derivatives of the entropy and provides an operational taxonomy for any system size~\cite{schnabel2011microcanonical,bachmann2014novel,qi2018classification,koci2017subphase,koci2015confinement,sitarachu2020exact,sitarachu2022evidence,sitarachu2020phase}. Related discussions emphasize that microcanonical and canonical descriptions can differ sharply in small systems and near first-order-like coexistence regions, where negative heat capacities and backbending phenomena can occur \cite{dunkel2006phase}. This literature establishes a clear benchmark for any geometric theory: if geometry is to be taken as primary, it must reproduce not only asymptotic singularities but also the finite-$N$ ``shapes'' that MIPA formalizes.

Finally, long-range interacting systems provide both a motivation and a stress test for geometric microcanonical approaches. Nonadditivity leads to ensemble inequivalence, negative heat capacities, and nonconcave entropies, making canonical-based reasoning unreliable precisely where the physics is richest \cite{barre2001inequivalence,campa2009statistical,bouchet2010thermodynamics}. In the quantum domain, renewed experimental access to power-law interactions has triggered a modern classification of long-range universality and dynamical regimes \cite{defenu2023long}. In this setting, approaches that (i) operate directly in the microcanonical ensemble, (ii) control finite-size behavior, and (iii) connect thermodynamic response to intrinsic geometric observables of energy shells and/or configuration manifolds are positioned to provide genuinely new predictive tools rather than post-hoc geometric interpretations.

Taken together, these strands show a consistent evolution: from early dynamical signatures of criticality to geometric identities for microcanonical derivatives, to topological and landscape-based criteria, to macrostate geometries, and to finite-size/long-range microcanonical methodologies. The remaining gap that motivates current research is a \emph{closed pipeline} from microscopic geometry to thermodynamics: a scheme in which geometric properties of $\Sigma_E$ determine the entropy and its derivatives in a controlled way, reproduce finite-size ``shapes'' in the sense of MIPA, and remain valid in regimes where canonical assumptions fail. The present work continues the program already initiated in Ref.~\cite{di2021topology,di2022geometrictheory} and is designed to address this gap by formulating thermodynamic observables as emergent from purely geometric ensemble averages and by validating the reconstruction across paradigmatic short- and long-range models.

\section{Symplectic Structure and the Thermodynamic Question}

\subsection{The fundamental question}

The central problem of statistical mechanics and thermodynamics is to determine how many states are accessible to a physical system under specified macroscopic conditions, such as fixed energy $E$, volume $V$, and particle number $N$. This counting, encoded in the volume of the accessible region of phase space, is the foundation from which all thermodynamic state variables emerge. The entropy $S(E,V,N)$ is the logarithm of this volume, and from it follow temperature, pressure, and the entire thermodynamic framework.

The question is geometric at its core: given the constraints imposed by energy conservation: \\

\textit{What is the measure---the size---of the set of accessible microstates? How can we measure it?}\\

Answering this question allows us to connect microscopic dynamics to macroscopic thermodynamics.

In what follows, we examine whether the Hamiltonian structure alone, specifically the symplectic geometry that governs classical dynamics, is sufficient to answer this question. We will find that it is not, and that the obstruction reveals the necessity of introducing a metric tensor.

\subsection{The symplectic framework}

Hamiltonian mechanics is formulated on a phase space $(\Lambda, \omega)$, where $\Lambda$ is a $2N$-dimensional manifold (with $N$ degrees of freedom) and $\omega$ is the canonical symplectic form. In canonical coordinates $\bm  x:=(q^i, p_i)_{i\in[1,N]}$, the symplectic form reads
\begin{equation}
    \omega = dp_i \wedge dq^i
\end{equation}
where we adopt the Einstein summation convention. The Hamiltonian $H : \Lambda \to \mathbb{R}$ generates a vector field $\bm X_H$, the Hamiltonian vector field, defined implicitly by the equation
\begin{equation}
    \iota_{\bm X_H} \omega = dH,
\end{equation}
where $\iota$ denotes interior contraction. In coordinates, this yields Hamilton's equations:
\begin{equation}
\dot{q}^i = \frac{\partial H}{\partial p_i}, \quad \dot{p}_i = -\frac{\partial H}{\partial q^i}.
\end{equation}

The symplectic structure has two fundamental properties. First, it is preserved by the Hamiltonian flow: $\mathscr{L}_{\bm X_H} \omega = 0$, where $\mathscr{L}$ denotes the Lie derivative~\cite{arnol2013mathematical,petersen2006riemannian}. Second, it generates a natural volume form on the entire phase space $\Lambda$ through
\begin{equation}
    d\mu_\Lambda = \frac{\omega^N}{N!}.
\end{equation}
This is the Liouville measure, which is also preserved by Hamiltonian dynamics: $\mathscr{L}_{\bm X_H} d\mu_\Lambda = 0$ (Liouville's theorem). The symplectic form thus provides both the equations of motion and a volume measure on the full phase space.

\subsection{Conservation laws and the energy shell}

The symplectic structure encodes not only the equations of motion but also the conservation laws that constrain the system's evolution (see App.~\ref{sec:details-symplectic-formalism} for further details). Any smooth function $f : \Lambda \to \mathbb{R}$ defines a unique Hamiltonian vector field $\bm X_f$ through
\begin{equation}
\iota_{\bm X_f} \omega = df.
\end{equation}
Given two smooth functions $f$ and $g$, the symplectic form yields the Poisson bracket
\begin{equation}
\{f, g\} := \omega(\bm X_f, \bm X_g) = \bm X_g(f) = -\bm X_f(g).
\end{equation}
Along the Hamiltonian flow generated by $\bm X_H$, the time derivative of any observable $f$ is
\begin{equation}\label{def:conservation-law}
\dot{f} = dH(\bm X_f) = \omega(\bm X_H, \bm X_f) = \{f, H\}.
\end{equation}
A function $f$ is a \textit{constant of motion} if and only if $\{f, H\} = 0$, in which case $\bm X_f$ generates a symmetry of the Hamiltonian flow.

In particular, energy itself is conserved:
\begin{equation}
\dot{H} = dH(\bm X_H) = \omega(\bm X_H, \bm X_H) = 0.
\end{equation}
The Hamiltonian vector field is tangent to the level sets of $H$. For a given energy $E$, the accessible region of phase space is the energy shell
\begin{equation}
\Sigma_E = \{\bm x \in \Lambda : H(x) = E\}.
\end{equation}
This is a $(2N-1)$-dimensional hypersurface. The system cannot leave $\Sigma_E$: its microstate evolves within this submanifold, tracing out a trajectory determined by $\bm X_H$.

The symplectic structure identifies $\Sigma_E$ perfectly as the locus of motion. The fundamental thermodynamic questions now become geometric: 
\begin{itemize}
    \item How do we define the size of $\Sigma_E$? 
    \item How do we measure the size of this hypersurface using the structures at our disposal?
\end{itemize}

\subsection{The obstruction: why the symplectic structure cannot measure $\Sigma_E$}

To measure a volume, we need a volume form, a nowhere-vanishing differential form of appropriate degree. For the $(2N-1)$-dimensional manifold $\Sigma_E$, we need a $(2N-1)$-form. However, the symplectic structure $\omega$ does not provide this due to three main obstructions:

\textit{First, dimensional incompatibility.} The symplectic form $\omega$ is a $2$-form. Its powers are: $\omega$ (a $2$-form), $\omega^2$ (a $4$-form), $\omega^{N-1}$ (a $(2N-2)$-form), and $\omega^N$ (a $2N$-form). No power of $\omega$ yields a $(2N-1)$-form. The symplectic structure naturally produces forms of even degree, while $\Sigma_E$ requires a form of odd degree. This is not a technical inconvenience but a fundamental mismatch of parity.

\textit{Second, non-uniqueness of transverse projection.} One might attempt to contract the $2N$-form $d\mu_\Lambda$ with a vector field $\bm\xi_\eta$ to obtain a $(2N-1)$-form:
\begin{equation}
    d\sigma_{\bm\xi_\eta} = \iota_{\bm\xi_\eta} d\mu_\Lambda \big|_{\Sigma_E}.
\end{equation}
For this to define a measure on $\Sigma_E$, the vector $\bm\xi_\eta$ must be transverse to $\Sigma_E$, satisfying $dH(\bm\xi_\eta) \neq 0$. However, this single scalar condition does not determine $\bm\xi_\eta$ uniquely. In a $2N$-dimensional tangent space, the equation $dH(\bm\xi_\eta) = a$ (fixing the normalization through some constant $a\in\mathbb{R}$) defines a $(2N-1)$-dimensional affine hyperplane of solutions. There is no canonical choice of $\bm\xi_\eta$ provided by the symplectic structure alone. Different choices yield different measures on $\Sigma_E$, with no fundamental principle to prefer one over another. Indeed, it is well-known that orthogonality is not a symplectic concept. The symplectic form $\omega$ defines a notion of ``symplectic orthogonality'' $\omega(\bm u,\bm v) = 0$, but this is fundamentally different from the geometric orthogonality required to define a normal (see App.~\ref{ssec:non-unique-transverse}).

\textit{Third, coisotropy and degeneracy.} The deepest obstruction is that $\Sigma_E$ is a coisotropic submanifold with respect to $\omega$. The symplectic complement of its tangent space,
\begin{equation*}
(T_{\bm x} \Sigma_E)^{\perp_\omega} = \{\bm w \in T_{\bm x} \Lambda : \omega(\bm w,\bm v) = 0 \; \forall \bm v \in T_{\bm x} \Sigma_E\},
\end{equation*}
is one-dimensional and generated by $\bm X_H$. But $\bm X_H$ is tangent to $\Sigma_E$, not transverse. The symplectic structure provides no natural transverse direction. Equivalently, the restriction $\omega|_{\Sigma_E}$ is degenerate: the Hamiltonian vector field $\bm X_H$ lies in its kernel, since $\omega(\bm X_H, \bm v) = dH(\bm v) = 0$ for all $\bm v \in T\Sigma_E$. A degenerate form cannot be used to construct a volume form (see App.~\ref{ssec:degeneracy} and \ref{ssec:coisotropy-complement}).

These three obstructions—dimensional, non-uniqueness, and degeneracy—are independent manifestations of a single fact: the symplectic structure does not provide, alone, a canonical way to induce a measure on coisotropic submanifolds. To measure $\Sigma_E$, we need additional structure.

\subsection{The statistical mechanics resolution: postulating the ensemble}

Historically, statistical mechanics resolves the problem of weighting accessible states by introducing an external structure: the ensembles. They follow from the observation that any function of the Hamiltonian $\rho(H)$ is a constant of motion, which suggests defining statistical weights as functions of $H$. 
In fact, if we pose $f = \rho(H)$, then $df = \rho'(H)\,dH$ and thus $\iota_{\bm X_\rho}\omega = \rho'(H)\,\iota_{\bm X_H}\omega$, implying $\bm X_\rho = \rho'(H) \bm X_H$. Therefore, Eq.~\eqref{def:conservation-law} yields
\begin{equation}
    \{f, H\} = \omega(\bm  X_\rho, \bm  X_H) = \rho'(H)\,\omega(\bm X_H, \bm  X_H) = 0,
\end{equation}
where the last equality follows from the antisymmetry of $\omega$. Any function of the Hamiltonian is conserved along the Hamiltonian flow.
This leads to the introduction of ensemble measures of the form
\begin{equation}
    d\mu_\rho = \rho(H)\,d\mu_\Lambda,
\end{equation}
where $\rho(H)$ is a weight function. The two fundamental ensembles are the microcanonical, with $\rho_{\text{mc}} = \delta(H - E)$, and the canonical, with $\rho_{\text{can}} = e^{-\beta H}$.

The microcanonical ensemble postulates that all states at energy $E$ are equally weighted. The density of states is then defined as
\begin{equation}
\Omega(E) = \int_\Lambda \delta(H(\bm x) - E)\,d\mu_\Lambda(\bm x).
\end{equation}
This integral, when evaluated using the properties of the Dirac delta distribution, effectively counts the ``size'' of the energy shell $\Sigma_E$. The microcanonical entropy follows as a postulate: $S(E) = \ln \Omega(E)$~\cite{pathria2017statistical}.

\section{The Hamiltonian geometric foundation of thermodynamics}

The ensemble postulate is successful; it reproduces thermodynamics and agrees with experiments. Yet, it is introduced as an independent statistical ingredient added to the Hamiltonian structure from outside. The ensemble is not derived from the symplectic structure but is assumed alongside it. This raises a conceptual question: 
\begin{quote}
Can the statistical weight of accessible states be internalized within a natural framework rather than postulated?
\end{quote}

This is the guiding insight that justifies the foundation of the geometric framework.

\subsection{The necessity of a metric tensor}

The answer to the previous question requires recognizing that the ensemble postulate, while operationally successful, addresses a deeper geometric necessity. The threefold obstruction identified above is not merely a difficulty---it reveals that the concept of ``counting accessible states'' itself presupposes geometric structures that the symplectic form cannot provide: a notion of measure, volume, and orthogonality. The metric tensor is not an arbitrary addition, but the minimal structure required to make thermodynamic questions well-posed.
We are then guided by a central recognition: \\

\textit{The thermodynamic content of a Hamiltonian system---specifically, the possibility of defining the set of accessible states---presupposes the existence of a measure on phase space. This measure, in turn, requires a geometric structure (metric tensor) compatible with the symplectic form.}\\

The introduction of a metric tensor is not a postulate but represents a \textit{recognition} of what is already implicit in the Hamiltonian structure. Thermodynamic quantities presuppose a measure, which presupposes a metric.
Our aim now is to make this recognition explicit and constructive, and this act represents the foundation of the geometric formulation of thermodynamics.
The metric tensor, $\eta$, (that we specify soon) is then introduced as an axiom alongside the symplectic structure $(\Lambda,H,\omega)$.

\subsubsection{Resolving the obstruction: what is needed to measure hypersurfaces}

Our scope now is to show how to answer the original thermodynamic question: \textit{how large is $\Sigma_E$?} To do so requires resolving the obstructions identified above, but two essential ingredients are fundamentally missing to measure hypersurfaces:
\paragraph*{1) A canonical direction identified by a vector, say $\bm\xi_\eta$, ``transverse'' (orthogonal) to $\Sigma_E$.} As already discussed, orthogonality is not a symplectic concept. It necessarily requires a \textit{scalar product}: a symmetric, positive-definite bilinear form. This is precisely what a metric tensor $\eta$ provides: it defines the concept of \textit{normal vector}, namely, a direction that is orthogonal to every tangent vector of $\Sigma_E$.
\paragraph*{2) A measure induced on $\Sigma_E$ by $\bm\xi_\eta$.} A metric tensor on $\Lambda$ naturally induces a metric on any $\Sigma_E$. Denoting with $d\mu_E$ the induced metric on $\Sigma_E$, this can be obtained by $\iota_{\bm\xi_\eta}d\mu_{\Lambda}=d\mu_E$ and it determines an intrinsic volume form which is coordinate-independent and measures the geometric area of the hypersurface.

Then, the metric tensor does not add arbitrary structure; it provides the minimal geometric framework needed to distinguish normal from tangent, to induce a measure on $\Sigma_E$, and to relate energy increments to thermodynamic observables. The geometric foundation of thermodynamics relies on these essential ingredients, which we rigorously develop below.

\subsection{Orthogonal direction to the energy shells}

The natural metric tensor that can be considered on phase space is the Euclidean one that in canonical coordinates $(q^i, p_i)$ reads:
\begin{equation}\label{eq:euclidean-metric}
    \eta = \delta^{ij} dp_i \otimes dp_j + \delta_{ij} dq^i \otimes dq^j.
\end{equation}
It defines a scalar product on each tangent space $T_{\bm{x}}\Lambda$; i.e., for any pair of vectors $v,\,w\in T_{\bm{x}}\Lambda$, we have
\begin{equation}
    \eta(v,w)=\langle v,w\rangle\,.
\end{equation}
This is the simplest Riemannian structure compatible with the canonical symplectic form $\omega$. Requesting the compatibility for the pair $(\omega,\eta)$ implies the existence of a (1,1) tensor $J$ (a matrix) such that $J^2=-\mathrm{Id}$ and $\eta(Ju,Jv)=\eta(u,v)$, which yields
\begin{equation}\label{def:omega-eta-compatibility}
    \omega(u,v)=\eta(Ju,v)\,.
\end{equation}
Additionally, by construction, the presence of a metric induces isomorphisms between the tangent $T_{\bm x}\Lambda$ and cotangent $T^*_{\bm x}\Lambda$ spaces of the phase space, i.e., a one-to-one correspondence between vectors in $T_{\bm x}\Lambda$ and covectors in $T^*_{\bm x}\Lambda$. 

This correspondence is crucial and yields the first important consequence. The differential $dH$---a one-form---can be mapped to a vector field, namely, the gradient of $H$, denoted by $\nabla_{\!\eta} H$, and is defined implicitly by the condition
\begin{equation}\label{def:grad}
    \eta(\nabla_{\!\eta} H,\,Y)=dH(Y)
    \qquad \forall\, Y\in T_{\bm x}\Lambda.
\end{equation}
Note that the gradient depends on the metric tensor; for that reason, we use the notation $\nabla_{\!\eta}$. Usually, since the metric tensor is defined once and for all, one drops such a subscript; here, instead, we need to keep track of this information for future purposes, so we will maintain it in our notation.  

These properties of the metric tensor allow us to determine the orthogonal vector to $\Sigma_E$. In fact, setting $Y=\bm{X}_H$ in Eq.~\eqref{def:grad}, we have
\begin{equation}
    \label{def:eta-pmega-relation}
    \eta(\nabla_{\!\eta} H,\bm{X}_H)=dH(\bm{X}_H)=\omega(\bm{X}_H,\bm{X}_H)\,,
\end{equation}
By construction, the third term vanishes $\omega(\bm{X}_H,\bm{X}_H)=0$, the second equality indeed corresponds to the energy conservation law:
\begin{equation}\label{eqn:energy-conservation}
        \dot H = dH(\bm{X}_H)=\omega(\bm{X}_H,\bm{X}_H)=0.
\end{equation}
which translates into an orthogonality condition $\bm{X}_H\perp\nabla_{\!\eta} H$; indeed: 
\begin{equation}\label{eq:orth}
    dH(\bm{X}_H)=\eta(\nabla_{\!\eta} H,\bm{X}_H)=0
    \quad\Rightarrow\quad
    \langle \nabla_{\!\eta} H,\bm{X}_H\rangle_{\!\eta}=0,
\end{equation}
The gradient points normal to the energy shells; the Hamiltonian flow (identified locally by $\bm{X}_H$) moves tangentially to them.

\begin{figure}
    \centering
    \includegraphics[width=0.8\linewidth]{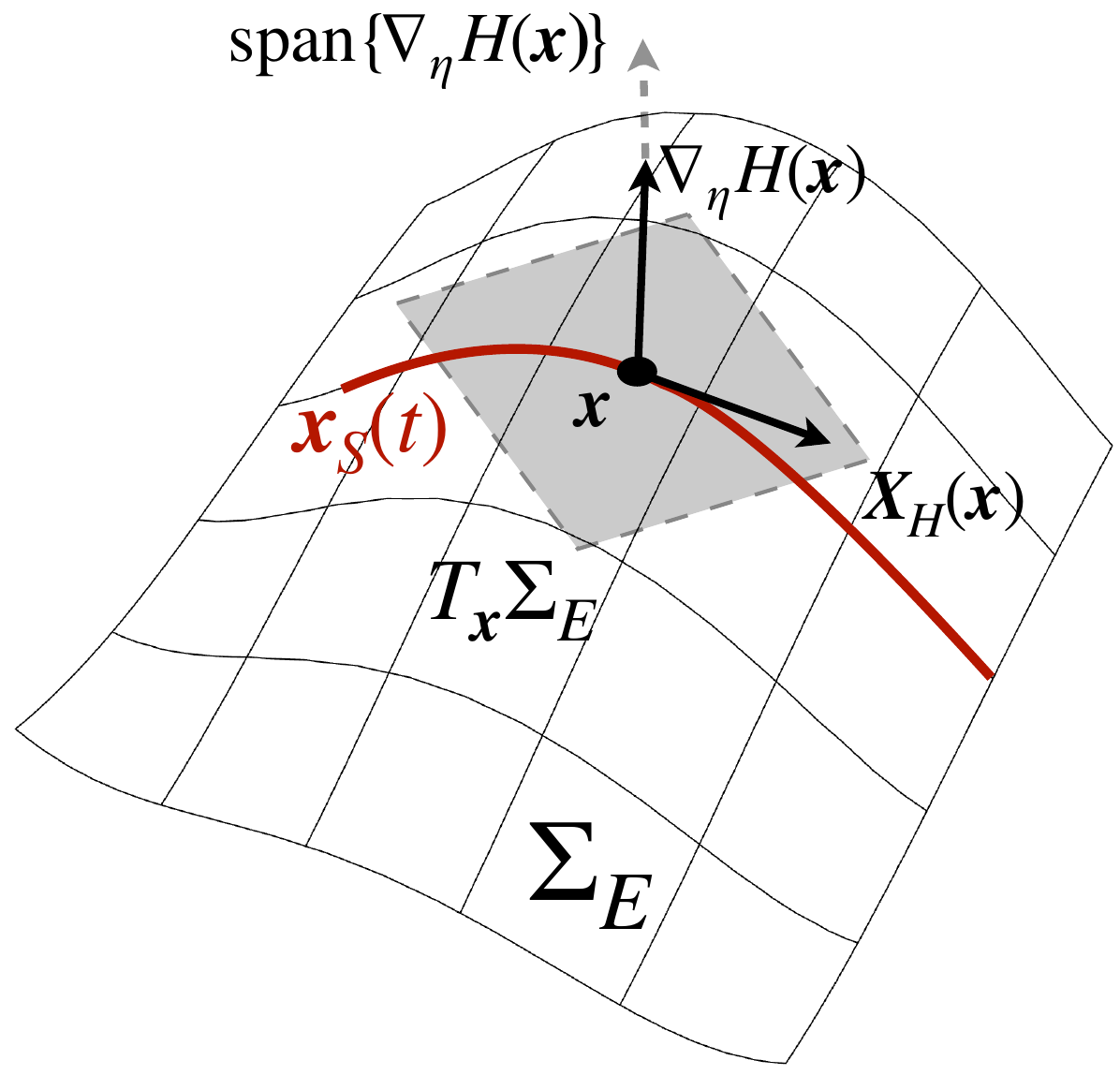}
    \caption{\textbf{Pictorial representation of the orthogonality relation $\nabla_\eta H\perp\bm{X}_H$}. The Hamiltonian flow represented by the red curve $\bm{x}_S(t)$ follows the direction of the Hamiltonian vector field $\bm{X}_H$ which lies in turn on the energy level set $\Sigma_E$ at each time.  }
    \label{fig:orthonormality}
\end{figure}

This orthogonality thus induces a canonical (natural) decomposition at each point $\bm x \in \Lambda$:
\begin{equation}\label{eq:tangent-decomp-nabla}
    T_{\bm x} \Lambda = \mathrm{span}\{\nabla_{\!\eta} H(\bm x)\} \oplus T_x \Sigma_E.
\end{equation}
The gradient $\nabla_{\!\eta} H$ provides the unique transverse direction, while $T_{\bm x} \Sigma_E$ is the $(2N-1)$-dimensional subspace tangent to the energy shell (see Fig.~\ref{fig:orthonormality}). 

\subsubsection{The energy flow and clock}

Intuitively, the gradient of $H$ indicates the direction in which energy increases. 
Thermodynamics naturally parameterizes the fundamental quantities, such as the density of states and entropy, as functions of energy. Indeed, thermodynamic observables follow from variations (differentiation) of entropy with respect to the energy variables, for instance, the inverse temperature $\beta:=\partial_E S(E)$, etc.

Therefore, for physical reasons, we need to determine the energy flow; that is, the flow of diffeomorphism, which is a mapping between hypersurfaces $\Sigma_E\to\Sigma_{E+d E}$ under the variation of energy $E\to E+dE$. The main purpose is to determine the vector field that generates such a flow. One could be tempted to identify this vector field with $\nabla_\eta H$. However, $\nabla_\eta H$ only identifies the direction of increasing values of energy; but it does not move each point on $\Sigma_E$ onto a point in $\Sigma_{E+dE}$ under the variation $E\to E+dE$.

This can be easily seen as follows. Let us consider the flow generated by $\nabla_{\!\eta} H$, namely, the map $\bm{Y}(\epsilon)=\Phi(\epsilon,\bm{x}_0)$ with $\bm{x}_0\in\Sigma_{E_0}$ such that: 
\begin{itemize}
    \item $\bm{Y}(0)=\Phi(0,\bm{x}_0)=\bm{x}_0$,
    \item it is generated by $\nabla_{\!\eta} H$, i.e.:
\[
    \frac{d\bm{x}}{d\epsilon}(\epsilon)=\frac{\partial\Phi}{\partial\epsilon}(\bm{x}_0,\epsilon)=\nabla_{\!\eta} H(\bm{x}(\epsilon))
\]
\end{itemize}
In infinitesimal form, we have $\bm{x}(\epsilon)=\bm{x}_0+\nabla_{\!\eta} H(\bm{x}_0)\,d\epsilon$ then
\begin{equation}\label{def:energy-step}
    \begin{split}
    E=H(\bm{x}_0&+\nabla_{\!\eta} H(\bm{x}_0)\,d\epsilon )\\
    &=H(\bm{x}_0)+\nabla_{\!\eta} H(\bm{x}_0)\cdot \nabla_{\!\eta} H(\bm{x}_0)\,d\epsilon\\
    &=E_0+\|\nabla_{\!\eta} H(\bm{x}_0)\|_{\!\eta}^2\;d\epsilon
\end{split}
\end{equation}
Now, posing $dE:=E-E_0$ implies that
\begin{equation}\label{def:relation-dE-depsilon}
        dE=\|\nabla_{\!\eta} H(\bm{x}_0)\|_{\!\eta}^2\;d\epsilon
\end{equation}
In other words, the energy step, $dE$, that selects the target energy manifold $\Sigma_{E_0+dE}$ depends on the point $\bm{x}_0\in\Sigma_{E_0}$. Now, since $\epsilon$ is the parameter, we have the freedom to fix it, say, $d\epsilon=1$, rather than $dE_{\bm{x}}=\|\nabla_{\!\eta} H(\bm{x})\|_{\!\eta}^2$.
In general, for two points $\bm{x}\neq\bm{y}$ on $\Sigma_{E_0}$, we have $\|\nabla_{\!\eta} H(\bm{x})\|_{\!\eta}\neq\|\nabla_{\!\eta} H(\bm{y})\|_{\!\eta}$, and therefore $dE_{\bm{x}}\neq dE_{\bm{y}}$ yields $E_{\bm{x}}\neq E_{\bm{y}}$. Then, two points on the same $\Sigma_E$ are mapped onto different hypersurfaces.\\

Our scope is to determine the vector that generates the energy flow. Then, we can modify the vector $\nabla_{\!\eta} H$ to introduce a vector field $\bm\xi_\eta\in\text{span}\{\nabla_{\!\eta} H\}$ such that the parametric step coincides with the energy step, namely, $d\epsilon=dE$.

\paragraph*{Determination of the generator of energy motion.} 

We can choose a vector field parallel to $\nabla_{\!\eta} H$ such that $\bm\xi_\eta^\alpha(\bm{x}):=\alpha(\bm{x})\nabla_{\!\eta} H(\bm{x})$ with $\alpha>0$ everywhere. We then determine the function $\alpha$ so that we obtain a parametrization for the flow such that $d\epsilon=dE$. We will see later that this choice does not affect the geometry of the energy hypersurfaces, which remain unchanged. \\
Let us define a motion $\bm{x}(\epsilon):=\Phi^{\alpha}(\bm{x}_0,\epsilon)$, such that differentiating with respect to $\epsilon$ yields
\[
    \frac{d\bm{x}}{d\epsilon}=\frac{\partial\Phi^\alpha}{\partial \epsilon}(\bm{x}_0,\epsilon)=:\bm{\xi}_\eta^\alpha(\bm{x}(\epsilon))\,.
\]
In order to determine $\alpha$, we consider the infinitesimal displacement
\begin{equation}
    \label{eqn:infinitesimal-displacement}
    \bm{x}(\epsilon)=\bm{x}_0+\bm\xi_\eta^{\alpha}(\bm{x}_0)\,d\epsilon
\end{equation}
together with
\begin{equation}\label{def:energetic-step}
        dE=\nabla_{\!\eta} H\cdot\bm\xi_\eta^\alpha\;d\epsilon= dH(\bm{\xi}_\eta^\alpha)\,d\epsilon
\end{equation}
The condition $dE=d\epsilon$ follows from above by imposing
\[
    dH(\bm\xi_\eta^\alpha)=1
\]
which gives
\[
    \alpha\langle\nabla_{\!\eta} H,\nabla_{\!\eta} H\rangle_{\!\eta}=1\implies \alpha=\frac{1}{\|\nabla_{\!\eta} H\|_\eta^2}
\]
In conclusion, the energy flow is generated by the vector field 
\[
    \bm\xi_\eta^{\alpha}:=\frac{\nabla_{\!\eta} H}{\|\nabla_{\!\eta} H\|_{\!\eta}^2}
\]
whose norm is not unitary but $\|\bm\xi_\eta^\alpha\|=1/\|\nabla_{\!\eta} H\|_{\!\eta}$. From now on, we drop the superscript and write $\bm\xi_\eta\equiv\bm\xi_\eta^\alpha$.\\

\paragraph*{Invariance of the hypersurfaces' geometry.} Let us now check that the introduction of $\bm\xi_\eta$ with a general $\alpha$ does not modify the geometry of hypersurfaces. 

Let us proceed by considering the infinitesimal displacement
\begin{equation}\label{def:geometric-step}
        d\bm{x}:=\bm{x}(\epsilon)-\bm{x}_0=\bm\xi_\eta^\alpha(\bm{x}_0)\,d\epsilon=\alpha(\bm{x}_0)\nabla_{\!\eta} H\;d\epsilon.
\end{equation}
Then, the geometric displacement $d\ell:=\|\bm{x}(\epsilon)-\bm{x}_0\|_{\!\eta}$, between two points under the flow $\Phi^{\alpha}$ gives
\begin{equation}\label{def:geometric-displacement}
    d\ell=\|\bm\xi_\eta^\alpha\|_{\eta}\;d\epsilon=\alpha\|\nabla_{\!\eta} H\|_{\!\eta}\,d\epsilon
\end{equation}
that, combined with Eq.~\eqref{def:energetic-step}, gives:
\begin{equation}
\begin{split}
    dE=\alpha(\bm{x})\|\nabla_{\!\eta} H\|_{\!\eta}^2\,d\epsilon=&\|\nabla_{\!\eta} H\|_{\eta}\,d\ell\\
    &\implies \frac{dE}{d\ell}=\|\nabla_{\!\eta} H\|_{\eta}\ . 
\end{split}
\end{equation}
In other words, the change of energy (Eq.~\eqref{def:energetic-step}) or geometric displacement (Eq.~\eqref{def:geometric-displacement}) with respect to $\epsilon$ (which is the chosen parametrization) is affected by $\alpha$. However, the fundamental physical content is encoded in how energy changes with respect to geometric displacement, and this is independent of $\alpha$. The factor $\alpha$ simply describes how we move from a hypersurface to another, but the invariance of the ratio $dE/d\ell$ represents the intrinsic foliation structure, and it is not affected by our choice of $\alpha$. \\

\paragraph*{Clocks.}
This invariance implies that there are infinitely many choices of $\alpha$. Each choice represents a different \emph{clock}, and we can recognize a few of them:
\begin{equation}
    \text{Clock}:=dH(\bm{\xi}_\eta)
\end{equation}
Convenient choices are provided by 
\begin{enumerate}
    \item \textbf{Direct clock}. $\alpha=1$, then $\bm\xi_\eta^{(\rm dir)}=\nabla_{\!\eta} H$ with $dE=\|\nabla_{\!\eta} H\|_{\!\eta}^2\,d\epsilon$;
    \item \textbf{Geometric clock}. $\alpha=1/\|\nabla_{\!\eta} H\|_{\!\eta}$, then $\bm\xi_\eta^{(\rm inv)}:=\nabla_{\!\eta} H/\|\nabla_{\!\eta} H\|_{\!\eta}$ with $dE=\|\nabla_{\!\eta} H\|_{\eta}\,d\epsilon$;
    \item \textbf{Energetic clock}. $\alpha=1/\|\nabla_{\!\eta} H\|_{\!\eta}^2$, then $\bm\xi_\eta^{(\rm en)}:=\nabla_{\!\eta} H/\|\nabla_{\!\eta} H\|_{\!\eta}^2$ with $dE=d\epsilon$.
\end{enumerate}
Without much ado, it is easy to see that the direct and geometric block does not map all the points of a $\Sigma_E$ onto all points of $\Sigma_{E+dE}$, but in general, they are mapped onto different hypersurfaces since the energy step is not constant $dE$ (see, for instance, Fig.~\ref{fig:geometric-clock} in App.~\ref{sec:geometric-clock}). The resulting flows are not energetic. \\

\begin{figure}[H]
    \centering
    \includegraphics[width=1\linewidth]{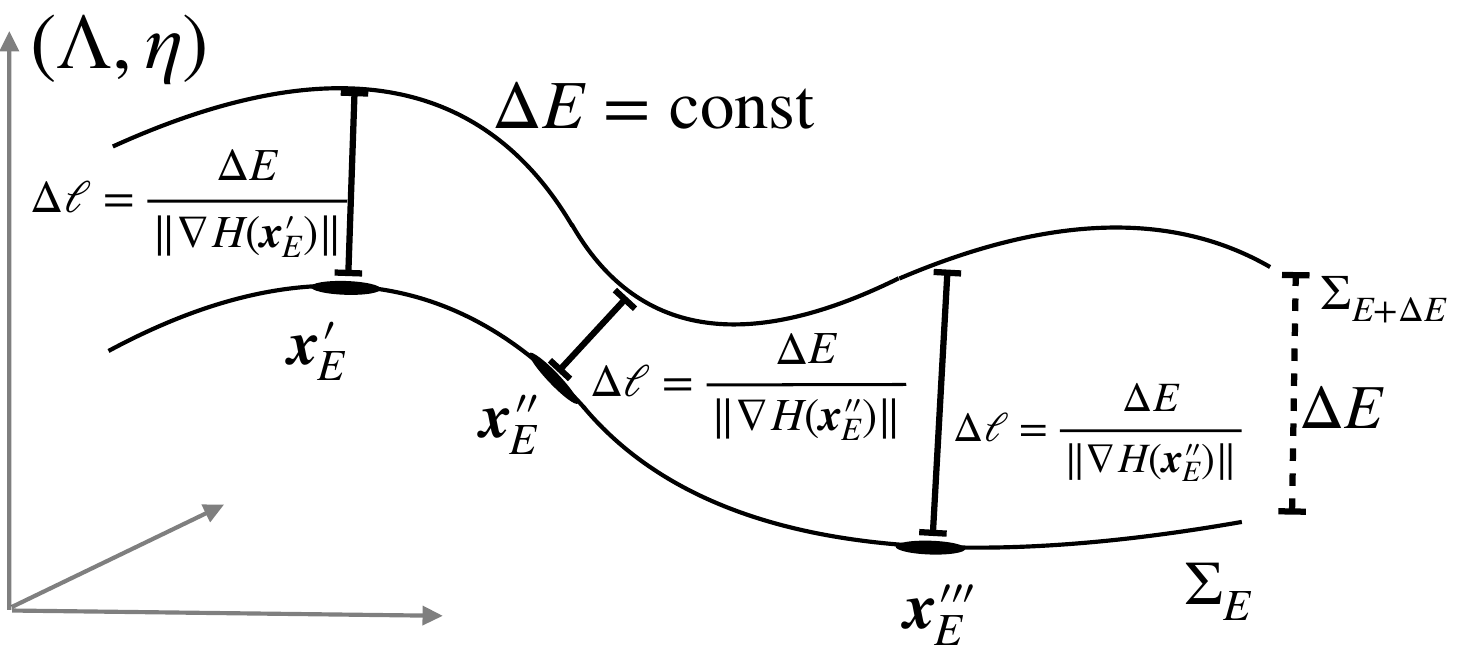}
    \caption{\textbf{Energetic Clock.} The parameter $\epsilon$ directly measures energy: $dE = d\varepsilon$. We have the freedom to fix the energy step, say, $dE=1$, so that the geometric step varies point by point: $d\ell = 1/\|\nabla H\|$. This is an energy flow since we fix the target hypersurface (through the energy step $dE=1$) and vary $d\ell$ point by point in order to map all points of $\Sigma_E$ onto $\Sigma_{E+dE}$.}
    \label{fig:energetic-clock}
\end{figure}

In conclusion, the only way to define an energy flow is thus provided by the energetic clock (see Fig.~\ref{fig:energetic-clock}), which we will denote as $\bm\xi_\eta$ from now on. This clock decomposes the tangent space of $\Lambda$ into
\begin{equation}
    \label{eq:tangent-decomp}
    T_{\bm{x}}\Lambda=\text{span}\{\bm\xi_\eta(\bm x)\}\oplus T_{\bm x}\Sigma_E\,.
\end{equation}
Therefore, the phase space itself is foliated with respect to the parameter $E$ and can be expressed as follows:
\[
    \Lambda=\bigcup_{E}\Sigma_E\,.
\]

\subsection{Thermodynamic dynamics as evolution of energy shells}

\begin{figure*}
    \centering
    \includegraphics[width=0.9\linewidth]{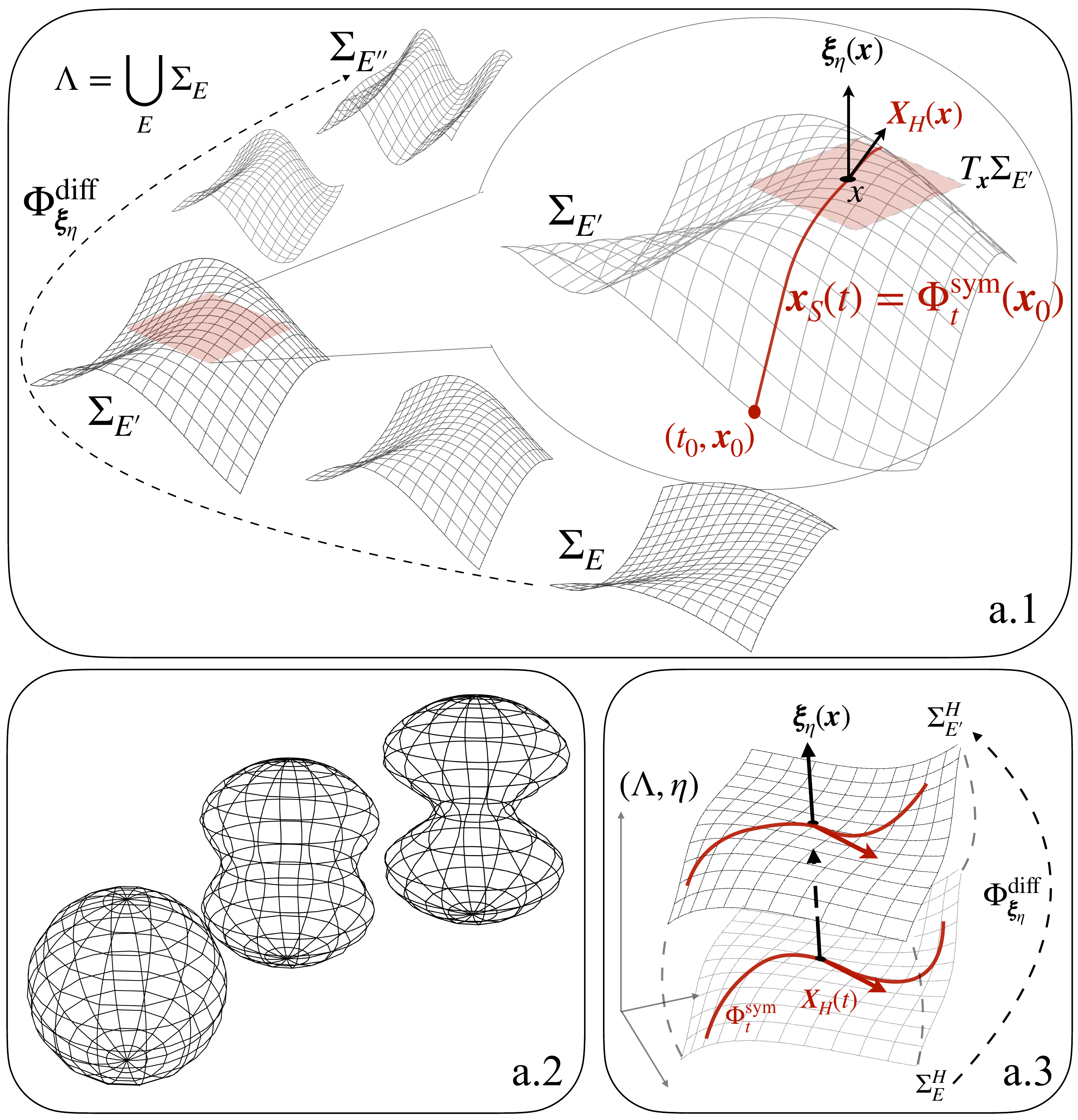}
    \caption{\textbf{Phase-space flows on the energy foliation $\Lambda=\bigcup_E \Sigma_E$.}
(\textbf{a.1}) \emph{Energy flow} $\Phi^{\rm diff}_{\bm\xi_\eta}$ generated by $\bm\xi_\eta=\nabla_{\!\eta} H/\|\nabla_{\!\eta} H\|_\eta^2$ that maps a energy hypersurfaces onto another. Once the initial condition
$\bm x_0$ is fixed, it remains on the selected energy level set $\Sigma_E$, i.e.\ $H(\bm x_0)=E$ for all $t$. Then, on this level set, a \emph{Hamiltonian (symplectic) flow} $\Phi^{\rm sym}_t$ is established: the trajectory
$\bm x_S(t)=\Phi^{\rm sym}_t(\bm x_0)$ is an integral curve of the Hamiltonian vector field $\bm X_H$ and, $H(\bm x_S(t))=H(\bm x_0)$
The introduction of the phase-space metric $\eta$ turns energy conservation into the orthogonality condition
$\eta(\bm X_H,\bm\xi_\eta)=0$.
(\textbf{a.2}) \emph{Pictorial deformation of the energy hypersurfaces} as $E$ varies (e.g., round $\to$ distorted $\to$ necked),
illustrating the qualitative geometric change that is captured by the energy-derivatives of geometric observables.
(\textbf{a.3}) \emph{Geometric (diffeomorphic) flow} $\Phi^{\rm diff}_{\bm\xi_\eta}$, transversal to $\Phi^{\rm sym}_t$ and generated by
$\bm\xi_\eta$, which maps one leaf into a neighboring one, $\Sigma_E\mapsto\Sigma_{E'}$,
providing the controlled displacement along the energy direction in $(\Lambda,\eta)$.}
    \label{fig:flows}
\end{figure*}

The geometric structure obtained in Eq.~\eqref{eq:tangent-decomp} provides a direct meaning of thermodynamics in terms of evolving hypersurfaces and how their geometry evolves along the energy direction. The key insight is the separation of two types of dynamics in the phase space, each acting on different geometric directions (Fig.~\ref{fig:flows}).

The first is the \textbf{symplectic Hamiltonian flow}:
\begin{equation}
    \Phi^{\mathrm{sym}}_t : \Sigma_E \to \Sigma_E, \qquad \frac{d\Phi^{\mathrm{sym}}_t}{dt}(\bm{x}(t)) = \bm{X}_H(\bm{x}(t)),
\end{equation}
generated by the Hamiltonian vector field $\bm{X}_H$ defined by $\iota_{\bm{X}_H} \omega = dH$. This flow is tangent to each $\Sigma_E$ and represents the \textit{microscopic dynamics}: it conserves both the energy $H(\bm{x}) = E$ and the Liouville measure $d\mu_\Lambda = \omega^n / n!$. The flow $\Phi^{\mathrm{sym}}_t$ describes how the system evolves over time within a given $\Sigma_E$, following trajectories dictated by Hamilton's equations.

The second is the \textbf{thermodynamic geometric flow}:
\begin{equation}
    \Phi^{\mathrm{diff}}_{\bm{\xi}_\eta} : \Sigma_E \to \Sigma_{E'}, \qquad \frac{d\Phi^{\mathrm{diff}}_{\bm{\xi}_\eta}}{d
    E} = \bm{\xi}_\eta\,,
\end{equation}
defined by the vector field $\bm{\xi}_\eta$ satisfying $dH(\bm{\xi}_\eta) = 1$. This flow acts as the generator of diffeomorphisms between energy hypersurfaces, being transverse to the foliation $\Lambda = \bigcup_E \Sigma_E$. Geometrically, $\bm{\xi}_\eta$ measures how the shape and intrinsic geometry of the energy shells change with energy $E$. This vector generates the \textit{thermodynamic dynamics}: the evolution of geometric quantities—volume, curvature, and entropy—along the energy direction.

The two flows are orthogonal with respect to the metric $\eta$, and together, they define a natural decomposition of the tangent bundle $T\Lambda$, separating \textit{microscopic} motion (Hamiltonian evolution within a leaf) from \textit{macroscopic} or \textit{energy} motion (geometric evolution across leaves).\\

The scope of the upcoming sections is to make this picture concrete: thermodynamics is not imposed as an additional statistical layer on mechanics—it emerges from the intrinsic geometry of the energy foliation. The metric $\eta$ is the minimal structure required to make this emergence rigorous: it manifests the geometric picture where mechanics and thermodynamics coexist as complementary aspects of the same Hamiltonian system.

\subsection{The measure induced on the energy level sets}

The $\bm\xi_\eta$-decomposition of $T_{\bm x}\Sigma_E$ naturally defines adapted coordinates $(E, y^\alpha)$ where $E = H(\bm x)$ parameterizes the energy and $\{y^\alpha\}_{\alpha=1}^{2N-1}$ on $\Sigma_E$. Then, we introduce a local chart $\Psi : (E, \bm{y}) \mapsto \bm{x}$ on $\Sigma_E$ that defines a local frame on $T_{\bm{x}}\Sigma_E$
\begin{equation}\label{def:frame-hypersurfaces}
    \bm{e}_E:=\partial_E \Psi\equiv\bm\xi_\eta,\; \bm{e}_\alpha:=\partial_\alpha \Psi\in T\Sigma_E,\; dH(\bm e_{\alpha})=0.
\end{equation}
The phase-space metric $\eta$, which in canonical coordinates $(q^i,p_i)$, reads
\[
    \eta=\delta_{ij}\,dq^i\!\otimes dq^j+\delta^{ij}\,dp_i\!\otimes dp_j.
\]
in adapted coordinates $(E,y)$, it is rewritten as follows:
\[
\eta_{AB}\;=\;\eta\big(\partial_A \Psi,\partial_B \Psi\big),\qquad A,B\in\{E,\alpha\}.
\]

Then, using $dH(\partial_\alpha x)=0$ and $\bm\xi_\eta=\nabla_{\!\eta} H/\|\nabla_{\!\eta} H\|_\eta^{2}$,
\begin{align}\label{eqn:components-eta-adapted}
    \eta_{E\alpha}&=\eta(\partial_E \Psi,\partial_\alpha \Psi)=\eta(\bm\xi_\eta,\partial_\alpha \Psi)=\frac{dH(\partial_\alpha \Psi)}{\|\nabla_{\!\eta} H\|_\eta^{2}}=0,\nonumber\\
    \eta_{EE}&=\eta(\bm\xi_\eta,\bm\xi_\eta)=\frac{1}{\|\nabla_{\!\eta} H\|_\eta^{2}},\\
    \eta_{\alpha\beta}&=\eta(\partial_\alpha \Psi,\partial_\beta \Psi),\nonumber
\end{align}
The tangential components, $\eta_{\alpha\beta}$, are generally non-vanishing and represent the \textit{induced area measure} on $\Sigma_E$. we obtain the induced metric, which we recall
\[
    \sigma^{\eta}_{\alpha\beta}(E,y):=\eta_{\alpha\beta},
\]
We will come back to this object soon. Therefore, the Euclidean metric in adapted coordinates takes the block–diagonal form
\begin{equation}\label{def:eta-adapted-coordinates}
        \eta\;=\;\frac{dE\otimes dE}{\|\nabla_{\!\eta} H\|_\eta^{2}}\;+\;\sigma^{\eta}_{\alpha\beta}(E,y)\,dy^\alpha\!\otimes dy^\beta\,.
\end{equation}
Then, the Liouville volume element in the adapted frame following from $\bm\xi_\eta$-decomposition is derived from $\det \eta = \chi^2 \det \sigma_E$, with $\chi:=1/\|\nabla_{\!\eta} H\|$, and reads:
\begin{equation*}\label{eq:vol-adapted}
\begin{split}
    d\mu_\Lambda &= \sqrt{\det \eta} \, dE \, dy^1 \wedge \cdots \wedge dy^{2N-1}\\
    &= \chi \sqrt{\det \sigma_E} \, dE \, dy^1 \wedge \cdots \wedge dy^{2N-1} = \frac{d\sigma_E}{\|\nabla_{\!\eta} H\|_\eta} \wedge dE,
\end{split}
\end{equation*}
where $d\sigma_E := \sqrt{\det \sigma_E} \, dy^1 \wedge \cdots \wedge dy^{2N-1}$ is the induced area element on $\Sigma_E$.
 
This leads to the identification of the natural measure induced on $\Sigma_E$ by the energy parametrization:
\begin{equation}
     \label{def:induced-measure}d\mu^\eta_E:=d\mu_{\Lambda}\Big|_{\Sigma_E}=\frac{d\sigma^{\eta}_E}{\|\nabla_{\!\eta} H\|_\eta}.
\end{equation}
Note that the projection of the Liouville measure on $\Sigma_E$ can also be written in terms of the unit normal vector $\bm n:=\nabla_{\!\eta} H/\|\nabla_{\!\eta} H\|$:
\[
    d\mu_E^\eta\equiv\iota_{\bm\xi_\eta}d\mu_\Lambda=\chi\iota_{\bm n}d\mu_\Lambda\,.
\]
Now, the quantity $\iota_{\bm n}d\mu_\Lambda$ coincides with the definition of area measure on $\Sigma_E$ when it is identified with the unit normal vector $\bm{n}$, i.e., $d\sigma_E^\eta:=\iota_{\bm n}d\mu_\Lambda$, and
\[
    \text{area}^{\eta}(\Sigma_E)=\int_{\Sigma_E}\!\!d\sigma_E^\eta=\int_{\Sigma_E}\!\!\!\sqrt{\det \sigma_E^\eta}\,dy^1\wedge\ldots\wedge dy^{2N-1}\,.
\]
Therefore, the physical requirement to consider an energy flow $\Phi^\text{diff}_{\bm\xi_\eta}$ with energy clock $dE(\bm\xi_\eta)=1$ leads to the definition of a measure on $\Sigma_E$ that does not coincide with the area $\text{area}(\Sigma_E)$ but with
\begin{equation}\label{eq:density-of-states}
    \Omega_\eta(E) = \int_{\Sigma_E} d\mu^\eta_E = \int_{\Sigma^\eta_E} \frac{d\sigma^\eta_E}{\|\nabla_{\!\eta} H\|_{\!\eta}}.
\end{equation}
Remarkably, using the coarea formula in the integral above, we obtain
\begin{equation}
\begin{split}\label{eqn:coarea-boltzmann}
    \Omega_\eta(E) &= \int_{\Sigma_E} \frac{d\sigma^\eta_E}{\|\nabla_{\!\eta} H\|_\eta}\\
    &=\int_{\Lambda}\delta(H(\bm{x})-E)d\mu_\Lambda(\bm{x})\equiv\Omega_B(E)\,,
\end{split}
\end{equation}
which coincides with the Boltzmann density of states. This result embodies a conceptual inversion of the standard statistical mechanics framework. Ordinarily, one postulates an ensemble $\rho_H$ first, and then—if desired—adds geometric structure and passes from right to left of Eq.~\eqref{eqn:coarea-boltzmann}. Here, we reverse the logical order: the metric tensor is introduced axiomatically alongside the fundamental Hamiltonian and symplectic structure, and the microcanonical ensemble emerges \emph{necessarily} as a purely geometric consequence.

\section{Thermodynamic equivalence of geometric description}

A fundamental question is still open: \textit{is the Euclidean metric tensor, $\eta$, the only geometric structure that defines the microcanonical measure?} In other words, \textit{does there exist another metric tensor that naturally induces the microcanonical ensemble?}

The answer to this question reveals fundamental properties of the geometric formulation. Indeed, the Hamiltonian flow conserves the microcanonical measure $d\mu_E$ (Eq.~\eqref{def:induced-measure}), but does not impose a specific structure for the metric tensor itself. 

This can be easily shown by observing that the conservation of the Liouville measure (in terms of the Lie derivative with respect to the Hamiltonian vector field) reads
\[
    \mathscr{L}_{\bm X_H}d\mu_\Lambda=\mathscr{L}_{\bm X_H}\left(\frac{d\sigma^\eta_E}{\|\nabla_{\!\eta} H\|_{\eta}} \wedge dE\right)=0
\]
Then, 
\[
    \mathscr{L}_{\bm X_H}\left(\frac{d\sigma^\eta_E}{\|\nabla_{\!\eta} H\|_\eta}\right) \wedge dE+\frac{d\sigma^\eta_E}{\|\nabla_{\!\eta} H\|_\eta} \wedge \mathscr{L}_{\bm X_H}\left(dE\right)=0
\]
but $\mathscr{L}_{\bm X_H}dE=d(\iota_{\bm X_H}dE)+\iota_{\bm X_H}d(dE)=0$ since the flow carries no component in the energy direction (namely, $\iota_{\bm{X}_H}dE=0$ and $d(dE)=0$). Therefore:
\[
    \mathscr{L}_{\bm X_H}\left(\frac{d\sigma^\eta_E}{\|\nabla_{\!\eta} H\|_\eta}\right)=\mathscr{L}_{\bm X_H}d\mu^{\eta}_{E}=0\,.
\]
This implies that the induced measure is preserved, independent of which metric is used (provided that it is compatible with $\omega$). Therefore, the symplectic structure does not \textit{rigidly} determine the metric tensor. This means that if there exists another metric tensor, $g$, with an induced measure $d\mu^g_E$ on $\Sigma_E$ such that
\begin{equation}
    \Omega_g(E)=\int_{\Sigma_E} d\mu^g_E =  \int_{\Sigma_E} d\mu^{\eta}_E = \Omega_\eta(E).
\end{equation}
than they are thermodynamically equivalent.
In other words, the choice of the metric possesses \textit{gauge freedom}, constrained only by the requirement that it induces the same induced measure on $\Sigma_E$. 
Therefore, the metric $g$ is said to be \textit{thermodynamically equivalent} to $\eta$.

We thus found a geometric gauge symmetry of thermodynamics resembling a principle:
\begin{quote}
\textbf{Thermodynamic covariance}. \textit{The thermodynamic content of a Hamiltonian system is independent of the geometric representation of phase space, provided that the natural measure on energy shells is preserved.}
\end{quote}
\noindent
Operationally, a Riemannian metric $g$ on phase space $\Lambda$ is \textit{thermodynamically equivalent to $\eta$}, denoted $g \sim \eta$, if and only if
\begin{equation}\label{eq:gauge-equivalence-def}
    \frac{d\sigma^g_E}{\|\nabla_{\!g}H\|_g} = \frac{d\sigma^\eta_E}{\|\nabla_{\!\eta} H\|_\eta}\,.
\end{equation}

The equivalence class $[\eta] = \{g : g \sim \eta\}$ consists of all metrics that induce the same microcanonical measure. All microcanonical observables---$S(E)$, and its energy derivatives---depend only on the equivalence class $[\eta]$, not on which representative is chosen. The Euclidean metric $\eta$ plays the role of a \textit{reference gauge}: it is the canonical choice that resolves the symplectic obstruction, but any $g \in [\eta]$ describes the same thermodynamics.

We now exploit this gauge freedom to construct an explicit representative $g \in [\eta]$ with particularly convenient properties, called the unit-norm gauge. In this gauge, the energy-flow generator coincides exactly with the gradient of the Hamiltonian $\nabla_g H$ and it has unit length.

\subsubsection{The unit-norm gauge}

The explicit representative $g \in [\eta]$ can be built by introducing a suitable change of coordinates for the metric in Eq.~\eqref{eqn:components-eta-adapted} that normalizes the generator of diffeomorphisms to unity.
Following Ref.~\cite{gori2018topological}, we define the transformation:
\begin{equation}\label{eq:rescaling-coords}
    dx^{0} = \chi\, dE, \quad
    dx^{\alpha} = \chi^{-\frac{1}{n-1}}\, dy^{\alpha}, \quad \alpha\in[1,2N-1],
\end{equation}
where $\chi := \|\nabla_{\!\eta} H\|_\eta^{-1}$. In these new coordinates $(x^0,x^1,\ldots,x^{2N-1})$, the metric tensor is (bi-)conformally related to $\eta$:
\begin{equation}\label{eq:biconformal-components}
    g_{00} = \chi^{-2}\,\eta_{00}=1, \qquad
    \sigma^g_{ij} = \chi^{\frac{2}{n-1}}\, \sigma^\eta_{ij},
\end{equation}
yielding the equivalent form
\begin{equation}\label{eq:unit-norm-metric}
    g = dx^{0}\otimes dx^{0} + \sigma^g_{ij}\, dx^{i}\otimes dx^{j}.
\end{equation}
This defines a new metric tensor $g$ that we call the \textit{unit-norm gauge}. Importantly, this transformation satisfies two properties \cite{gori_configurational}. 

\textbf{A) Thermodynamic equivalence.} The Riemannian volume form is preserved:
\begin{equation}\label{eq:volume-invariance}
\begin{split}
    d\mathrm{vol}^{g}
    &= \sqrt{\det\sigma^g}\; dx^{0}\,dx^{1}\cdots dx^{2N-1} 
    \\
    &= \chi\,\sqrt{\det \sigma^\eta}\; dE\,dy^{1}\cdots dy^{2N-1} 
    = d\mathrm{vol}^{\eta},
\end{split}
\end{equation}
and correspondingly, the induced area form on $\Sigma_E$,
\begin{equation}\label{eq:area-invariance}
\begin{split}
    d\mu^{g}_E
    = \sqrt{\det \sigma^g}\; &dx^{1}\cdots dx^{n}
    \\
    &= \chi\,\sqrt{\det \sigma^\eta}\; dy^{1}\cdots dy^{n}
    = d\mu^{\eta}_E.
\end{split}
\end{equation}
Therefore, $g \sim \eta$ in the sense of Eq.~\eqref{eq:gauge-equivalence-def}: both metrics induce the same microcanonical measure $d\mu_E$ on every energy shell.

\textbf{B) Unit normalization.} The vector field $\bm{\xi}_\eta=\partial_{E}$ associated with variations in energy is rescaled according to $\bm\xi_g=\chi^{-1}\bm\xi_\eta$, so that 
\begin{equation}\label{eq:unit-normal-property}
    g(\bm\xi_g,\bm\xi_g) = \chi^{-2}\,\eta(\bm{\xi}_\eta,\bm{\xi}_\eta) = 1.
\end{equation}
More precisely, this transformation makes the gradient of $H$ equal to $\bm\xi_g$. To see that, let us consider the definition of the gradient in the coordinates 

\[
    \nabla_g H=g^{00}\frac{\partial H}{\partial x^0}\partial_{x^0}+g^{ij}\frac{\partial H}{\partial x^i}\partial_{x^j}\,.
\]
However, independently of the metric tensor ($g$ remains compatible with $\omega$):
\[
    g(\nabla_g H,\bm{X}_H)=dH(\bm{X}_H)=0\,,
\]
therefore, any components of $\nabla_g H$ on $T_{\bm x}\Sigma_E$ must vanish: $\partial H/\partial x^i=0$ which leads to
\[
    \nabla_gH=g^{00}\frac{\partial H}{\partial x^0}\partial_{x^0}\,.
\]
Finally, since $g^{00}=g_{00}=1$, we notice that 
\[
    \frac{\partial H}{\partial x^0}=dH(\partial_{x^0})=1\;\;\implies \nabla_g H\equiv\bm\xi_g\equiv\partial_{x^0}.
\]

In conclusion, in the unit-norm gauge, the gradient of the Hamiltonian and the energy-flow generator are identical, with norm:
\begin{equation}
    \|\nabla_g H\|_g = g(\bm\xi_g,\bm \xi_g) = 1\,.
\end{equation}
This leads to a crucial point: the microcanonical measure simplifies to:
\begin{equation}
    d\mu_E^g = \frac{d\sigma_E^g}{\|\nabla_{\!g} H\|_g} \equiv d\sigma_E^g\,.
\end{equation}
thus coinciding with the area measure of $\Sigma_E$ and we have:
\begin{quote}
\textbf{Geometric Microcanonical Equivalence}. \textit{Within the thermodynamically equivalent metric class, it is possible to find a geometric representation $g$ such that counting accessible states is equivalent to measuring the occupied geometric area. Entropy follows as an intrinsic geometric property:}
\begin{equation}\label{def:geometric-entropy}
    S_g(E):=\ln\mathrm{area}^{g}(E)\overset{g\in[\eta]}{\simeq}\ln\Omega_B(E)=S_B(E).
\end{equation}
\end{quote}
This result suggests that geometry itself dictates the full thermodynamic behavior.
Intuitively, the thermodynamic properties of a physical system and the emergence of phase transitions are encoded within the geometry of the energy hypersurfaces and detected through the energy flow $\Phi^{\rm diff}_{\bm\xi_g}$.

\section{Geometric thermodynamics as evolution of energy shells}

We have found that microcanonical thermodynamics emerges from the intrinsic geometric structure of Hamiltonian systems. We now show that entropy satisfies a deterministic equation in the energy variable subjected to a force term that is purely geometric. This equation allows us to determine the presence of phase transitions and, in general, to investigate the thermodynamic properties of a given physical system.

\subsection{Variations of leaf volume in the unit-normal gauge}

The phase space is now equipped with $(H,\omega,g)$ with smooth, oriented, closed energy hypersurfaces
$\Sigma_E=\{x\in\Lambda\,:\,H(x)=E\}$ whose induced area form is
\begin{equation}
  d\mu^g_E \;=\; \iota_{\bm\xi}\,d\mu_\Lambda,
\end{equation}
From now on, in order to lighten the notation, we denote $\bm\xi\equiv\bm\xi_g$.

Thus, we define the $N-1$-dimensional area with respect to the metric tensor $g$ to be:
\begin{equation}\label{def:volume_rescaled_metric}
    \text{area}^{g}(E)=\int_{\Sigma_{E}}d\mu^{g}_E.
\end{equation}
which coincides with the density of states; therefore, entropy is given by (Boltzmann constant $k_B=1$)
\begin{equation}
    S(E):=\ln\text{area}^g(E)\;.
\end{equation}
Microcanonical observables correspond to energy derivatives of entropy, namely, $\partial_E^k S(E)$, we can recast such observables in terms of derivatives of volume $vol^g(E)$. In particular, we define the geometric curvature functions (GCFs):
\begin{equation}
    \Upsilon^{(k)}_g(E):=\frac{\partial^{k}_E \text{area}^g(E)}{\text{area}^g(E)}\,.
\end{equation}
which are recursive structures that will appear in the following sections.
Our current scope is to compute the first and second variations of volume.

\subsubsection{First variation of volume}

We notice that $\partial_E \text{area}^g(E)$ enters $\Upsilon^{(1)}_g(E)$. Then, we have 
\[
    \partial_E \text{area}^g(E)=\int_{\Sigma_{E}}\mathscr{L}_{\bm\xi}d\mu^{g}_E.
\]
that rewrites $\mathscr{L}_{\bm\xi} d\mu^g_E = \mathscr{L}_{\bm\xi}(\iota_{\bm\xi} d\mu^g_{\Lambda})$. Using the commutator identity 
\[
    [\mathscr{L}_X,i_Y]=i_{[X,Y]}\implies \mathscr{L}_X i_Y =i_Y\mathscr{L}_X +i_{[X,Y]}\ .
\]
Now, setting $X=Y=\bm\xi$, we have $[\bm\xi,\bm\xi]=0$; hence $\mathscr{L}_{\bm\xi}\iota_{\bm\xi}=\iota_{\bm\xi}\mathscr{L}_{\bm\xi}$ and
\begin{equation}
  (\mathscr{L}_{\bm\xi} \iota_{\bm\xi})\, d\mu^g_{\Lambda}
  \;=\; i_{\bm\xi}\,(\mathscr{L}_{\bm\xi} d\mu^g_{\Lambda})
  \ .
\end{equation}
By the definition of the Lie derivative \cite{petersen2006riemannian}, for any vector field $X$ on $\Lambda$, it reads:
\begin{equation}\label{eq:Lie-dV}
  \mathscr{L}_X d\mu^g_\Lambda \;=\; (\mathrm{div}_g X)\, d\mu^g_\Lambda.
\end{equation}
so that
\begin{equation}\label{eq:two-terms}
  \mathscr{L}_{\bm\xi} d\mu^g_E
  = (\mathrm{div}_g \bm\xi)\, \iota_{\bm\xi} d\mu^g_{\Lambda}=(\mathrm{div}_g \bm\xi)\,d\mu^g_E\ .
\end{equation}
In conclusion, equation~\eqref{eq:two-terms} reduces to
\begin{equation}\label{eq:Lie-dmug-div}
  \mathscr{L}_{\bm\xi} d\mu^g_E
  \;=\; (\mathrm{div}_g \bm\xi)\, d\mu^g_E\ .
\end{equation}
In the unit-normal gauge, the vector field $\bm\xi$, coinciding with the normal to $\Sigma_E$, satisfies $\nabla_{\bm\xi}\bm\xi=0$ since $g(\bm\xi,\bm\xi)=1$. 
Hence, the integral curves of $\bm\xi$ are geodesics of the ambient manifold $(\Lambda,g)$ as observed in Ref.~\cite{gori_configurational}: the flow generated by $\bm\xi$ moves orthogonally between neighboring energy hypersurfaces without intrinsic acceleration. 
Geometrically, the condition $\nabla_{\bm\xi}\bm\xi=0$ implies that the normal direction is preserved rigidly along the flow, so that variations of the volume element are fully determined by the intrinsic geometry of each leaf, encoded in its curvature properties, as we will see in a moment. No ``extrinsic bending'' contributes to the evolution: the changes of $\text{area}^g(E)$ arise solely from the intrinsic deformation of the energy manifolds.  

\begin{figure*}
    \centering
    \includegraphics[width=0.9\linewidth]{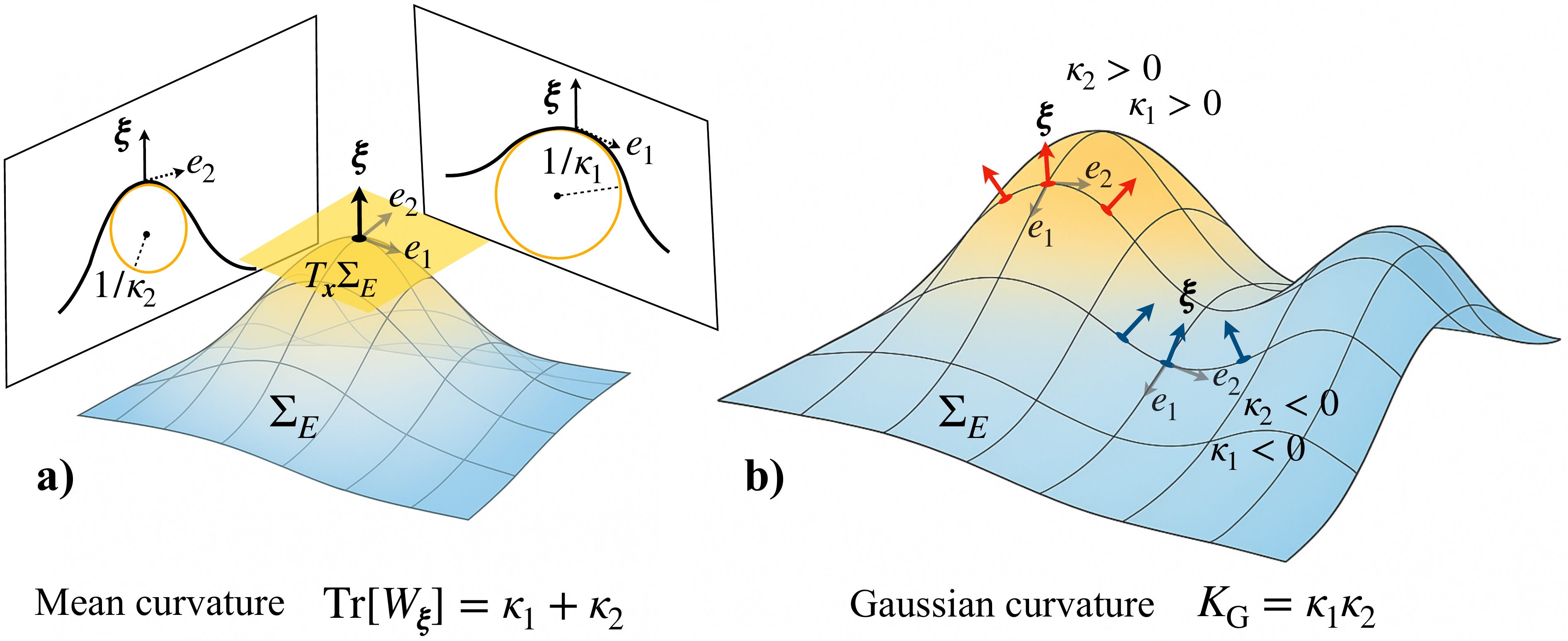}
\caption{Pictorial meaning of the Weingarten (shape) operator $W_{\bm\xi}$ and its principal curvatures for a two-dimensional hypersurface $\Sigma_E$ embedded in an ambient manifold. \textbf{(a)} At a point $\bm x\in\Sigma_E$, with unit normal $\bm\xi$ and tangent plane $T_{\bm x}\Sigma_E$, the principal directions $\bm e_1,\bm e_2$ diagonalize $W_{\bm\xi}$ and define the principal curvatures $\kappa_{1,2}$ via the normal sections; the corresponding osculating radii are $1/\kappa_1$ and $1/\kappa_2$. \textbf{(b)} Two representative local geometries: an elliptic (convex) point (on the left) with $\kappa_1,\kappa_2>0$ and a hyperbolic (saddle) point (on the right) with opposite signs. The invariants $\text{Tr}[W_{\bm\xi}]=\kappa_1+\kappa_2$ (mean curvature) and $K_G=\kappa_1\kappa_2$ (Gaussian curvature) summarize the local curvature signature.}
    \label{fig:weingarten}
\end{figure*}

Therefore, the purely geometric interpretation of the factor $\text{div}_g\bm\xi$ is provided in terms of the shape or Weingarten operator, which is defined by
\begin{equation}
  W_{\bm\xi}(X) := \,\nabla_X \bm\xi\;,\qquad\forall X\in T_{\bm{x}}\Sigma_{E}\,.
\end{equation}
For a reader unfamiliar with this structure, see the pictorial illustration of the Weingarten operator (for a 2D surface) in Fig.~\ref{fig:weingarten}.
To see this, consider the orthonormal basis $\{\bm e_i\}_{i=1}^{2N-1}$ on $T_{\bm{x}}\Sigma_E$ as that in Eq.~\eqref{def:frame-hypersurfaces}; the trace of the Weingarten operator is rewritten as:
\begin{equation}\label{eq:Hg-div}
\begin{split}
    \text{Tr}^g[W_{\bm\xi}]=\sum_i g( \bm e_{i},&W_{\bm\xi}(\bm e_i))\\
    &=\sum_i g( \bm e_{i},\nabla_{\bm e_i}\bm\xi)=\text{div}_g\bm\xi\,.
\end{split}
\end{equation}
Note that the trace is defined with respect to the metric tensor $g$. In this case, for any tensor $A$ of rank-2, $\text{Tr}^g[A]=g_{ij}A^{ij}=g(e_i,Ae_i)$ is considered a contraction.
Therefore, combining \eqref{eq:Lie-dmug-div} with \eqref{eq:Hg-div}, we conclude that
\begin{equation}\label{eq:1-variation}
  \mathscr{L}_{\bm\xi} d\mu^{g}_E \;=\text{Tr}^g[W_{\bm\xi}]\, d\mu^{g}_E\;.
\end{equation}
This means that the first variation of volume is read as follows:
\begin{align}
  \partial_E\, \text{area}^g(E)
  =\int_{\Sigma_E} \mathrm{Tr}[W_{\bm\xi}]\, d\mu^{g}_E\,.
\end{align}
Note that the trace can be written in terms of the gradient and Hessian with respect to the metric tensor $\eta$:
\begin{equation}
    \mathrm{Tr}^g[W_{\bm\xi}]
    =\frac{\Delta^{\eta} H}{\|\nabla^{\eta} H\|^{2}_{\eta}}
     -2\,\frac{\langle\nabla^{\eta} H,(\text{Hess}^{\eta}\,H)\,\nabla^{\eta} H\rangle_{\eta}}{\|\nabla^{\eta} H\|_{\eta}^{4}}\;.
\end{equation}
From now on, in order to make the notation lighter, we will drop all the superscripts $^\eta$, and unless explicitly mentioned, we will intend for all these structures to be defined with respect to $\eta$. We also drop the superscript $g$ in the trace; thus, $\text{Tr}\equiv\text{Tr}^g$.

\subsubsection{Second variation of volume}

Starting from Eq.~\eqref{eq:1-variation},
the second derivative reads
\begin{widetext}
\begin{align}\label{eqn:intermediate-second-variation}
    \partial_{E}^{2}{\rm area}^{g}(E)
   & =\int_{\Sigma_{E}}\mathscr{L}_{\bm\xi}\!\left(\mathrm{Tr}[W_{\bm\xi}]\,d\mu^g_E\right)\nonumber\\
    &\qquad\qquad\qquad\qquad=\int_{\Sigma_{E}}\bigg\{\Big(\mathscr{L}_{\bm\xi}\mathrm{Tr}[W_{\bm\xi}]\Big)d\mu^g_E+\mathrm{Tr}[W_{\bm\xi}]\mathscr{L}_{\bm\xi}\!\big(d\mu^g_E\big)\bigg\}\\
    &\qquad\qquad\qquad\qquad\qquad\qquad\qquad\qquad=\int_{\Sigma_{E}}\Big[\mathrm{Tr}[\nabla_{\bm\xi}W_{\bm\xi}]
      +\big(\mathrm{Tr}[W_{\bm\xi}]\big)^{2}\Big]\,d\mu^g_E,\nonumber
\end{align}
\end{widetext}
where we used $\mathscr{L}_{\bm\xi}\mathrm{Tr}[W_{\bm\xi}]=Tr[\nabla_{\bm\xi}W_{\bm\xi}]$. This term is subject to the equation \cite{gromov2019four,di2022geometrictheory}
\begin{equation}\label{eqn:Riccati-W}
    \text{Tr}[\nabla_{\bm\xi}W_{\bm\xi}]=-\text{Tr}[W_{\bm\xi}^2]-\text{Ric}_{\bm\xi}\,,
\end{equation}
where $\text{Ric}_{\bm\xi}$ is the Ricci curvature along the vector field $\bm\xi$ and is defined by $\text{Ric}_{\bm\xi}=\sum_ig(e_i,R(e_i,\bm\xi)\bm\xi)$ with $R$ being the Riemann curvature tensor. Now, the Ricci curvature can be written as \cite{gromov2019four}:
\begin{equation}
    2\,\text{Ric}_{\bm\xi}=R^g_{\Lambda}-R^g_{\Sigma_{E}}-\text{Tr}[W_{\bm\xi}^{2}]+\text{Tr}[W_{\bm\xi}]^{2}\,,
\end{equation}
and that, plugged into Eq.~\eqref{eqn:Riccati-W}, yields
\begin{align}
    \text{Tr}[\nabla_{\bm\xi}W_{\bm\xi}]=&-\text{Tr}[W_{\bm\xi}^2]\\
    &-\frac12\bigg(R^g_{\Lambda}-R^g_{\Sigma_{E}}-\text{Tr}[W_{\bm\xi}^{2}]+\text{Tr}[W_{\bm\xi}]^{2}\bigg)\nonumber\\
    =&-\frac12\text{Tr}[W_{\bm\xi}^2]-\frac12\text{Tr}[W_{\bm\xi}]^{2}+\frac12\left(R^g_{\Sigma_{E}}-R^g_{\Lambda}\right),\nonumber
\end{align}
Finally, plugging this relation into the last equality of Eq.~\eqref{eqn:intermediate-second-variation} produces
\begin{align*}
    \partial_{E}^{2}{\rm area}^{g}(E)
    =\frac12\int_{\Sigma_{E}}\!\!\Big[\text{Tr}[W_{\bm\xi}]^{2}&-\text{Tr}[W_{\bm\xi}^2]\nonumber\\
    &\quad+R^g_{\Sigma_{E}}-R^g_{\Lambda}\Big]d\mu^{g}_E,
\end{align*}
Notice that the right-hand side of the equation above can be related to the second-order GCF; we have:
\begin{equation}
    \Upsilon^{(2)}_g(E)=\frac12\int_{\Sigma_{E}}\!\!\left[\text{Tr}[W_{\bm\xi}]^{2}-\text{Tr}[W_{\bm\xi}^2]+R^g_{\Sigma_{E}}-R^g_{\Lambda}\right]d\rho^g_E,
\end{equation}
where
\[
    d\rho^g_E:=\frac{d\mu^{g}_E}{\text{area}^g(E)} .
\]
The explicit calculation of $\Upsilon^{(2)}_g$ in terms of the gradient and Hessian of $H$ with respect to the metric tensor $\eta$ is reported in Appendix~\ref{app:calculation-2-GCF}.

We are now ready to write the dynamical equations for the entropy function. As we have seen, every derivative of entropy is linked to a specific combination of GCFs. For $k=1$, we have
\begin{equation}
\begin{split}\label{def:upsilon-1}
    \partial_E S_g(E)&=\Upsilon^{(1)}_g(E)=\langle \text{Tr}[W_{\bm\xi}]\rangle_{E},
\end{split}
\end{equation}
where
\[
    \langle O\rangle_E:=\int_{\Sigma_E}O\;d\rho^g_E\, ,
\]
for any observable $O$ defined on the phase space.
Then, for $k=2$:
\begin{equation}
\begin{split}\label{def:upsilon-2}
    \partial^2_E S_g(E)&=\Upsilon^{(2)}_g(E)-(\Upsilon^{(1)}_g(E))^2\,,
\end{split}
\end{equation}
while for $k=3$, we have:
\begin{equation*}
\begin{split}\label{def:upsilon-3}
    \partial^3_E S_g(E)&=\Upsilon^{(3)}_g(E)-3 \Upsilon^{(1)}_g(E)\Upsilon^{(2)}_g(E)+2(\Upsilon^{(1)}_g(E))^3\,,
\end{split}
\end{equation*}
We see that by fixing a given order, we can combine this hierarchy of equations and put them into a closed form. For instance, for $k=2$, we combine \eqref{def:upsilon-1} with \eqref{def:upsilon-2} and we get
\begin{equation}\label{eqn:EFE-2}
    \partial^2_E S_g(E)+(\partial_E S_g(E))^2=\Upsilon^{(2)}_g(E)\,.
\end{equation}
while for $k=3$, we get
\begin{equation*}\label{eqn:EFE-3}
    \partial^3_E S_g(E)-3\partial_E S_g(E)\,\partial^2_E S_g(E)+(\partial_E S_g(E))^3=\Upsilon^{(3)}_g(E)\,.
\end{equation*}
We call these equations entropy flow equations (EFEs) of order $k$. Equation~\eqref{eqn:EFE-2} is a second-order EFE.
Then, by induction, for a generic order $k$, we have
\begin{equation*}
\begin{split}
    \partial^k_E S_g(E)+\mathscr{P}_k[\partial_E S_g(E),\ldots,\partial^{k-1}_E S_g(E)]=2(\Upsilon^{(1)}_g(E))^3,
\end{split}
\end{equation*}
where $\mathscr{P}_k$ indicates a non-linear combination of entropy derivatives that produces a generalized Riccati equation.
The physical meaning behind this class of EFEs is profound. First, they should be interpreted as deterministic equations in closed form for the entropy function which is subjected to a source term $\Upsilon^{(k)}_g$ of purely geometric character. The entropy function and its derivatives are determined by the knowledge of the energy shell's geometry and by the set of ``initial values'' $\{\partial_ES(E_0),\ldots,\partial_E^{k-1}S(E_0)\}$ for some arbitrary $E_0$. 

Interestingly, they also reflect an intrinsic hierarchical property of thermodynamics that can be interpreted within the \textit{general principle of minimal sensitivity} \cite{stevenson1981optimized,stevenson1981resolution}
, which has been combined with the microcanonical
inflection-point analysis \cite{qi2018classification}.
Collective phenomena and phase transitions do not result from a single law; rather, they manifest in the order of the entropy derivative, which is more sensitive to changes in its energy profile. This means that low order phase transitions can be detected in the first or second order EFE, while for higher transitions, the lower EFE will be insensitive to these phenomena.

Here, we will focus on the second-order EFE \eqref{eqn:EFE-2} and show for emblematic models how this equation can fully capture phase transitions of second-order. Our strategy consists of solving the second-order EFE exactly (through numerical integration) and comparing the results, i.e., $S(E),\partial_ES$ and $\partial^2_E S$, with those obtained from numerical simulations using thermodynamic methods to demonstrate that the geometric approach is not an approximation of real thermodynamics.

Before going in this direction, we first proceed with an intermediate step. Through an analytical calculation, we show how geometry contains information about phase transitions and what underlying geometric mechanism triggers the transition.

\section{Geometric change in mean curvature for the 1D XY mean field}
\label{sec:geometric-change-XY-MF}

Consider the Hamiltonian of the 1D XY mean-field 
\begin{equation}
    H(\theta,p)=\sum_{i=1}^N \frac{p_i^2}{2}+\frac{J}{2N}\sum_{i,j}\bigl[1-\cos(\theta_i-\theta_j)\bigr],
\end{equation}
with magnetization 
\begin{equation}
    \bm M=\frac1N\sum_i(\cos\theta_i,\sin\theta_i)=(M\cos\phi,M\sin\phi).
\end{equation}

The key goal of this investigation is to show that the mean curvature can be rewritten in terms of $M$ (magnetization strength), and in the thermodynamic limit, it is expressed as follows:
\[
    \mathrm{Tr}W_{\bm\xi
    }(M)=C^{\infty}_0(\epsilon)+C^{\infty}_2(\varepsilon)\,M^2+O(M^4)
\]
where $\epsilon=E/N$ is the specific energy and 
\[
    C^{\infty}_2(\varepsilon)=\frac{J}{(2\varepsilon-J)^2}\Big(\varepsilon-\frac{3J}{4}\Big),
\]
The coefficient $C^{\infty}_2$ conceptually represents the curvature of the energy shell with respect to $M$. Its sign diagnoses the regime:
$C^{\infty}_2>0$ (sheet locally convex: high energy), $C_2=0$ (sheet locally flat: critical), $C^{\infty}_2<0$ (sheet locally concave: low energy). In particular, we see that imposing $C^{\infty}_2(\varepsilon_c)=0$ we obtain $\varepsilon_c=3J/4$, which is the known critical energy for the 1D XY mean-field model~\cite{antoni1995clustering,pettini2007geometry}.

Let us consider the Weingarten formula (here, we drop the subscript $\eta$),
\begin{equation}
    \mathrm{Tr}\,W_{\bm\xi}=\frac{\Delta H}{\|\nabla H\|^2}-\frac{2\,\nabla H^{\!\top}(\mathrm{Hess}\,H)\nabla H}{\|\nabla H\|^4},
\end{equation}
and define $\Theta_i=\theta_i-\phi$, so one has the identities (see App.~\ref{app:complex-magn}):
\begin{align}
\sum_j\cos(\theta_i-\theta_j)&=N\,M\cos\Theta_i,\\
\sum_j\sin(\theta_i-\theta_j)&=N\,M\sin\Theta_i,\\
\sum_{i,j}\cos(\theta_i-\theta_j)&=N^2 M^2.
\end{align}
Then, we have 
\begin{equation}
    \begin{split}
        \partial_{\theta_i}V&=\frac{J}{N}\sum_j \sin(\theta_i-\theta_j)=J\,M\sin\Theta_i,\\
    \partial^2_{\theta_i}V&=\frac{J}{N}\sum_j \cos(\theta_i-\theta_j)=J\,M\cos\Theta_i,\\
    \partial_{\theta_i}\partial_{\theta_j}V&=-\frac{J}{N}\cos(\theta_i-\theta_j)\quad (i\neq j).
    \end{split}
\end{equation}
Then, 
\[
    \Delta H=\sum_i \partial_{p_i}^2H+\sum_i \partial_{\theta_i}^2 V= N + J N M^2,
\]
\begin{equation}\label{def:nabla-H}
    \begin{split}
    \|\nabla H\|^2=\sum_i (\partial_{p_i}H)^2 &+ \sum_i (\partial_{\theta_i}V)^2\\
    &=K+J^2 M^2 \sum_i \sin^2\Theta_i,
\end{split}
\end{equation}

The, we simply rewrite $\mathrm{Tr}\,W_{\bm\xi}$ into three blocks (momenta, diagonal angular, and angular interactions) in the explicit form:
\begin{align}
\label{def:block-P}
    \sum_i\kappa_{p_i}=\frac{N}{\|\nabla H\|^2}-\frac{2K}{\|\nabla H\|^4},\;\;\; K:=\sum_{i=1}^N p_i^2,
\end{align}
\begin{align}
\label{def:block-D}
    \sum_i\kappa^{\text{diag}}_{\theta_i}=\frac{JN M^2}{\|\nabla H\|^2}-\frac{2J^3M^3\sum_{i=1}^N \sin^2\Theta_i\cos\Theta_i}{\|\nabla H\|^4},
\end{align}
\begin{align}
\label{def:block-I}
    \sum_{i\neq j}\kappa^{\mathrm{int}}_{ij}=\frac{2J^3 M^2}{N\,\|\nabla H\|^4}\sum_{i\ne j}\sin\Theta_i\sin\Theta_j\cos(\theta_i-\theta_j).
\end{align}
In doing so, the trace writes
\begin{equation}
    \text{Tr}[W_{\bm\xi}]=\sum_i\Big(\kappa_{p_i}+\kappa^{\text{diag}}_{\theta_i}+\sum_{j\neq i}\kappa^{\mathrm{int}}_{\theta_i\theta_j}\Big)\,.
\end{equation}
The $\kappa_{p_i}$-term contains only factors associated with momentum degrees of freedom, while $\kappa^{\text{diag}}_{\theta_i}$ encompasses factors related to the diagonal components of $\text{Hess}\,V_{ii}$, and finally, $\kappa^{\mathrm{int}}_{ij}$ pertains to those with $\text{Hess}\,V_{ij}$ out-diagonal. This decomposition is still exact.

The next step is to analyze the \emph{onset of order} from the disordered branch, $M=0$.
On that branch $M=0$ is a symmetry point; consequently, all \emph{linear} contributions in $M$ to $\mathrm{Tr}[W_{\bm\xi}]$ vanish by symmetry.
Therefore, the \emph{first nontrivial} information is at order $M^2$.
Expanding consistently up to $O(M^2)$ yields the coefficient $C^{\infty}_2(\varepsilon)$ that governs the local curvature of $\mathrm{Tr}[W_{\bm\xi}]$ along the collective direction $M$ and encodes the geometric competition.

\subsection{Controlled small–$M$ expansion}

Therefore, we wish to provide an expansion of the blocks in Eqs.~\eqref{def:block-P}-\eqref{def:block-I} in terms of the collective parameter magnetization, $M$. This means that we must coherently introduce it within those blocks. To this end, we recognize three angular functions within $\|\nabla H\|$ in Eq.~\eqref{def:nabla-H}, and within $\kappa_{\theta_i}^{\rm diag}$ and $\kappa_{ij}^{\rm int}$ in Eqs.~\eqref{def:block-D} and ~\eqref{def:block-I}, respectively. They are: 
\[
\begin{split}
    S_2:=\sum_{i=1}^N \sin^2&\Theta_i,\qquad
    S_{21}:=\sum_{i=1}^N \sin^2\Theta_i\,\cos\Theta_i,\\
    S_{\mathrm{int}}&:=\sum_{i\neq j}\sin\Theta_i\sin\Theta_j\cos(\Theta_i-\Theta_j).
\end{split}
\]
On the disordered branch ($M\to 0$), one can parametrize the angular statistics by a von Mises density $f(\Delta)\propto e^{h\cos\Delta}$ with small $h$.
Using the standard Bessel expansions (all details are given in Appendix~\ref{app:angular-averages}), one obtains
\begin{align}
    S_2&=N\Big(\frac12-\frac14 M^2\Big)+O(N M^4),
    \label{def:S_2}\\
    S_{21}&=N\Big(\frac14 M - \frac{1}{24}M^3\Big)+O(N M^5),
    \label{def:S_12}\\
    S_{\mathrm{int}}&=\frac{N^2}{4}+O(N^2 M^2)+O(N).
    \label{def:S_int}
\end{align}
With these, and introducing the kinetic density $c:=K/N$, the gradient and Laplacian of $H$ read:
\begin{align}
\label{def:nabla-H-expansion}
    G:=\|\nabla H\|^2 &= N + J^2 M^2 S_2\approx N\Big(c+\frac{J^2}{2}M^2\Big) ,\\
    \nonumber
    \Delta H &= N + J N M^2.
\end{align}
We now expand each block, building the $\text{Tr} W_{\bm\xi}$ up to $O(M^2)$, keeping the leading order in $N$. In practice, we plug Eqs.~\eqref{def:S_2}-\eqref{def:nabla-H-expansion} into Eqs.~\eqref{def:block-P}-\eqref{def:block-I}. 

Let us start with
\[
\begin{split}
    \sum_i\kappa_{p_i}=\frac{N}{G}-\frac{2K}{G^2}\ .
\end{split}
\]
Using the Binomial series $(1+x)^{\alpha}\simeq1+\alpha x+O(x^{2})$, for the denominator $G$, assuming $M\approx 0$, the first term reads
\[
\begin{split}
    \frac{N}{G}=\frac{1}{c(1+J^2M^2/2c)}&\approx\frac{1}{c}\left(1-\frac{J^2}{2c}M^2+O(M^4)\right)\\
    &\simeq\frac{1}{c}-\frac{J^2}{2c^2}M^2.
\end{split}
\]
we notice that both terms above are leading in the limit $N\to\infty$; therefore, we take both. The second term, instead, is negligible in the thermodynamic limit:
\[
    \begin{split}
        \frac{2K}{G^2}&=\frac{2Nc}{c^2N^2(1+J^2M^2/2c)^2}\\
        &\quad\quad\quad=\frac{2}{Nc}\left(1-\frac{2J^2}{c}M^2+O(M^4)\right)\\
        &\quad\quad\quad\quad\quad\quad\simeq O\left(\frac{1}{N}\right)\left(1+O(M^2)\right)\ .
    \end{split}
\]
The diagonal component reads
\[
\begin{split}
    \sum_i\kappa_{\theta_i}^{\rm diag}&=\frac{JNM^2}{G}-\frac{2J^3M^3S_{12}}{G^2}\approx\frac{JNM^2}{G}\\
    &\approx \frac{JNM^2}{Nc}\left(1-\frac{J^2M^2}{2c}\right)=\frac{JM^2}{c}+O(M^4)
\end{split}
\]
where we used the fact that $S_{21}=O(N M)$, and therefore it is negligible at order $M^2$ and in the thermodynamic limit:
\[
    \frac{2J^3M^3S_{12}}{G^2}=\frac{J^3}{2c^2}\frac{M^4}{N}+\mathcal{O}\left(\frac{M^6}{N}\right)\,.
\]
Finally, the interaction term is 
\[
\begin{split}
    \sum_{ij}\kappa_{ij}^{\rm int}=\frac{2J^3 M^2}{N\,G^2} \, S_{\mathrm{int}}&=\frac{2J^3 M^2S_{\mathrm{int}}}{N\,N^2(c+\frac{J^2}{2}M^2)^2}\\
    &=\frac{2J^3 M^2}{N^3 c^2}\Big(\frac{N^2}{4}\Big)=O\!\Big(\frac{M^2}{N}\Big).
\end{split}
\]
that is, it is subleading at order $M^2$ in the large–$N$ limit.

Collecting all of the $O(1)$ and $O(M^2)$ terms, the trace reads
\begin{align}
    \mathrm{Tr}W^{\infty}_{\bm\xi}(c,M)&=C^{\infty}_0(c)+C^{\infty}_2(c,J)M^2\\
    &=\frac{1}{c}+\Big(\frac{J}{c}-\frac{J^2}{2c^2}\Big)\,M^2 .
\end{align}
On the disordered branch, the potential energy $V=J(1-M^2)/2$ reduces to $V(M\approx0)\sim J/2$, and the energy density is
\[
    \varepsilon=\frac{H}{N}=\frac{K}{2N}+\frac{J}{2}=\frac{c}{2}+\frac{J}{2}\quad\Longrightarrow\quad c=2\varepsilon - J.
\]
Therefore, the coefficient, $C_2^{\infty}$, as a function of energy is
\begin{equation}
\begin{split}
    C_2(\varepsilon)=\frac{J}{2\varepsilon-J}&-\frac{J^2}{2(2\varepsilon-J)^2}
    \\
    &\qquad=\frac{2J}{(2\varepsilon-J)^2}\left(\varepsilon-\frac{3J}{4}\right) .
\end{split}
\end{equation}
This coefficient encodes the geometric information that triggers the transitions and thus deserves to be discussed.
We notice that for $\varepsilon=J/2$, $C_2$ diverges; this energy value represents the lower bound of the disordered branch ($M=0$). Hence, it only marks the onset of the feasibility of the disordered branch. For $\varepsilon>J/2$, the disordered configuration becomes energetically allowed and dominates above the transition. 

More importantly, let us focus on the zero of the $C_2$ coefficient. We find that a unique zero of $C_2(\varepsilon)$ is determined by
\[
\varepsilon_c=\frac{3J}{4},
\]
corresponding to the known energy value at which the second-order phase transition occurs in the mean field 1D XY model.

\section{Geometric Signatures of Phase Transitions}

We now show the geometric mechanism behind the phase transition in the 1D XY mean field model. In order to detect, understand, and visualize such a geometric change, we study the $\Sigma_E$'s shapes across the transition.

To this end, we expand the potential along with the tangent direction $\eta=\theta^\star +\delta\theta_i\,\bm{e}_i$ assuming tangential displacements, i.e., $\langle\delta\theta,\nabla H\rangle=\langle\delta\theta,\nabla V\rangle=0$; then, 
\begin{equation}\label{def:expansion-V}
    V(\theta)\simeq V(\theta^\star)+\langle\delta\theta,\text{Hess}V\,\delta\theta\rangle+O(\|\delta\theta\|^3)\,,    
\end{equation}
where
\begin{align}
    \text{Hess}V=D&-\frac{J}{N}(\bm{x}\otimes\bm{x}+\bm{y}\otimes\bm{y}),\nonumber\\
    D_{ii}=J\Big(M\cos\Theta_i-\frac{1}{N}\Big)\delta_{ij}&,\quad x_i=\cos\theta_i,\quad y_i=\sin\theta_i.\nonumber
\end{align}
where $\bm{x}\otimes\bm{x}$ can be expressed in component form as $\bm{x}\bm{x}^{\top}$ and similarly for $\bm y$.
Then, $x_ix_j+y_iy_j=\cos(\theta_i-\theta_j)$. In general, this structure can be diagonalized, thus obtaining an orthonormal basis $\{e_k(\theta^\star)\}$ such that 
\[
    \text{Hess}V(\theta^\star)e_k=\lambda_k e_k\,.
\]
Then, denoting $E(\theta^\star):=[e_1,\dots,e_N]$ as the orthogonal matrix of eigenvectors, we can map the system into normal coordinates such that 
\[
    q_k=e_k^\top \delta\theta\,,
\]
while the momenta are already diagonal. 
Then, the second-order term in the local expansion of the potential reads
\[
\begin{split}
    \delta\theta^\top & \text{Hess}V(\theta^\star)\delta\theta\\
    &=\delta\theta^\top E E^{T}\text{Hess} V(\theta^\star)EE^T\delta\theta\\
    &=(E^T\delta\theta)^\top (E^{T}\text{Hess} V(\theta^\star)E)(E^T\delta\theta)=\sum_k q^2_k\lambda_k
\end{split}
\]
In conclusion, the potential $V$ along the tangent directions rewrites as a constant $V(\theta^\star)$ plus a sum of quadratic terms in normal mode coordinates, modulated by the eigenvalues. Plugging this form into the Hamiltonian, we obtain 
\begin{equation}
    H_{\text{diag}} \simeq \frac12\sum_{k=1}^N p_k^2\;+\;\frac12\sum_{k=1}^N \lambda_k\,q_k^2\;.
\end{equation}
This expression describes the shape of $\Sigma_E$ locally.  
Notice that the structure of the Hessian is rank-2 diagonal due to the term $x_ix_j+y_iy_j$; therefore, we have at most two ``collective'' eigenvalues that deviate from the diagonal elements in $D_{ii}$. 

The section of $\Sigma_E$ in the $(q_k,p_k)$ plane is an ellipse for $\lambda_k>0$, a cylinder for $\lambda_k=0$, or a hyperbola for $\lambda_k<0$. This structure allows us to identify the directions responsible for changes in geometry. This structural property allows one to isolate at most two collective eigenvalues deviating from the bulk spectrum. In other words, we can diagonalize the Hessian. As shown in App.~\ref{app:hessianV-diagonalization}, writing $\text{Hess}V=D-\alpha UU^\top$, with $\alpha=J/N$ and $U:=[x,y]\in\mathbb R^{N\times 2}$, the Sylvester determinant identity gives
\begin{align}
        \det(\text{Hess}V-\lambda I)=&\det(D-\lambda I)\\
    &\cdot\det\!\big(I_2-\alpha\,U^\top(D-\lambda I)^{-1}U\big).\nonumber
\end{align}
Defining
\begin{align}
    G(\lambda)&:=U^T(D-\lambda I)^{-1}U\\
    &=
\begin{pmatrix}
    x\otimes(D-\lambda I)^{-1}x & x\otimes(D-\lambda I)^{-1}y\\
    y\otimes(D-\lambda I)^{-1}x & y\otimes(D-\lambda I)^{-1}y
\end{pmatrix},\nonumber
\end{align}
the characteristic equation reads
\begin{equation*}
    \det(\text{Hess}V-\lambda I)=\det(D-\lambda I)\det(I_2-\alpha G(\lambda))=0.
\end{equation*}
This relation must be intended carefully. The spectrum of $\text{Hess}V$ is indeed contained within the composition of $D-\lambda I_N$ and $I_2-\alpha G(\lambda)$. In general, $G(\lambda)$ possesses eigenvalues that are not the eigenvalues of the spectrum of $\text{Hess}V$. These $G(\lambda)$-eigenvalues are found among the zeros of $\text{det}(D-\lambda I_N)$, and only in the composition above are they precisely canceled, thus leaving only the correct spectrum of $\text{Hess}V$. 

In general, this secular equation cannot be solved analytically; thus, we analyze it regime by regime in energy.

\begin{figure*}
    \centering
    \includegraphics[width=1\linewidth]{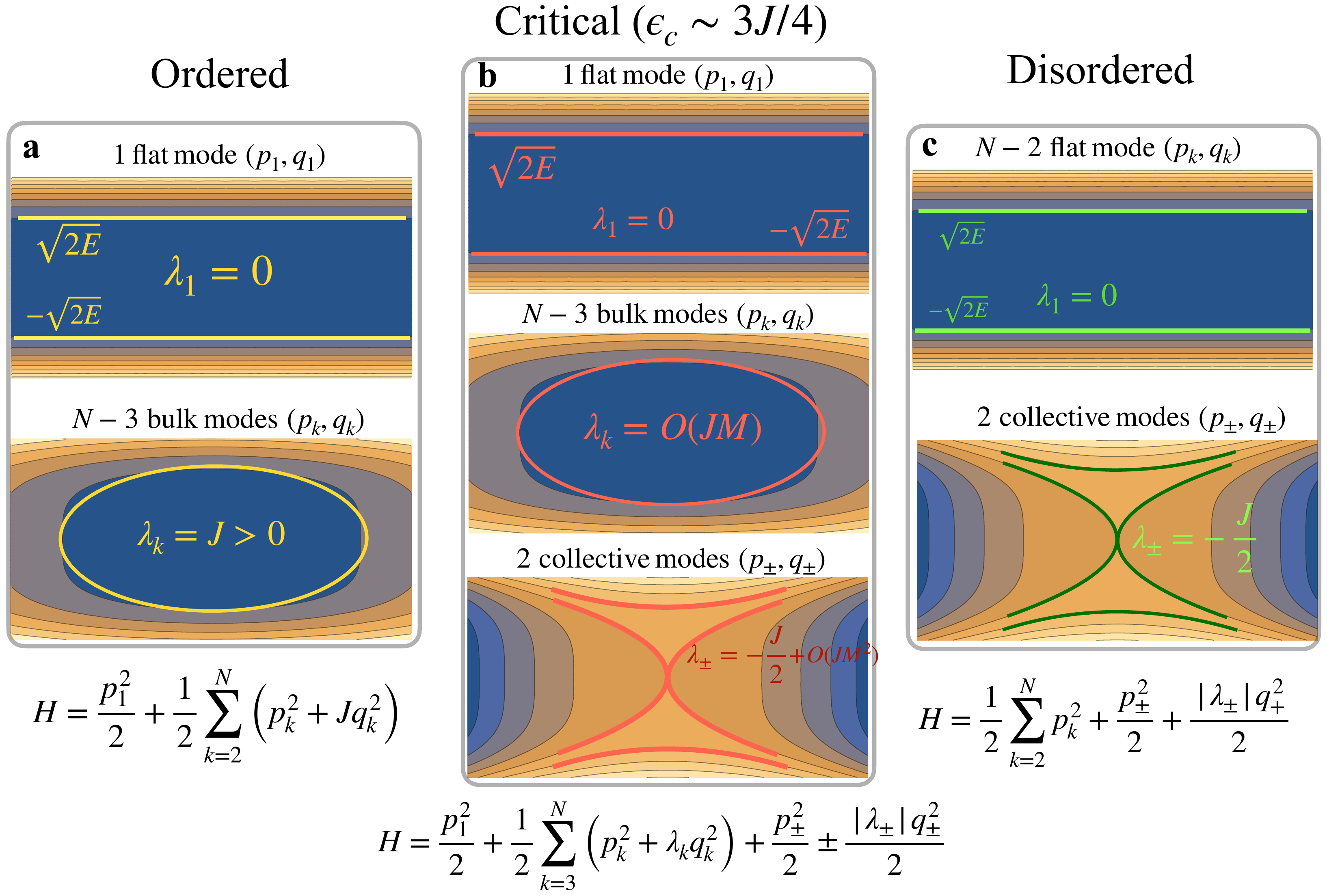}
\caption{\textbf{Pictorial visual representation of the geometric change triggering the second-order phase transition in the 1D XY mean-field model.}
Schematic decomposition of the Hamiltonian in canonical normal coordinates into: (i) a \emph{flat} mode $(p_1,q_1)$ with $\lambda_1=0$ (top boxes), (ii) a set of weakly mixed \emph{bulk} modes $(p_k,q_k)$ (middle boxes), and (iii) two \emph{collective} modes $(p_\pm,q_\pm)$ (bottom boxes).
In the ordered phase (left) the bulk sector is elliptic with positive curvature $\lambda_k=J>0$, while the collective sector is absent and the reduced sections are compact.
At criticality (center, $\epsilon_c\simeq 3J/4$) positive and negative curvature coexist: the bulk eigenvalues soften to $\lambda_k=\mathcal{O}(JM)$, whereas the collective eigenvalues become negative, $\lambda_\pm=-J/2+\mathcal{O}(JM^2)$, producing a ``neck'' (one--sheet hyperboloid) in the reduced sections.
In the disordered phase (right) the bulk becomes flat ($N-2$ flat modes) and the nontrivial curvature is concentrated in the collective hyperbolic sector with $\lambda_\pm=-J/2$.
The horizontal lines at $\pm\sqrt{2E}$ indicate the kinematic bounds from the energy constraint in the corresponding reduced coordinates.}
    \label{fig:change}
\end{figure*}

\subsubsection*{Low energy (ordered)}
Here all $\Theta_i \simeq0$, then $\theta_i\simeq\phi$ meaning that they are all oriented along the collective direction. Then $D\simeq J 1\!\!1_{N}$, while $xx^\top+yy^\top\simeq  1\!\!1_{2}$. Hence
\[
\text{Hess}V=J 1\!\!1_{N}-\frac{J}{N}\,\mathbf 1\otimes\mathbf 1,
\]
where $\mathbf{1}:=\{1,1,\ldots,1\}^{\top}$. The spectrum is then $\{0,J,\dots,J\}$. The Hamiltonian reads:
\[
H_{\text{low}}\simeq\frac{p_1^2}{2} +\frac{1}{2}\sum_{k=2}^N \Big(p_k^2+J q_k^2\Big)
\]
Thus, one neutral mode (uniform rotation) and $N-1$ convex directions: geometrically, an elliptic cylinder.

\subsubsection*{High energy (disordered)}
For $M\simeq0$, $D\simeq0$. Then, the Hessian
\[
    \text{Hess}V=-\alpha UU^\top\, ,
\]
is a rank-2 matrix. Its two nonzero eigenvalues are $-J/2$ (on average), the rest being $0$. The Hamiltonian reads
\[
    H_{\text{high}}=\frac{p_\pm^2}{2}+\frac{\lambda_\pm }{2}q_+^2+\sum_{k=2}^N \frac{p_k^2}{2}
\]
Then, the geometry consists of two saddles and a flat bulk: a hyperboloid of two sheets in reduced sections.

\subsubsection*{Near-critical regime}

For small but nonzero $M$, the diagonal entries $d_i = J M \cos\Theta_i$ are weak, and
\[
\mathrm{Hess}V = D - \alpha U U^\top, 
\qquad \alpha = \frac{J}{N},
\]
can be viewed as a rank–2 collective perturbation on top of an almost flat bulk. The secular equation shows that: (i) there is always one exact zero eigenvalue, $\lambda_0 = 0$, associated with global rotations; (ii) two collective eigenvalues $\lambda_\pm(M) = -J/2 + \mathcal O(M^2)$ emerge from the rank–2 structure, with $\lambda_-(M)$ remaining strongly negative, whereas $\lambda_+(M)$ moves towards zero and
changes sign across the transition; (iii) the remaining $N-3$ eigenvalues form a bulk
$\{\lambda_k\}_{k=2}^{N-1}$ of order $\mathcal O(JM)$ that interlaces the diagonal entries $\{d_i\}$ and contains both positive and negative values. In normal–mode coordinates $\{q_k,p_k\}$ associated with the eigenvectors of $\mathrm{Hess}V$, the local Hamiltonian decomposes as
\begin{equation}
H_{\text{crit}}
= 
\frac{p_0^2}{2}
+  \frac{p_{\pm}^2}{2} + \frac{\lambda_{\pm}}{2}\, q_{\pm}^2
+ \sum_{k=2}^{N-1} \Big( \frac{p_k^2}{2} + \frac{\lambda_k}{2}\,q_k^2\Big).
\end{equation}
This coexistence of positive and negative curvatures, concentrated in a low-dimensional collective sector and in a weakly mixed bulk, produces a neck geometry
(one-sheet hyperboloid in reduced sections) at the transition. This geometric change is illustrated in Fig.~\ref{fig:change}.

\section{Entropy flow equations, thermodynamics and phase transition}

In this section, we investigate and solve non-perturbatively the second-order EFE for two emblematic systems: the 2D $\phi^2$ model with nearest interactions and the 1D XY model with long-range interactions. We show how the EFE is able to exactly reproduce the thermodynamic content of these systems. We show that by computing the microcanonical observables, i.e., derivatives of the entropy function, using a thermodynamic method and a geometry-independent method~\cite{pearson1985laplace} (see below) and solving the EFE. Moreover, we also provide an analysis of phase transitions in the microcanonical ensemble. The scope of this section, however, is not to determine phase transitions in the thermodynamic limit, but to validate the geometric approach; therefore, any scaling size analysis will be presented elsewhere.

%%%%%%%%%%%%%%%%%%%

\subsection{The 2D $\phi^4$ model with nearest-neighbor interactions}

\begin{figure*}
    \centering
    \includegraphics[width=0.9\linewidth]{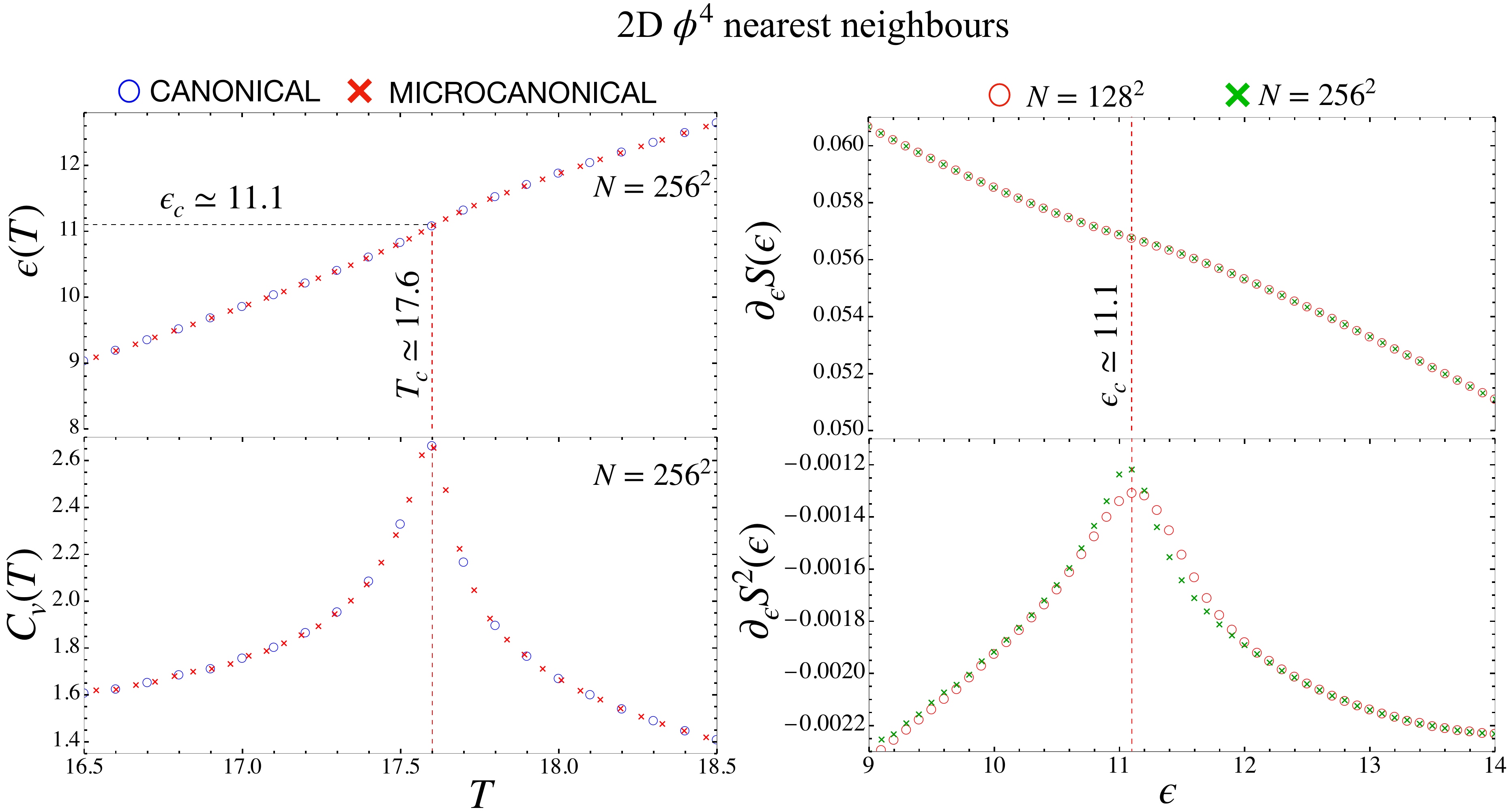}
\caption{\textbf{2D $\phi^{4}$ model with nearest-neighbour coupling.}
\textbf{Left panels:} comparison between canonical (blue circles) and microcanonical (red crosses) estimates for a system of size $N=256^{2}$. Top: caloric curve $\epsilon(T)$; the horizontal dashed line marks the critical energy density $\epsilon_{c}\simeq 11.1$. Bottom: specific heat $C_{v}(T)$; the vertical dashed line indicates the temperature $T_{c}\simeq 17.6$ at which $C_{v}$ peaks, corresponding (through the inverted caloric curve) to $\epsilon_{c}$.
\textbf{Right panels:} microcanonical observables as entropy derivatives versus energy density, for $N=128^{2}$ (red circles) and $N=256^{2}$ (green crosses): inverse temperature $\partial_{\epsilon}S(\epsilon)$ (top) and curvature $\partial_{\epsilon}^{2}S(\epsilon)$ (bottom). The vertical dashed line marks $\epsilon_{c}\simeq 11.1$, where $\partial_{\epsilon}S$ shows an inflection and $\partial_{\epsilon}^{2}S$ develops a pronounced maximum, sharpening with increasing system size.}
    \label{fig:thermo-phi4}
\end{figure*}

We consider the 2D $\phi^4$ model with nearest neighbor interactions, whose Hamiltonian is
\begin{equation}
\begin{split}\label{def:hamiltonian_phi4}
    H:=\sum_{\bm{i}}\Bigg[\frac{\pi^{2}_{\bm{i}}}{2}+\frac{\lambda}{4!}\phi^{4}_{\bm{i}}&-\frac{\mu^{2}}{2}\phi^{2}_{\bm{i}}+\frac{J}{4}\!\!\sum_{\langle\!\langle\bm{k}\rangle\!\rangle}(\phi_{\bm{i}}-\phi_{\bm{k}})^{2}\Bigg]
\end{split}
\end{equation}
and we choose values 
\[
    \lambda=3/5,\qquad \mu^2=2,\qquad J=1.
\]
The label $\bm{i}:=(i_1,i_2)$ is the two-dimensional array of integer numbers used for labeling the sites; finally, we denote $\langle\!\langle\bm{k}\rangle\!\rangle$ as the sum over the set of the nearest neighbors of the $\bm{i}$-th site.\\
This model is known to undergo a second-order phase transition at the energy density value $\epsilon_{c}=E/N\approx 11.1$, as already observed in Refs. \cite{bel2020geometrical,gori2018topological,di2022geometrictheory}. 

\subsubsection{Thermodynamic analysis}

In the canonical ensemble, relevant observables include the caloric curve defined by $E=E(T)$ and computed through the canonical average:
\[
    E(T)=\frac{1}{Z(T)}\int e^{-\beta H(\pi,\phi)}H(\pi,\phi
    )~D\pi\,D\phi\,,
\]
and the specific heat is given by 
\[
    C_v(T)=\partial_T E(T)=\frac{\langle H^2\rangle_{T}-\langle H\rangle^2_T}{T^2}\,.
\]
In the microcanonical ensemble, we consider the entropy function
\[
    S(E)=\int \delta(H(\pi,\phi)-E)\,D\pi\,D\phi\,,
\]
The average of a given observable is computed by:
\[
    \langle O\rangle_{E}=\int O(\pi,\phi)~\delta(H(\pi,\phi)-E)~D\pi\,D\phi\,.
\]
The microcanonical observables are defined by $\partial^k_ES(E)$. For instance, the microcanonical temperature reads:
\[
    T^{-1}_{\text{micro}}(E)=\frac{\partial S}{\partial E}(E)\,.
\]
The specific heat is
\begin{equation}\label{def:specific-heat-micro}
        C^{\text{micro}}_v(E)=-\frac{(\partial_E S)^2}{\partial^2_E S}=-\frac{1}{T_{\text{micro}}^2 (E)\partial^2_E S(E)}\,.
\end{equation}
The thermodynamic method for estimating the derivatives of entropy is provided by Pearson-Halicioglu-Tiller (PHT) \cite{pearson1985laplace} (see App.~\ref{app:PHT_method}). The first and second order derivatives are computed through averages of polynomials of the kinetic energy. Defining the total number of degrees of freedom with $M=N^2$ and introducing the specific kinetic energy $k=K/M$ and total energy $\epsilon=E/N$, we have
\begin{equation}
\label{}
\begin{split}
     \partial_{\varepsilon}S(\varepsilon)= \left(\frac{1}{2}-\frac{1}{M}\right)&\langle k^{-1}\rangle_{E}\,, \\
    \partial^{2}_{\varepsilon}S(\varepsilon)
    =M\bigg[\bigg(\dfrac{1}{2}-\dfrac{1}{M}\bigg) &\bigg(\dfrac{1}{2}-\dfrac{2}{M}\bigg)\langle k^{-2} \rangle_{E}
    \\
    - &\bigg(\dfrac{1}{2}-\dfrac{1}{M}\bigg)^2 \langle k^{-1}\rangle^2_{E} \bigg]\,,
\end{split}
\end{equation}

Any detail about the numerical implementation of the sampling method is reported in App.~\ref{sec:numerical-sampling}. In order to show the reliability of our microcanonical sampling method, we present, in Fig.~\ref{fig:thermo-phi4}, a comparison of the caloric curves and specific heat in the canonical and microcanonical ensembles. Note that in order to compare them, we inverted the relation between energy and temperature in the microcanonical ensemble so as to have $\epsilon=\epsilon(T_{\text{micro}})$ vs $\epsilon_{\text{can}}=\epsilon_{\text{can}}(T)$. For the specific heat in the microcanonical ensemble, we combine data for $C_v^{\text{micro}}$ with the caloric curve in order to have $C_v^{\text{micro}}(T_{\text{micro}})$. We find a very accurate agreement between the canonical and microcanonical estimations of these observables. The specific heat admits a peak at $T_c\simeq 17.6$ corresponding to $\epsilon_c\simeq 11.1$. Therefore, we observe the behavior of microcanonical observables. We plot in Fig.~\ref{fig:thermo-phi4} $\partial_\epsilon S$ and $\partial_\epsilon S$. We notice that there is an inflection point in $\partial_\epsilon S$ and a negative-valued maximum in $\partial^2_\epsilon S$, precisely at $\epsilon\simeq 11.1$. This is fully compatible with the microcanonical inflection point analysis (MIPA) by Bachmann \cite{qi2018classification,bachmann2014thermodynamics,bachmann2014novel}, which identifies phase transitions (or precursors at finite-size) in those points where a specific inflection point emerges. More precisely, a phase transition of order $n$ occurs if and only if $\partial^k_E S$ admits an inflection point such that $\partial^{k+1}_E S$ has either a negative-valued maximum or a positive-valued minimum. In this case, the 2D $\phi^4$ model undergoes a second-order phase transition that gives rise to a divergence in the thermodynamic limit; usually observed in the specific heat. It should be stressed that the divergence of the specific heat can be fully justified by examining the behavior of $\partial_E^2 S$ in the thermodynamic limit. By inspection of Fig.~\ref{fig:thermo-phi4}, we see that the maximum/peak of $\partial_E^2 S$ moves towards zero for increasing system sizes; that is, $\lim_{N\to\infty}\partial_E S\to0$. Now, from Eq.~\eqref{def:specific-heat-micro}, this means that the specific heat diverges $C_v\to\infty$, with $\partial_E^2S$ in the denominator.

\subsubsection{EFE and geometric approach}\label{secsec:EFE-phi4}

\begin{figure*}
    \centering
    \includegraphics[width=0.99\linewidth]{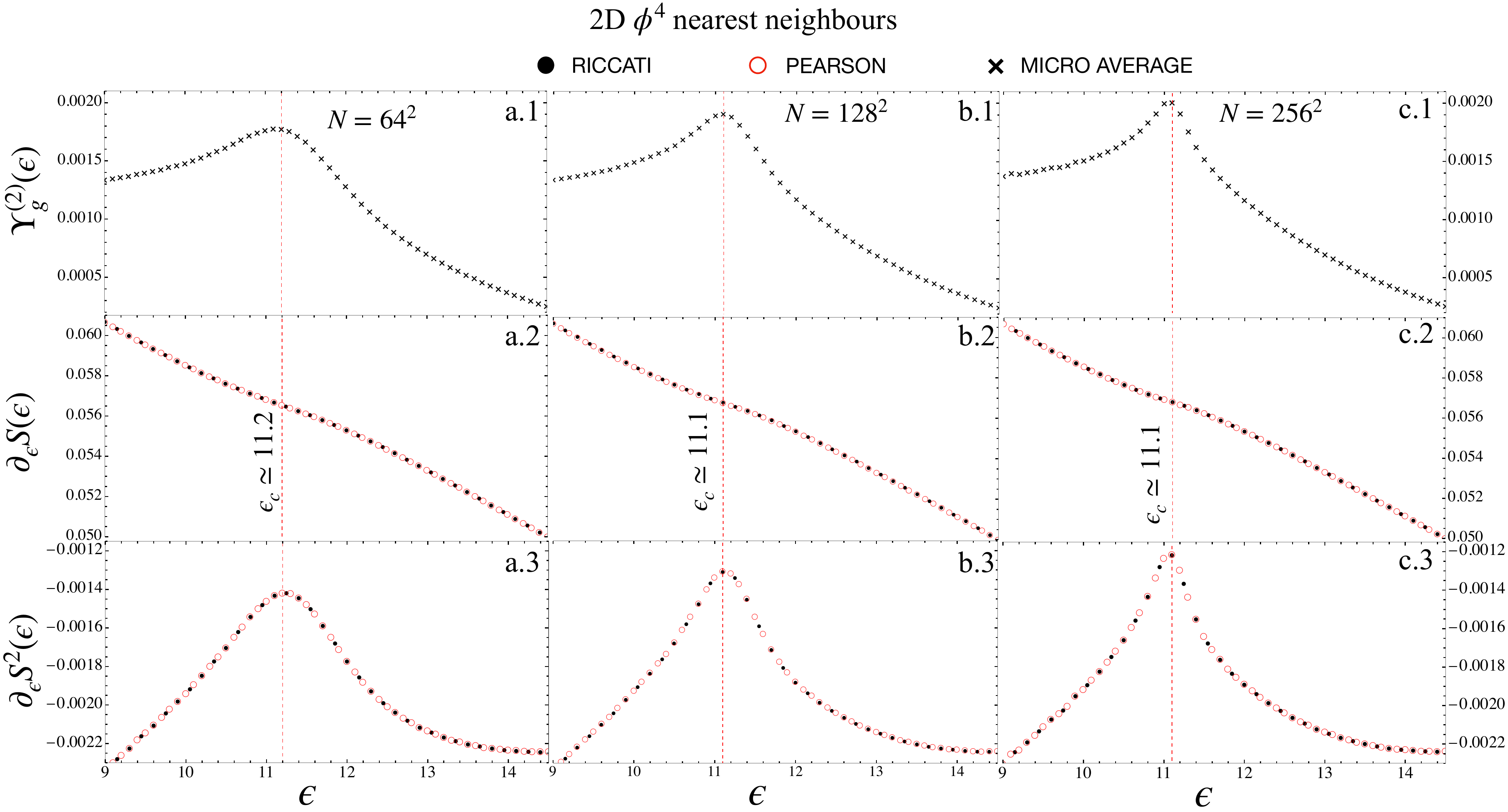}
\caption{\textbf{2D nearest-neighbor $\phi^4$ model: geometric input and reconstructed microcanonical observables.}
Panels (a.1–c.1) show the measured geometric quantity $\Upsilon^{(2)}_g(\epsilon)$ entering the EFE.
Panels (a.2–c.2) and (a.3–c.3) report, respectively, $\beta(\epsilon)=\partial_\epsilon S_g(\epsilon)$ and
$\partial_\epsilon^2 S_g(\epsilon)$: crosses are direct microcanonical estimates, filled black circles are obtained by solving the Riccati form of the EFE using $\Upsilon^{(2)}_g(\epsilon)$ as input, and open red circles denote the PHT-based reconstruction.
Columns correspond to $N=64^2,\,128^2,\,256^2$. The vertical red dashed line marks the critical
energy density $\epsilon_c\simeq 11.1$--$11.2$; the peak in $\partial_\epsilon^2 S_g$ becomes sharper and drifts toward $\epsilon_c$ with increasing $N$, consistently indicating the finite-size approach to the underlying change in the geometry of $\Sigma_E$.}
    \label{fig:efe-phi4}
\end{figure*}

We consider the second-order EFE
\begin{equation}\label{def:EFE-phi4}
    \partial^2_E S(E)+(\partial_E S(E))^2=\Upsilon^{(2)}_g(E)\,.
\end{equation}
Here, entropy is regarded as an unknown function that must be determined by solving the equation with given initial conditions $(S(E_0),\partial_ES(E_0))$. This means that the GCF $\Upsilon^{(2)}_g(E)$ is the source function whose energy-profile must be known. This function is given in Eq.~
\eqref{def:estimation-GCF} and computed through numerical simulation in the microcanonical ensemble (see App.~\ref{sec:numerical-sampling}). We plot $\Upsilon^{(2)}_g$ in Fig.~\ref{fig:efe-phi4}, panels \textbf{a.1}, \textbf{b.1} , and \textbf{c.1}. Interestingly, it manifests a pattern, i.e., a peak at $\epsilon=11.1$ where the transition is known to occur.

We numerically solve Eq.~\eqref{def:EFE-phi4} using Mathematica software with the function \texttt{NDSolve}, imposing \texttt{ExplicitRungeKutta} with \texttt{DifferenceOrder} 9 (see App.~\ref{sec:efe-solution-procedure}). The initial conditions are imposed by selecting the minimum value of energy used in the numerical simulations. In panels \textbf{a.2}-\textbf{a.3}, \textbf{b.2}-\textbf{b.3} and \textbf{c.2}-\textbf{c.3} of Fig.~\ref{fig:efe-phi4}, we show the comparison between PHT and Riccati estimations for $\partial_E S$ and $\partial_E^2 S$. We obtained excellent agreement, both quantitative and qualitative.

\subsection{The 1D XY long-range model}\label{ssec:1D-LR-XY}

The 1D XY long-range model is described by the lattice Hamiltonian 
\begin{equation}
    H(\theta,p)=\sum_{i=1}^N \frac{p_i^2}{2}
+\frac{J}{2N^{1+\sigma}}\sum_{i,j}\frac{1-\cos(\theta_i-\theta_j)}{|i-j|^{1+\sigma}},
\end{equation}
where the interactions are between all lattice field variables $\theta_i$. Here, $i$ is an integer label, and we use first-imagine periodic boundary conditions, i.e., $\theta_{N+1}=\theta_{1}$. The normalization $N^\alpha$ represents the Kac factor defined by
\[
    N^\alpha=\sum_{ij}\frac{1}{|i-j|^{1+\sigma}}\,.
\]
To put our microcanonical results into context, we consider the classical one-dimensional XY chain with pair couplings decaying as
\begin{equation}
  J(r) \propto \frac{1}{r^\alpha}
  = \frac{1}{r^{1+\sigma}}, 
  \qquad \alpha = 1+\sigma.
\end{equation}
In this model with Kac-type rescaling, the canonical partition function can be treated analytically exactly for $\alpha<1$.
In $d=1$, this implies that for $\sigma<0$, the model is canonically equivalent to the mean-field limit $\sigma=-1$ and exhibits a continuous ferromagnetic-paramagnetic transition at $T_c = J/2$ and $\varepsilon_c = 3J/4$, with mean-field critical exponents independent of $\alpha$ \cite{Campa2000Canonical,Giansanti2000Universal,antoni1995clustering}.

\begin{figure*}
    \centering
    \includegraphics[width=1\linewidth]{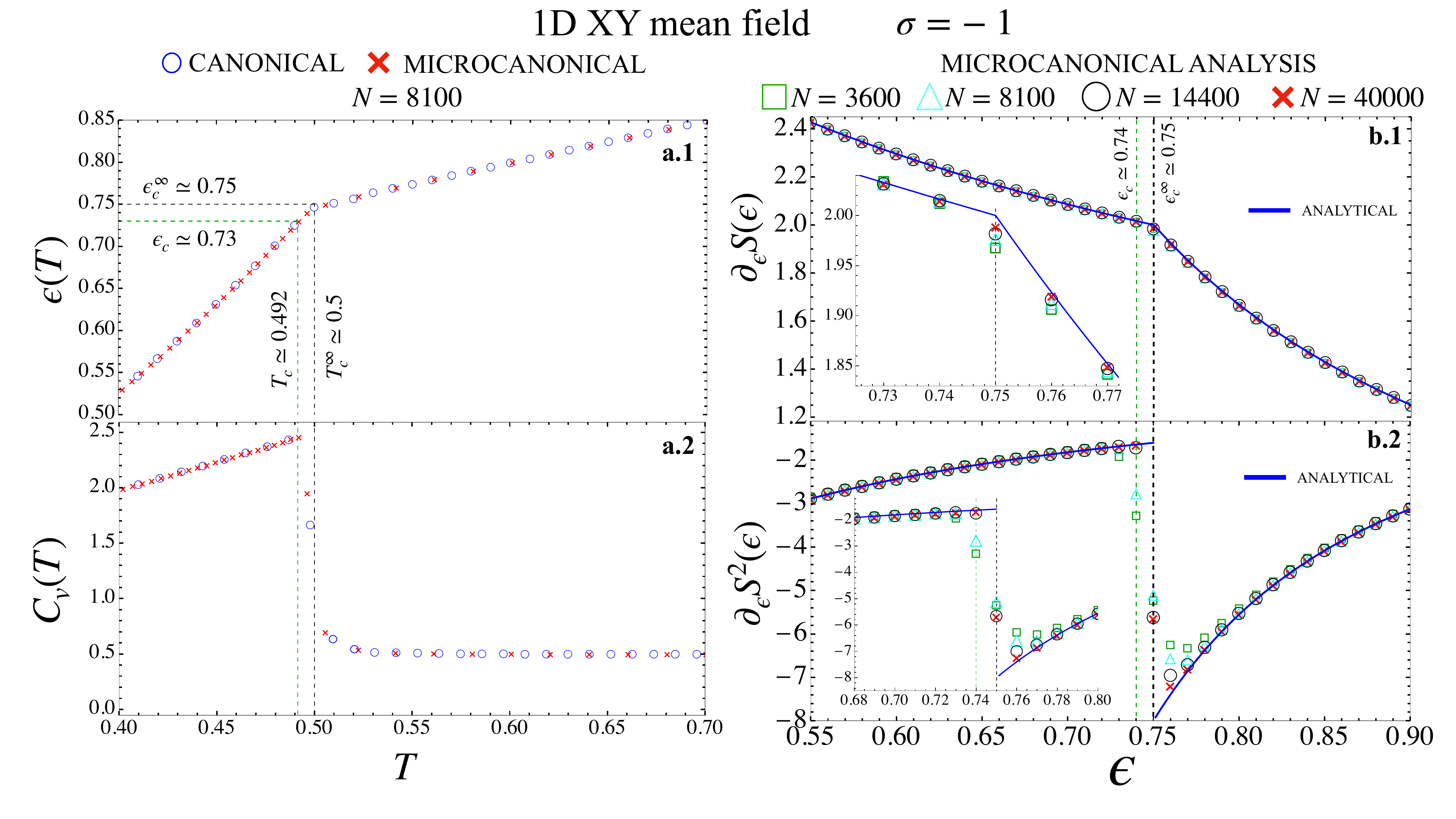}
\caption{\textbf{Mean-field 1D XY model ($\sigma=0$).}
\textbf{(a)} Canonical-microcanonical comparison for $N=8100$:
\textbf{(a.1)} caloric curve $\epsilon(T)$ (blue circles: canonical; red crosses: microcanonical) and
\textbf{(a.2)} canonical specific heat $C_V(T)$.
The vertical lines mark the canonical critical temperature $T_c\simeq 0.492$ and the infinite-size critical temperature $T^{\infty}_c\simeq 0.5$ known from analytical methods; the corresponding critical energies are indicated in panel \textbf{(a.1)}, with $\epsilon_c^{\infty}\simeq 0.75$ and $\epsilon_c^{\rm finite}\simeq 0.73$.
\textbf{(b)} Microcanonical analysis of the entropy derivatives for increasing sizes
($N=3600,8100,14400,40000$): \textbf{(b.1)} $\partial_\epsilon S(\epsilon)$ and \textbf{(b.2)} $\partial_\epsilon^2 S(\epsilon)$.
Symbols are microcanonical data at different different sizes, while the solid blue curves are the analytical results.
The vertical lines indicate the infinite-size critical energy $\epsilon_c^\infty= 0.75$ and the finite-size location $\epsilon_c\simeq 0.74$ for $N=40000$ (see insets for a magnification of the critical region), highlighting that the maximum of $\partial_\epsilon^2 S$ collapse onto the analytical curve for all $N$, whereas for small sizes $\partial_\epsilon S$ rounds off and bends down earlier.}
    \label{fig:thermo-xy-mf}
\end{figure*}

For $\sigma \ge 0$, the sum $\sum_r J(r)$ becomes convergent, and the interaction is effectively short-range in one dimension. In this regime, rigorous results by Bruno show that for monotonically decaying ferromagnetic couplings $J(r)\sim r^{-1-\sigma}$, spontaneous magnetic order at any $T>0$ is excluded whenever $\sigma \ge 1$ \cite{Bruno2001Absence}. Between these two limits, $1<\alpha<2$ (or $0<\sigma<1$), no rigorous theorem settles the existence or absence of a thermodynamic phase transition. 

From the broader long-range $O(N)$ literature with couplings $\propto r^{-(d+\sigma)}$, it is known that, for small $\sigma$, long-range tails can generate distinct universality classes and non-classical critical exponents, with a crossover to short-range behavior as $\sigma$ increases. Additionally, long-range fixed points branch off from the short-range ones at $\sigma_* = 2 - \eta_{\mathrm{SR}}$ \cite{Defenu2015FixedPoint,Defenu2017QuantumLR,Defenu2023RMP}.
Recent Monte Carlo and RG studies of long-range spin models in 1D and 2D further emphasize that the intermediate window $0<\sigma<1$ is precisely where strong cooperative effects and rich crossover phenomena are expected, even when the presence of a sharp, finite-temperature ordered phase is not established~\cite{Dyson1969Ising,Giachetti2022Villain}.

In this section, we focus on this regime and, in particular, on $0\le\sigma\le 1/2$, using the exactly solved mean-field case ($\sigma=-1$) as a reference. 
Within this interval, our goal is not to reach the infinite-size phase transition, but to detect the precursor through the MIPA approach and investigate how it behaves for increasing system sizes. This is necessary to confirm the physical reliability of the geometric approach. Moreover, knowing the exact solution for the 1D XY mean-field model, we can test and verify MIPA and the geometric approach against the exact microcanonical observables. This is extremely useful when we move to a genuine long-range system ($\sigma> -1$), where the lack of exact solutions does not allow us to confirm the reliability of these microcanonical approaches. 

\subsubsection{Mean field case}

The mean field case is recovered by posing $\sigma=-1$. Here, we analyze the thermodynamics of such a system and then solve the EFE equation after computing the GCF. In this case, we can exploit the fact that analytical calculations are feasible.
We then compute the second-order derivative of entropy analytically. We also show the derivatives of entropy obtained from numerical simulations, and finally, we solve the EFE.

\subsubsection*{Exact microcanonical curvature for the 1D mean-field XY model}

The canonical partition function for this model is exactly solvable, and the equilibrium magnetization is determined by the standard self-consistency equation
\begin{equation}
    \label{def:magnetization-self-consistent}
    M (\beta)= I_1(\beta M)/I_0(\beta M)\ ,
\end{equation}
with $I_n$ modified Bessel functions and
$\beta=1/T$~\cite{Campa2000Canonical,campa2009statistical}.
Exploiting ensemble equivalence for this model, the microcanonical entropy density $s(\varepsilon)$ can be obtained from the canonical free energy via Legendre transform, and its first derivative satisfies $ds/d\varepsilon=\beta(\varepsilon)$.

In App.~\ref{app:HMF_entropy_curvature}, we derive an explicit parametric expression for the second derivative $s''(\varepsilon)$ in terms of Bessel functions. Introducing the effective field $h=\beta M$ and
\begin{equation}
  A(h) = \frac{1}{2}\left(1 + \frac{I_2(h)}{I_0(h)}\right)
         - \left(\frac{I_1(h)}{I_0(h)}\right)^2,
\end{equation}
the canonical specific heat per particle can be written as
\begin{equation}
  c(\beta)
  = \frac{1}{2}
    + \frac{\beta^2 A(h)\,M^2}{1 - \beta A(h)},
\end{equation}
where $M=I_1(h)/I_0(h)$ and $h=\beta M$.

For the HMF/mean–field XY model, the energy per particle in the canonical ensemble is
\begin{equation}
  \varepsilon(\beta)
  = \frac{1}{2\beta} + \frac{1 - M(\beta)^2}{2},
\end{equation}
where the magnetization $M(\beta)$ is determined by Eq.~\eqref{def:magnetization-self-consistent}. In the paramagnetic phase ($M=0$, $\beta \le 2$), this reduces to
\begin{equation}\label{def:beta-ferromagnetic}
  \varepsilon(\beta) = \frac{1}{2\beta} + \frac{1}{2}
  \quad\Rightarrow\quad
  \beta(\varepsilon) = \frac{1}{2(\varepsilon - 1/2)},
  \qquad \varepsilon \ge \frac{3}{4}.
\end{equation}
In the ferromagnetic phase ($\beta>2$), $\beta(\varepsilon)$ is obtained
in parametric form from the system
\begin{equation}\label{def:beta-paramegnetic}
  M(\beta) = \frac{I_1(\beta M)}{I_0(\beta M)},
  \qquad
  \varepsilon(\beta) = \frac{1}{2\beta} + \frac{1 - M(\beta)^2}{2}.
\end{equation}

The second-order derivative of entropy then follows from
$s''(\varepsilon)=d\beta/d\varepsilon=-\beta^2/c(\varepsilon)$ as
\begin{equation}\label{def:s-secondo-analytical}
  s''\bigl(\varepsilon(\beta)\bigr)
  = -\,\frac{\beta^2}{
        \dfrac{1}{2}
      + \dfrac{\beta^2 A(h)\,M^2}{1 - \beta A(h)}
      },
\end{equation}
with $\varepsilon(\beta)=\tfrac{1}{2\beta} + \tfrac{1-M^2}{2}$.
This provides a fully analytic benchmark for the microcanonical curvature of a textbook second-order phase transition.
In particular, $s''(\varepsilon)$ exhibits a single, negative, and sharply localized maximum at the critical energy $\varepsilon_c = 3/4$, precisely where our geometric indicator $C_2(\varepsilon)$ and the MIPA criterion identify the onset of the phase transition. In the upcoming section, the expressions in Eqs.~\eqref{def:beta-ferromagnetic}-\eqref{def:s-secondo-analytical} are used and compared with our numerical data. These functions can be computed on Mathematica Software (see App.~\ref{ssec:mathematica-code}).

\subsubsection{Thermodynamics of the 1D XY mean field model}

For this model, we perform numerical simulations using a microcanonical Monte Carlo sampling method (see all the details in App.~\ref{sec:numerical-sampling}). 

We repeat the comparative analysis of the caloric curve and specific heat between canonical and microcanonical data, adopted for the $\phi^4$ model and reported in Fig.~\ref{fig:thermo-xy-mf} panels \textbf{a.1} and \textbf{a.2}. This ensures that the microcanonical sampling is correct. Notice that we do not expect any ensemble inequivalence, but this is just a confirmation that our microcanonical sampling method correctly estimates the thermodynamic properties of the 1D XY mean-field model. 
Following the standard detection of phase transitions, we identify a second order phase transition with the temperature associated with the peak of the specific heat (see Fig.~\ref{fig:thermo-xy-mf}, panel \textbf{a.2}). The phase transition is found at $T_{c}\simeq 0.942$ corresponding to $\epsilon_c\simeq0.73$ for $N=8100$. This approach has been applied (but not reported) to all the system sizes investigated in this work, and we found consistent agreement between the canonical and microcanonical locations of the peak in the specific heat. In particular, as the system size increases, the location of the peak moves toward $T_c^{\infty}=0.5$.

The first and second derivatives of entropy (computed using the PHT method in App.~\ref{app:PHT_method}) are reported in Fig.~\ref{fig:thermo-xy-mf}, panels \textbf{b.1} and \textbf{b.2}, for sizes $N=3600,\,8100,\,14400$ and $N=40000$. The comparison with the analytical curves in blue is also reported. 

The specific-heat analysis is consistent with the MIPA. Indeed, in correspondence with the specific heat's peak, $\partial_ES$ admits an inflection point (see panel \textbf{b.1}) and $\partial_E^2S$ a negatively valued maximum (see panel \textbf{b.2}). Obviously, the inflection point and the maximum of $\partial_E^2S$ are sensitive to the system's size $N$, but we can easily observe an interesting and subtle trend. For $N=3600$ (green squares), $\max_{\epsilon}\partial_\epsilon ^2S$ occurs at $\epsilon_c\simeq0.72$ while for $N=8100$ (cyan triangles) at $\epsilon_c\simeq0.73$, for $N=14400$ (black circles) at $\epsilon_c\simeq 0.74$ and finally for $N=40000$ (red crosses) at $\epsilon_c\simeq 0.74$. 

The numerical trend is unambiguous and is strongly supported by the analytical prediction (blue curve): for every system size, the locations of the maxima of $\partial_\epsilon^2 S$ lie on the analytical curve. For small sizes, however, the numerical $\partial_\epsilon S$ deviates from the analytical behavior at lower $\epsilon$, i.e., it bends downward ``earlier'' than it does for larger systems.

The $N$-limit trend that we observed indicates that $\epsilon^{N}_c\to\epsilon_c^{\infty}=3J/4=0.75$ for $N\to\infty$. This trend is strongly supported by the analytical black curve to which these maxima belong. The comparison with the analytical solution is strong evidence that MIPA consistently identifies (precursor) phase transitions already at finite size and coherently extrapolates the critical energy values depending on the system size. 

In other words, at any system size, $N$, the maximum of $\partial_\epsilon S^2$, say $\epsilon_\text{max}(N)$, falls onto the analytical curve. Thus, MIPA identifies this point as the precursor of the infinite-size phase transition. Increasing the system size, $N\to\infty$, results in $\epsilon_\text{max}(N)\to\epsilon_{c}^{\infty}=3J/4$ as predicted by the analytical curve. Remarkably, the transition is clearly identified by the jump in $\partial^2_\epsilon S$; however, we notice that the MIPA criterion is, in principle, also valid in the thermodynamic limit where $\epsilon_c^{\infty}$ is the maximum for $\partial_\epsilon S^2_{\infty}$. This analysis indicates that the second-order phase transition (in the sense of MIPA) is compatible with the infinite-size phase transition of second-order, as expected in the thermodynamic limit. This represents a side result of our analysis. Entropy derivatives, namely the microcanonical observables, retain all the information about the emergence of collective phenomena at any size. For a fixed system size, these collective phenomena are signaled by the derivatives of entropy through the presence of inflection points and maxima/minima, which we call \textit{precursors} or finite-size phase transitions. Therefore, the presence of a pattern in the derivatives of entropy cannot be dismissed a priori as cross-overs or non-universal behavior. 

\subsubsection{EFE equation for mean field model}

\begin{figure*}
    \centering
    \includegraphics[width=0.99\linewidth]{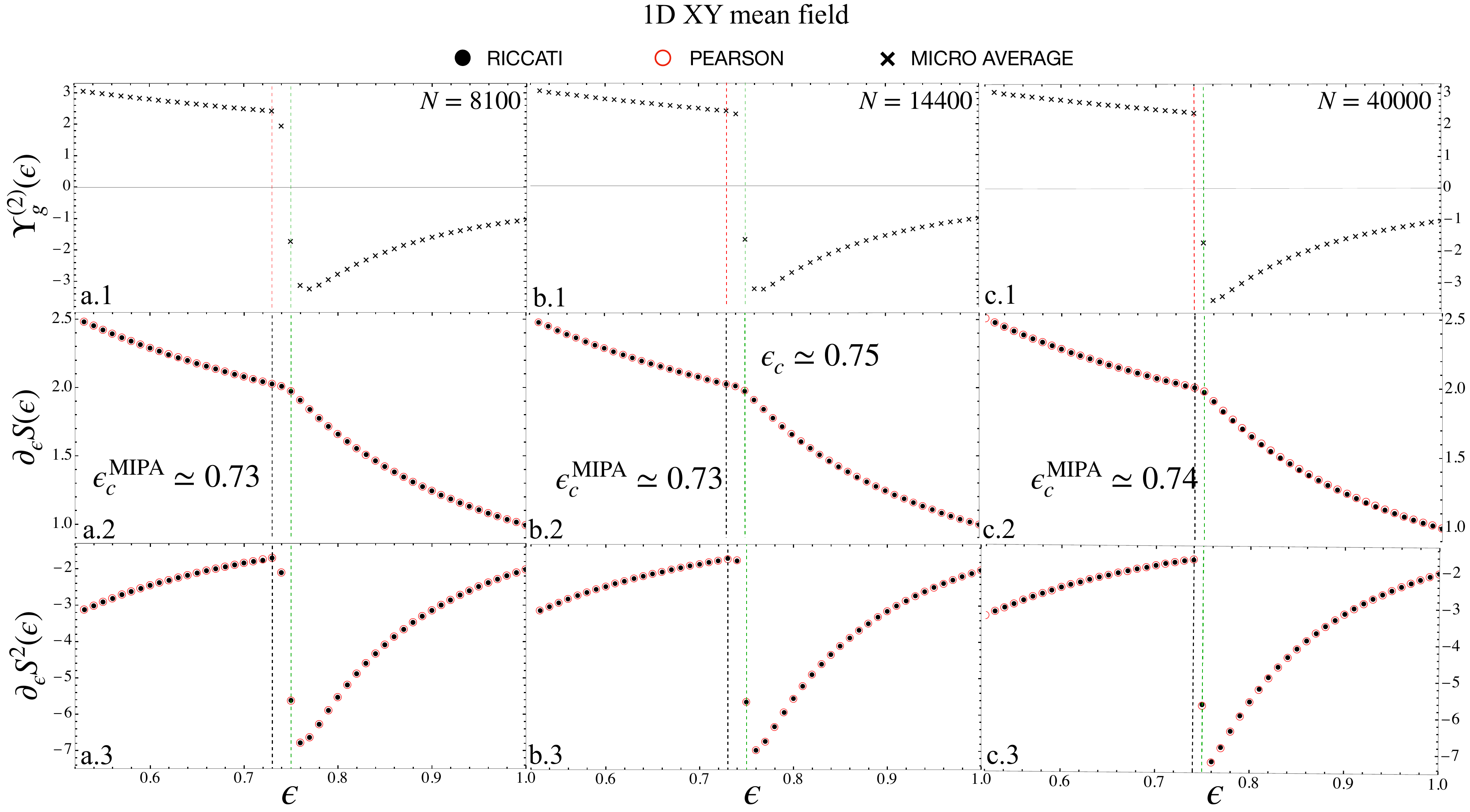}
\caption{\textbf{1D XY mean-field model: geometric input and reconstructed microcanonical observables.}  Columns correspond to system sizes $N=8100$ (a.1-a.3), $N=14400$ (b.1-b.3), and $N=40000$ (c.1-c.3). The first row shows the geometric quantity $\Upsilon^{(2)}_g(\epsilon)$ entering the EFE; the second row shows the inverse microcanonical temperature $\partial_\epsilon S(\epsilon)$; the third row shows $\partial_\epsilon S^{2}(\epsilon)$. Filled circles: reconstruction from the EFE scheme; open circles: PHT estimate; crosses: direct microcanonical averages for $\Upsilon_g^{(2)}$ used in the EFE. Vertical black dashed lines indicate the critical energy estimated through MIPA at different system sizes, vertical green dashed lines correspond to the infinite-size critical energy $\epsilon_c\simeq 0.75$. Note that MIPA estimates energy values that shift to $\epsilon_c^{\infty}$ for increasing system sizes.}
    \label{fig:efe-xy}
\end{figure*}

The procedure to solve the EFE is the same as the one adopted in Sec.~\ref{secsec:EFE-phi4}. We report the results in Fig.~\ref{fig:efe-xy}. The second-order GCF, $\Upsilon^{(2)}_g$, for the mean-field model is reported in panels \ref{fig:efe-xy}~\textbf{a.1} and \ref{fig:efe-xy}~\textbf{b.1} and \ref{fig:efe-xy}~\textbf{c.1}, respectively, for the sizes $N=8100,\,14400$ and $N=40000$. The solutions of the EFE (in terms of $\partial_\epsilon S$ and $\partial_\epsilon^2 S$) are compared with those from the PHT method, coinciding with the already ones reported in Fig.~\ref{fig:thermo-xy-mf}, panels \textbf{b.1} and \textbf{b.2}. Also, in the case of the mean field $(\sigma=-1)$, the qualitative and quantitative agreement is quite impressive.

\begin{figure}
    \centering
    \includegraphics[width=1\linewidth]{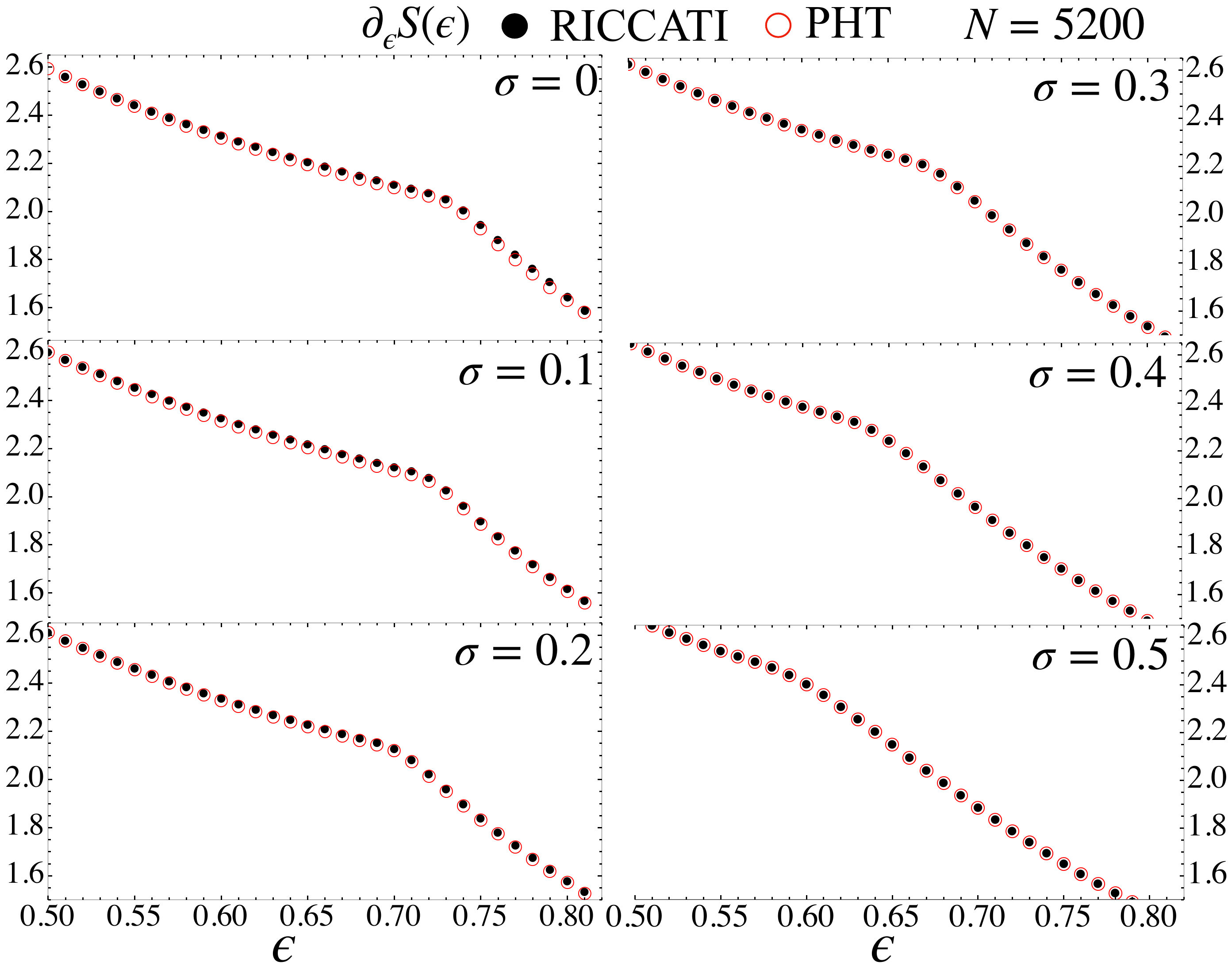}
%========================
% FIG. 11 (Weak LR regime)
%========================
\caption{\textbf{Weak long-range regime: cross-validation of the inverse microcanonical temperature.}
Microcanonical inverse temperature $\beta(\epsilon)=\partial_\epsilon S(\epsilon)$ for the 1D long-range XY model at several decay exponents $\sigma$ (one panel per $\sigma$, as labeled). Filled black circles: $\partial_\epsilon S(\epsilon)$ reconstructed from the geometric EFE/MIPA pipeline. Open red circles: independent PHT estimator computed directly from the microcanonical data set. The pointwise agreement throughout the energy window shows that the Riccati reconstruction reproduces the correct thermodynamic derivative also away from the mean-field point, i.e. for genuine $1/r^{-(1+\sigma)}$ interactions in the weak long-range regime. Note that an inflection point is visible in all of the $\sigma$-value investigated here. According with the microcanonical inflection point analysis, this produces a negative-valued maximum in $\partial^2_\epsilon S$ (see Fig.~\ref{fig:efe-second-der-xy-lr}) and therefore a finite-size second-order phase transition.}
    \label{fig:efe-first-der-xy-lr}
\end{figure}

\begin{figure}
    \centering
    \includegraphics[width=1\linewidth]{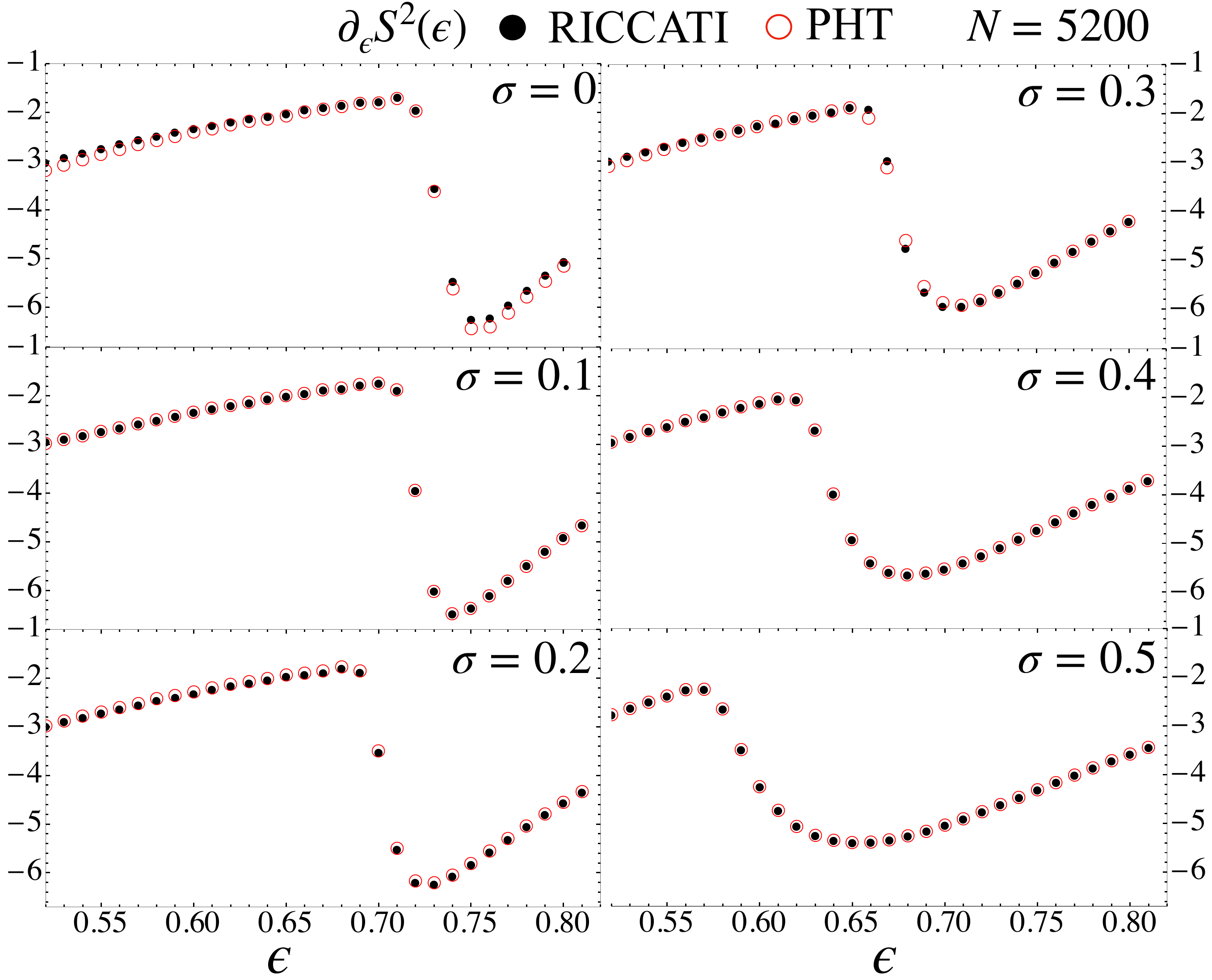}
%========================
% FIG. 12 (Weak LR regime)
%========================
\caption{\textbf{Weak long-range regime: cross-validation of the reconstructed thermodynamics in the weak long-range regime.}
Second derivative of the microcanonical entropy, $\partial_\epsilon^2 S(\epsilon)$, shown in the crossover window for $\sigma=0,0.1,\,0.2,\,0.3,\,0.4,\,0.5$ (panels). Filled symbols: EFE/MIPA reconstruction from geometric input on $\Sigma_E$; open symbols: independent PHT estimator computed directly from microcanonical data. The agreement across the full energy window and for all $\sigma$ demonstrates that the geometric reconstruction remains quantitatively reliable beyond the mean-field limit and does not rely on canonical assumptions. Note that a negative-valued maximum appears for all of the $\sigma$-values; this indicates the presence of a finite-size second-order phase transition according with the microcanonical inflection point analysis.}

    \label{fig:efe-second-der-xy-lr}
\end{figure}

%==========================================================
\subsection{Weak long-range regime: robustness of MIPA for long-range interactions}
\label{subsec:weak_LR}
%==========================================================

We now test the geometric microcanonical framework on the 1D long-range XY model with ferromagnetic couplings $J(r)\propto r^{-(d+\sigma)}$ and $d=1$ introduced in Sec.~\ref{ssec:1D-LR-XY}, in the regime where the interaction tail is weak in the sense that it no longer enforces mean–field–like collective behavior. In $d=1$, rigorous results exclude spontaneous magnetic order at any $T>0$ when the decay is fast enough (in the notation of this paper, this corresponds to $\sigma\ge 1$), while for intermediate exponents (roughly $0<\sigma<1$) the existence of a thermodynamic transition is not settled by general theorems.

Our aim here is to demonstrate that the geometric pipeline remains predictive and physically reliable when we leave the exactly solvable mean-field point and move to true $1/r^{-(1+\sigma)}$ interactions.
Concretely, we want to show that:
(i) the entropy reconstructed from the EFE yields microcanonical observables consistent with independent estimators;
(ii) the same geometric signatures introduced earlier continue to organize the thermodynamics when the interaction range is tuned.

\begin{figure*}
    \centering
    \includegraphics[width=0.85\linewidth]{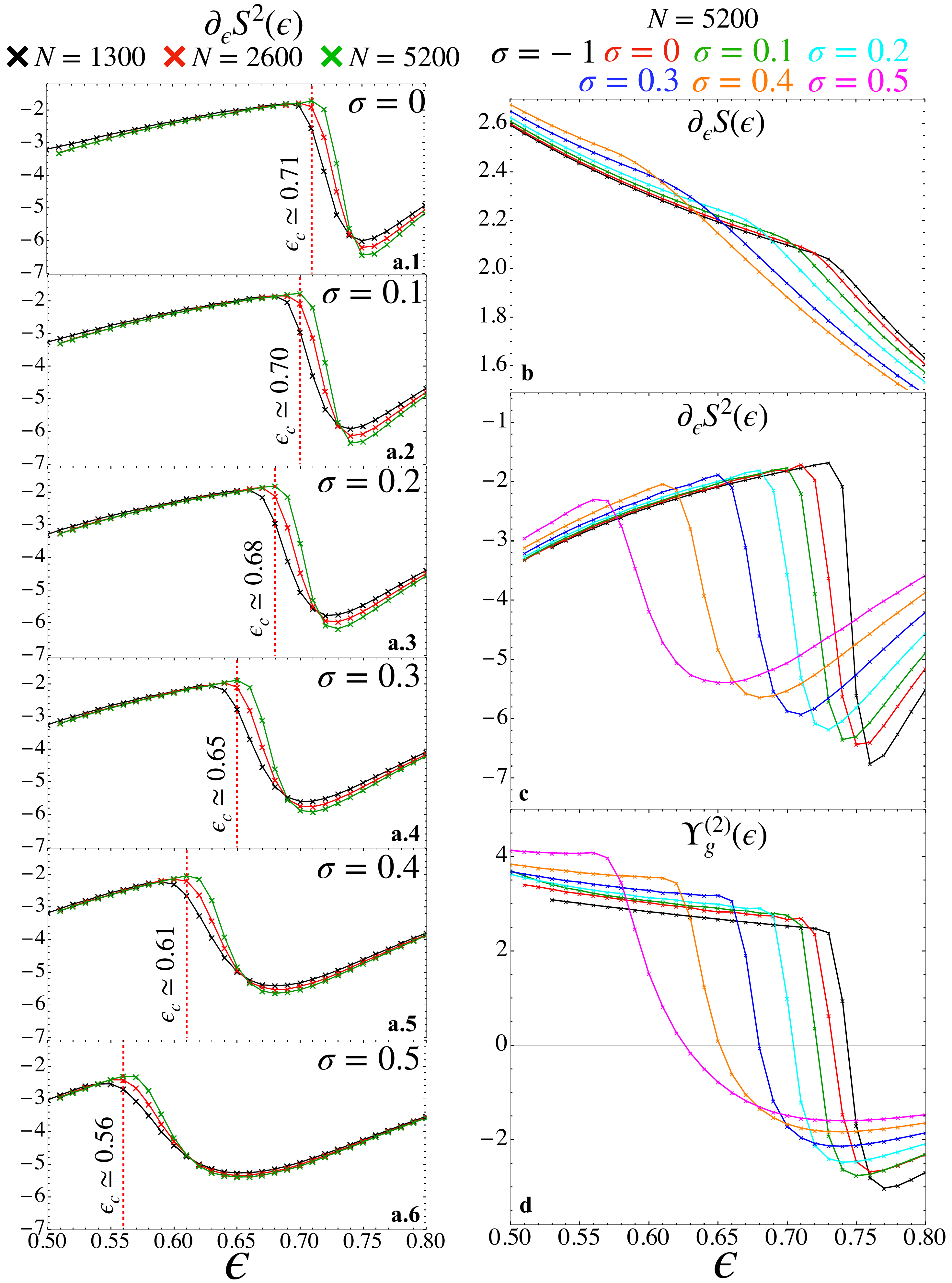}
%========================
% FIG. 10 (Weak LR regime)
%========================
\caption{\textbf{Weak long-range regime of the 1D long-range XY model: entropy derivatives and geometric response across interaction ranges.}
\textbf{(a.1--a.5)} Second energy derivative of the microcanonical entropy, $\partial_\epsilon^2 S(\epsilon)$, in the weak long-range regime for decay exponents $\sigma=0,\,0.1,\,0.2,\,0.3,\,0.4,\,0.5$ (one panel per $\sigma$). Symbols correspond to the Riccati/MIPA reconstruction for three system sizes $N=1300,2600,5200$ (as in the legend). The vertical dashed line marks the characteristic energy $\epsilon_c(\sigma)$ extracted from the MIPA morphology (annotated in each panel) and highlights the systematic drift of the crossover scale as the interaction tail is weakened.
\textbf{(b--d)} Fixed-size comparison at $N=5200$ showing how the thermodynamic morphology deforms with $\sigma$ (colors label $\sigma$ as in the legend): \textbf{(b)} inverse microcanonical temperature $\beta(\epsilon)=\partial_\epsilon S(\epsilon)$; \textbf{(c)} curvature/response channel $\partial_\epsilon^2 S(\epsilon)$; \textbf{(d)} purely geometric response indicator $\Upsilon_g^{(2)}(\epsilon)$. The close tracking between the thermodynamic response channel in \textbf{(c)} and the geometric observable in \textbf{(d)} shows that the same geometric mechanism organizing the mean-field case persists in the genuinely nonlocal $1/r^{1+\sigma}$ setting, while the progressive broadening and drift with $\sigma$ signal a smoother rather than a mean-field-like sharpening. We observe that for all of the $\sigma$-values simulated here, $\partial^2_\epsilon S$ admits a negative-valued maximum thus indicating a finite-size second-order phase transition according with the microcanonical inflection point analysis}
    \label{fig:thermo-LR-alpha}
\end{figure*}

\subsubsection{Our data and purposes}

As in the previous sections, the core object is the microcanonical entropy $S(\epsilon)$, reconstructed by solving the EFE driven by geometric averages on $\Sigma_E$.
From it, we form the microcanonical inverse temperature and curvature response $\beta(\epsilon)=\partial_\epsilon S(\epsilon)$ and $\partial_\epsilon^2 S(\epsilon)$. Note that $\partial_\epsilon^2 S(\epsilon)$ controls the specific heat and the local concavity of entropy.
In the weak long-range regime, we report, for several decay exponents $\sigma$ and increasing sizes (e.g., $N=1300,2600,5200$), a direct comparison between:
(a) the Riccati/MIPA reconstruction and
(b) the PHT (thermodynamic) estimator computed directly from microcanonical data (App.~\ref{sec:numerical-sampling}).
This comparison is displayed for $\partial_\epsilon S$ and $\partial_\epsilon^2 S$ in Figs.~\ref{fig:efe-first-der-xy-lr} and \ref{fig:efe-second-der-xy-lr}.
In addition, Fig.~\ref{fig:thermo-LR-alpha} shows the behavior of the geometric indicator $\Upsilon_g^{(2)}(\epsilon)$ introduced earlier, i.e., an ensemble average of purely geometric combinations (built from Weingarten/Hessian data) that act as a \emph{geometric surrogate} for the thermodynamic ``response'' channel.

By inspecting $\partial_\epsilon S$ and $\partial^2_\epsilon S$ in Figs.~\ref{fig:efe-first-der-xy-lr} and \ref{fig:efe-second-der-xy-lr}, we notice that, for all of the $\sigma$-values, an inflection point appears in $\partial_\epsilon S$ where the function passes from positive to negative curvature. This reflects a negative-valued maximum in $\partial^2_\epsilon S$. According to MIPA, this is the manifestation of a finite-size second-order phase transition (precursors).

\subsubsection*{Main observation 1: quantitative agreement survives for true long-range interactions}
The first key message from Fig.~\ref{fig:thermo-LR-alpha} is methodological and nontrivial: \emph{the Riccati reconstruction remains quantitatively consistent with an independent microcanonical estimator} even when the interaction is genuinely long--range (nonlocal in real space) and we are far from the exactly solved mean--field reference.
This matters because, in this regime, one cannot lean on closed--form solutions or on a canonical narrative; the agreement instead confirms that the information injected in the EFE---namely geometric data of the energy hypersurfaces---is sufficient to reproduce the actual microcanonical thermodynamics.

Operationally, in Fig.~\ref{fig:efe-first-der-xy-lr}, the curves of $\partial_\epsilon S(\epsilon)$ obtained from the Riccati route track the PHT curves across the full energy interval shown, and the same holds for $\partial_\epsilon^2 S(\epsilon)$ in Fig.~\ref{fig:efe-second-der-xy-lr}.
This is exactly the robustness statement that one wants from a ``geometry $\Rightarrow$ thermodynamics'' program: the reconstruction is not a delicate mean-field artifact.

\subsubsection*{Main observation 2: the method does \emph{not} create false phase transitions}

The second message is physical.
In the mean–field section ($\sigma=-1$), we saw the standard scenario: $\partial_\epsilon S$ develops a sharp inflection and $\partial_\epsilon^2 S$ a pronounced negative peak, whose location converges with $N$ to the known $\epsilon_c$ (and the finite–$N$ maxima fall on the analytic curve). This was the ``positive control'' of the method.

Here, the behavior is qualitatively different: in the weak long-range regime, the features in $\partial_\epsilon^2 S$ do \emph{not} display the same sharpening and $N$–stabilization that characterized the mean--field transition.
Instead, what appears is a behavior that one would usually classify as a \emph{smooth crossover structure} whose location drifts with the interaction range. It is not excluded that this smooth behavior transforms into a jump or non-analyticity in the thermodynamic limit. 
This is visible in Fig.~\ref{fig:thermo-LR-alpha} through the systematic shift of the characteristic energy scale (marked as $\epsilon_c$ in the panels) as $\sigma$ is increased: $\epsilon_c$ moves from $\simeq 0.71$ at $\sigma=0$ down to $\simeq 0.56$ at $\sigma=0.5$.
Indeed, at this level, our scope is not to determine the precise location of the phase transition and therefore the exact critical energy/temperature value, but to show that MIPA coherently detects the transition \emph{already} at finite-size and that it continues to do so as the system size increases. In parallel, our scope is to show that the geometric approach is able to predict the entire energy behavior of the thermodynamic observables.

This is precisely where the geometric approach is \emph{strong}: it provides a disciplined criterion, when combined with MIPA, to distinguish a finite-size precursor of an $N\to\infty$ phase transition and can determine the deep mechanism triggering the phase transition itself. In practice, the absence of mean-field-like stabilization as $N$ grows is the signature that the ``shape'' observed in $\partial_\epsilon^2 S$ is converging to a thermodynamic singularity.

\subsubsection*{Main observation 3: the infinite-size phase transition is already encoded in geometry}

A central conceptual point of this work is that \emph{thermodynamic response is geometrically pre-encoded in $\Sigma_E$}.
The weak long-range regime is an important stress test of this claim because the thermodynamic signal is weaker.
Figure~\ref{fig:thermo-LR-alpha} explicitly juxtaposes $\partial_\epsilon^2 S(\epsilon)$ with the geometric quantity $\Upsilon_g^{(2)}(\epsilon)$ (same energy window and same $\sigma$ set).
The fact that the characteristic structure (broad maximum/feature scale) is mirrored by $\Upsilon_g^{(2)}$ is the operative meaning of ``seeing the shape'': even when the system has not reached the sharp (infinite-size) transition, the \emph{same} geometric channel continues to track and organize the thermodynamic transition.

%%%%%%%%%%%%%%%%%%%%%%%%%%%

\section{Conclusions and Outlook}

We have established that thermodynamics can be formulated in a fully geometric fashion in terms of the structure $(\Lambda,H,\omega,\eta)$ without postulating \emph{a priori} the concept of a statistical ensemble. The microcanonical measure emerges naturally as the hypersurface element induced by the metric on energy manifolds, and entropy becomes a geometric quantity---the logarithm of the area of $\Sigma_E$---whose evolution is governed by the deterministic Entropy Flow Equation. This equation connects the derivatives of entropy directly to curvature invariants of the energy manifold (Weingarten operator, Hessian-derived scalars, Ricci curvatures), providing a self-contained relation between Hamiltonian dynamics and thermodynamic response. The explicit analytical and numerical investigations across paradigmatic models---the 1D mean-field XY model, the 2D $\phi^4$ lattice field theory, and the 1D XY chain with genuine long-range $1/r^{1+\sigma}$ interactions---demonstrate that phase transitions correspond to qualitative geometric reorganizations of energy hypersurfaces: elliptic (convex) geometries at low energy transform through mixed or hyperbolic (saddle) regions at criticality, leaving systematic fingerprints in the curvature sources that drive the entropy flow and in the concavity structure of $S(E)$ itself.

Combining the EFE with the microcanonical inflection-point analysis (MIPA), we establish a direct geometric foundation for phase transitions: each order of critical behavior classified by MIPA corresponds to a specific geometric profile of the source term $\Upsilon^{(k)}_g$ in the hierarchy of entropy flow equations, and the finite-size precursors identified by MIPA are shown to encode geometric deformations that converge to thermodynamic singularities as $N\to\infty$. The framework thus provides a deterministic bridge between geometry and criticality, revealing the \emph{mechanisms}---rather than merely the \emph{signatures}---that generate phase transitions at the microscopic level.

Beyond its conceptual implications, the present framework has broad scope and opens several directions of immediate physical relevance. First, it applies naturally to \emph{finite systems}, where the thermodynamic limit is neither achievable nor physically meaningful. Examples include complex biomolecules such as proteins and peptides~\cite{koci2017subphase,schnabel2011microcanonical,bachmann2014thermodynamics,aierken2020comparison}, atomic and molecular clusters, and---crucially---\emph{finite quantum platforms} in atomic, molecular, and optical (AMO) physics, such as trapped-ion arrays, Rydberg atom ensembles, and ultracold dipolar gases~\cite{defenu2023long}. In these systems, phase-transition-like phenomena (sharp changes in order parameters, diverging response functions, bistabilities) emerge at fixed, experimentally accessible system sizes $N\sim 10$--$10^3$, yet the conventional paradigm of thermodynamic-limit criticality does not apply. The geometric microcanonical approach is designed for precisely this regime: it operates systematically at finite $N$, encoding transition precursors and their $N$-scaling in the geometry of $\Sigma_E$ without requiring \emph{ad hoc} extrapolations or ensemble assumptions. Moreover, AMO platforms increasingly realize long-range interactions (dipolar, Rydberg, Coulomb) and constrained dynamics (gauge theories, fragmented Hilbert spaces), making the microcanonical description not merely convenient but often the only thermodynamically consistent starting point.

Second, the framework is essential for \emph{systems with long-range interactions}~\cite{campa2009statistical,dunkel2006phase,campa2014physics,dauxois2002dynamics,bouchet2010thermodynamics,defenu2023long,Giachetti2022Villain}, where ensemble inequivalence generically arises: the microcanonical and canonical ensembles predict different phase diagrams, different response functions, and even qualitatively distinct thermodynamic branches (e.g., negative heat capacities, non-concave entropies, backbending caloric curves)~\cite{barre2001inequivalence,dauxois2000violation}. In such cases, the microcanonical ensemble is thermodynamically \emph{more fundamental}~\cite{ellis2002nonequivalent,ellis2000large}: it is the only ensemble that consistently describes isolated systems and respects the additivity of entropy in the presence of non-additive interactions. Canonical or grand-canonical treatments may yield physically inconsistent predictions (e.g., ensemble-dependent phase boundaries), while renormalization-group approaches predicated on local interactions and scale invariance may break down entirely. The geometric formulation developed here provides the natural theoretical language for these systems, encoding thermodynamic response and transition structures directly in the curvature geometry of $\Sigma_E$, which remains well-defined and physically meaningful regardless of interaction range or ensemble equivalence.

Third, the formalism extends naturally to \emph{quantum systems}. The phase-space manifold $(\Lambda,\omega,H)$ generalizes to the projective Hilbert space equipped with the Fubini-Study metric and symplectic structure, and energy eigenstates foliate this space into stratified manifolds. This direction can be concretely taken and provides insights into the field of quantum thermodynamics.

Fourth, when combined with large-deviation theory, the geometric approach sheds light on the \emph{origin} of ensemble inequivalence~\cite{ellis2000large,ellis2002nonequivalent,touchette2005nonequivalent,ellis2004thermodynamic,touchette2003equivalence}: non-equivalent ensembles correspond to different geometric representations (different foliations, different metrics) of the same underlying phase space, and the breakdown of equivalence is encoded in how the induced measures on macrostate manifolds fail to concentrate uniformly. Finite-size corrections, metastable branches, and the thermodynamic stability hierarchy all acquire geometric interpretations in terms of subleading curvature terms and the topology of macrostate level sets.

Finally, the framework interfaces naturally with quantum information and cryptography, where the relevant ``phase space'' is the projective Hilbert space of pure states endowed with the Fubini-Study metric, and entanglement measures can be treated as macroscopic functionals on this curved manifold. In this spirit, we recently introduced a \emph{geometric entanglement entropy}~\cite{di2025geometric}: by viewing bipartite entanglement as a scalar field on projective Hilbert space, its level sets stratify the state manifold into constant-entanglement hypersurfaces, whose (weighted) log-area provides a microcanonical notion of degeneracy in entanglement space and can be computed explicitly in elementary spin examples. Building on the same viewpoint, we also proposed a geometric route to quantum cryptography \cite{di2025secret} in which the state manifold, its metric, and the allowed unitary moves are fully public, while secrecy is encoded in a hidden choice of entanglement foliation (selected by a key parameter): messages are conveyed through controlled trajectories that move upward, downward, or tangentially with respect to that foliation, leading to the notion of \emph{geometric entanglement codes}. 

\begin{acknowledgments}
The calculations presented in this paper were carried out using the HPC facilities of the University of Luxembourg~\cite{VBCG_HPCS14} {\small (see \href{http://hpc.uni.lu}{hpc.uni.lu})} and those of the Luxembourg national supercomputer MeluXina.
\end{acknowledgments}

\appendix

\section{Statistical mechanics as a symplectic Hamiltonian theory of thermodynamics}\label{sec:details-symplectic-formalism}

Let $(\Lambda,\omega)$ be a $2N$-dimensional phase space with a canonical symplectic form 
\[
    \omega=dp_i\wedge dq^i\,,
\] 
where the Einstein convention on repeated indices is introduced, and let \(H\) be a Hamiltonian function on $\Lambda$. 
Then, let us introduce a vector field $\bm{X}$ on a manifold $\Lambda$. By definition, this is an operator acting on a smooth function $f:\Lambda\rightarrow\mathbb{R}$ as a directional derivative along its integral curves $\gamma(t):=(\bm{p}(t),\bm{q}(t))$:
\begin{equation*}
\bm{X}[f] = \frac{d}{dt} f(\gamma(t))\Big|_{t=0},
\quad 
\frac{d\gamma(t)}{dt} = \bm{X}(\gamma(t)),\; \;\;\gamma(0)=\bm{x}_0.
\end{equation*}
Hence, in canonical coordinates $(q^i,p_i)$ the components of the Hamiltonian vector field 
\[
\bm{X}_H = a_i(q,p)\,\frac{\partial}{\partial q^i}
          + b_i(q,p)\,\frac{\partial}{\partial p_i}
\]
are identified with the time derivatives of the phase--space variables,
\begin{equation*}
    a_i=\dot q^i,\qquad b_i=\dot p_i.
\end{equation*}
The Hamiltonian flow is defined by the symplectic equation
\begin{equation*}
    \iota_{\bm{X}_H}\omega = dH, 
\end{equation*}

\textbf{\textit{The symplectic form generates the equations of motion.}}
Evaluating the contraction gives
\begin{equation*}
    \iota_{\bm{X}_H}\omega =  (\dot{p}_i\,dq^i - \dot{q}_i\,dp_i).
\end{equation*}
Equating this to 
$dH=\sum_i(\partial_{q^i}H\,dq^i+\partial_{p_i}H\,dp_i)$ yields
\begin{equation*}
    \dot q^i = \frac{\partial H}{\partial p_i},
    \qquad
    \dot p_i = -\,\frac{\partial H}{\partial q^i},
\end{equation*}

\textbf{\textit{Hamiltonian flow preserves the Liouville volume.}}
Another crucial property of the symplectic form is to give meaning to the volume integral through the Liouville volume form, which reads:
\begin{equation*}
  \mu_L = \frac{\omega^{\wedge N}}{N!}\,,
\end{equation*}
a nowhere-vanishing $2N$-form that orients $\Lambda$. The symbol $\wedge$ represents the wedge product of $\omega$ with itself.

The Hamiltonian flow preserves the Liouville volume form and the symplectic structure. This can be shown by introducing the the Lie derivative. For any differential form $\alpha$ and vector field $X$, the Lie derivative is defined by Cartan’s identity \cite{arnol2013mathematical}:
\begin{equation*}
\mathscr{L}_X \alpha = d(\iota_X \alpha) + \iota_X (d\alpha),
\end{equation*}
where $d$ is the exterior derivative, and $\iota_X$ denotes interior contraction.\\

\textbf{\textit{Hamiltonian flow preserves $\omega$.}}
Since $\omega$ is closed  $(d\omega=0)$ then applying Cartan’s identity to $\iota_{\bm X_H}\omega = dH$ yields
\begin{equation*}
    \mathscr{L}_{\bm X_H}\omega
    = d(\iota_{\bm X_H}\omega) + \iota_{\bm X_H}(d\omega)
    = d(dH) + \iota_{\bm X_H}(0)
    = 0.
\end{equation*}
Thus, the symplectic form $\omega$ is invariant under the Hamiltonian flow
$\Phi_t^H$, i.e.\ $(\Phi_t^H)^*\omega = \omega$.

\paragraph*{Hamiltonian flow preserves the Liouville form: Liouville’s theorem.}
Since the Lie derivative is a derivation with respect to the wedge product,
\begin{equation*}
  \mathscr{L}_{\bm X_H}\!\left(\omega^{\wedge n}\right)
  = n\,(\mathscr{L}_{\bm X_H}\omega)\wedge \omega^{\wedge(n-1)} \;=\; 0\,,
\end{equation*}
hence
\begin{equation*}
  \mathscr{L}_{\bm X_H}\mu_L \;=\; 0\,.
\end{equation*}
Therefore, the Liouville volume form $\mu_L$ is invariant under the Hamiltonian flow (Liouville’s theorem). In canonical coordinates $(q^i,p_i)$, with $\omega=\sum_{i=1}^n dq^i\wedge dp_i$, one has:
\begin{equation*}
  \mu_L \;=\; dq^1\wedge dp_1 \wedge \cdots \wedge dq^n\wedge dp_n\,.
\end{equation*}
The invariance $\mathscr{L}_{\bm X_H}\mu_L=0$ is equivalent to the fact that $X_H$ is divergence-free with respect to $\mu_L$. Indeed, the divergence of a vector field on a symplectic manifold is defined by
\begin{equation*}
   \mathscr{L}_{\bm{X}}\mu_L=(div\,X)\mu_L .
\end{equation*}
Now, with $\bm{X}=\bm{X}_H$, we have
\begin{equation*}
    div\,X_H=\partial_{q^i}\partial_{p_i}H+\partial_{p_i}(-\partial_{q^i}H)=0\,.
\end{equation*}

\paragraph*{Conservation laws.}
Any smooth function $f$ defines a unique Hamiltonian vector field $\bm X_f$ through
\begin{equation*}
  \iota_{\bm X_f}\omega = df.
\end{equation*}
Then, given two smooth functions $f$ and $g$, the symplectic form yields the Poisson bracket
\begin{equation*}
  \{f,g\} := \omega(\bm X_f,\bm X_g)
  = \bm X_g(f) = -\bm X_f(g).
\end{equation*}
Now, along the Hamiltonian flow
\begin{equation*}
  \dot{f} = dH(\bm{X}_f)=\omega(\bm{X}_H,\bm{X}_f) = \{f,H\}.
\end{equation*}
A function $f$ is a \emph{constant of motion} if and only if $\{f,H\}=0$, in which case $\bm X_f$ generates a symmetry of the Hamiltonian flow. 
If $f=F(H)$ is any smooth function of the Hamiltonian,
then $df=F'(H)\,dH$ and thus  $\iota_{\bm{X}_f}\omega=F'(H)\,\iota_{\bm{X}_H}\omega$, implying $\bm{X}_f=F'(H)\bm{X}_H$.
In conclusion,
\begin{equation}\label{def:integral-of-motion}
  \{f,H\}=\omega(\bm{X}_f,\bm{X}_H)
  =F'(H)\,\omega(\bm{X}_H,\bm{X}_H)=0,
\end{equation}
any function of the Hamiltonian is conserved along its own flow. More precisely, for any smooth function of the Hamiltonian, $f=F(H)$, one has $\bm X_f=F'(H)\bm X_H$ and $\{f,H\}=0$.  This observation leads to the introduction of the statistical ensembles: typically, microcanonical and canonical. They consist of introducing a weight, a function of the Hamiltonian, $\rho(H)$, in the Liouville volume form:
\[  
    d\mu_{\rho}=\rho(H)\,d\mu_\Lambda
\]
where $\rho_{\text{mc}}:=\delta(H-E)$ represents the microcanonical weight, and $\rho_{\text{mc}}:=e^{-\beta H}$ the canonical weight.

In conclusion, the symplectic form $\omega$ encodes the deterministic dynamics of a Hamiltonian system: it provides the equations of motion, conserves the Liouville measure, and generates invariants through the Poisson bracket. However, it does not generate or predict a statistical weight internally, and to give thermodynamic meaning, it is necessary to introduce ``external'' information encoded in the concept of ensembles compatible with Eq.~\eqref{def:integral-of-motion}.
In conclusion, statistical mechanics is founded on the collection of symplectic Hamiltonian systems and the postulate of statistical ensembles.

\section{Obstructions to the definition of measure on energy level sets}
\label{sec:obstruction}

In this section, we show and discuss the obstruction of the symplectic form in naturally inducing a measure on the energy hypersurfaces. For further details about the proofs and tools used in this section, we refer to Refs.~\cite{arnol2013mathematical,AbrahamMarsden,LeeSmooth,McDuffSalamon}.

\subsection{Degeneracy of the symplectic form on the energy hypersurfaces}\label{ssec:degeneracy}

We show that the restriction of the symplectic form $\omega$ to the tangent bundle of $\Sigma_E$ is degenerate, with the Hamiltonian vector field $\bm X_H$ in its kernel.\\

Let ${\bm x} \in \Sigma_E$ and consider the restriction $\omega|_{T_{\bm x} \Sigma_E}$, which we denote by $\omega_E$. For any vector $\bm v \in T_{\bm x} \Sigma_E$ (that is, $\bm v$ tangent to the level set $H = E$), we have, by definition:
\begin{equation*}
    dH(\bm v) = 0.
\end{equation*}

By the defining equation of the Hamiltonian vector field, $\iota_{\bm X_H} \omega = dH$, we have for any vector $v$:
\begin{equation*}
\omega(\bm X_H, \bm v) = dH(\bm v).
\end{equation*}

In particular, for $\bm v \in T_{\bm x} \Sigma_E$:
\begin{equation*}
    \omega_E(\bm X_H,\bm v) = \omega(\bm X_H, \bm v) = dH(\bm v) = 0.
\end{equation*}

Since this holds for all $\bm v \in T_{\bm x} \Sigma_E$, we have $\bm X_H \in \ker(\omega_E)$. The kernel is non-trivial; hence, $\omega_E$ is degenerate.

Moreover, $\bm X_H(x) \in T_{\bm x} \Sigma_E$ because
\begin{equation*}
    dH(\bm X_H) = \omega(\bm X_H, \bm X_H) = 0
\end{equation*}
by the antisymmetry of $\omega$. Thus, $\bm X_H$ is both tangent to $\Sigma_E$ and in the kernel of $\omega$ restricted to $\Sigma_E$. 

\subsection{The Symplectic Complement and Coisotropic Submanifolds}
\label{ssec:coisotropy-complement}

Here, we show that the symplectic complement of $T_{\bm x} \Sigma_E$ is one-dimensional and is generated by $\bm X_H$. Furthermore, $(T_{\bm x} \Sigma_E)^{\perp_\omega} \subseteq T_{\bm x} \Sigma_E$, making $\Sigma_E$ a coisotropic submanifold.\\

Consider the symplectic complement, which is defined as
\begin{equation*}
    (T_{\bm x} \Sigma_E)^{\perp_\omega} = \{\bm w \in T_{\bm x} \Lambda : \omega(\bm w, \bm v) = 0 \; \forall \bm v \in T_{\bm x} \Sigma_E\}.
\end{equation*}

We first show that $\bm X_H \in (T_{\bm x} \Sigma_E)^{\perp_\omega}$. For any $\bm v \in T_{\bm x} \Sigma_E$:
\begin{equation*}
\omega(\bm X_H, \bm v) = dH(\bm v) = 0,
\end{equation*}
as shown in Sec.~\ref{ssec:degeneracy}. Thus $\bm X_H \in (T_{\bm x} \Sigma_E)^{\perp_\omega}$.

To show that $\bm X_H$ generates the entire complement, we use a dimensional argument. For a subspace $W$ of a symplectic vector space $(V, \omega)$ of dimension $2N$, the symplectic complement satisfies:
\begin{equation*}
\dim(W) + \dim(W^{\perp_\omega}) = 2N.
\end{equation*}

In our case, $\dim(T_{\bm x} \Sigma_E) = 2N - 1$ (since $\Sigma_E$ is a codimension-1 hypersurface), therefore:
\begin{equation}
\dim((T_{\bm x} \Sigma_E)^{\perp_\omega}) = 2N - (2N-1) = 1.
\end{equation}

Since $\bm X_H$ is non-zero (assuming $H$ is a regular function, i.e., $dH \neq 0$ on $\Sigma_E$) and belongs to $(T_{\bm x} \Sigma_E)^{\perp_\omega}$, and this complement has dimension $1$, we conclude:
\begin{equation*}
(T_{\bm x} \Sigma_E)^{\perp_\omega} = \mathrm{span}\{\bm X_H\}.
\end{equation*}

Finally, we verify that $(T_{\bm x} \Sigma_E)^{\perp_\omega} \subseteq T_{\bm x} \Sigma_E$. We have already shown that $\bm X_H \in T_x \Sigma_E$ (since $dH(\bm X_H) = 0$). Thus, the symplectic complement is contained in the tangent space itself. This is precisely the defining property of a coisotropic submanifold \cite{McDuffSalamon}. 

\subsection{Non-uniqueness of Transverse Vectors}
\label{ssec:non-unique-transverse}
Finally, we show that the condition $dH(\bm\xi_\eta) = 1$ does not uniquely determine a vector field $\bm\xi_\eta$ transverse to $\Sigma_E$. The solution space has dimension $2N - 1$ at each point.\\

At each point $\bm x \in \Lambda$, we seek vectors $\bm\xi_\eta \in T_{\bm x} \Lambda$ that satisfy the linear equation
\begin{equation*}
dH(\bm\xi_\eta) = 1.
\end{equation*}

This is a single linear equation in a $2N$-dimensional vector space. The solution set is an affine hyperplane:
\begin{equation*}
S = \{\bm\xi_\eta \in T_{\bm x} \Lambda : dH(\bm\xi_\eta) = 1\}.
\end{equation*}

To find the dimension, we first note that the solutions to the homogeneous equation $dH(\bm\xi_\eta) = 0$ form the tangent space $T_{\bm x} \Sigma_E$, which has dimension $2N - 1$ \cite{LeeSmooth}.

The solution space $S$ to the inhomogeneous equation $dH(\bm\xi_\eta) = 1$ is a translation of this vector space (an affine subspace); hence, it also has dimension $2N - 1$.

Explicitly, if $\bm\xi_\eta^0$ is any particular solution (for example, $\bm\xi_\eta^0 = \partial_{p_1}/(\partial H/\partial p_1)$ if $\partial H/\partial p_1 \neq 0$), then the general solution is:
\begin{equation*}
\bm\xi_\eta = \bm\xi_\eta^0 + \bm v, \quad \text{where } \bm v \in T_{\bm x} \Sigma_E.
\end{equation*}

Since $T_{\bm x} \Sigma_E$ has dimension $2N - 1$, we have a $(2N-1)$-parameter family of transverse vectors.

Different choices of $\bm\xi_\eta$ yield different $(2N-1)$-forms $\iota_{\bm\xi_\eta} d\mu_\Lambda$ when restricted to $\Sigma_E$. Without additional structure (such as a metric to define a canonical normal direction), there is no geometric principle within the symplectic framework to select one element of $S$ over another. 

\vspace{1em}
\noindent We remark that these three obstructions establish that symplectic geometry, while sufficient to identify the energy shell $\Sigma_E$ and govern the dynamics upon it, does not provide the geometric data necessary to measure the volume of $\Sigma_E$. The introduction of a metric tensor resolves all three obstructions simultaneously: it eliminates the dimensional mismatch by providing the gradient $\nabla_{\!\eta} H$ as a canonical transverse direction; it removes the non-uniqueness by defining orthogonality; and it repairs the degeneracy by inducing a non-degenerate metric on $\Sigma_E$ itself.

\section{Geometric clock}
\label{sec:geometric-clock}

Let us consider Eqs.~\eqref{def:energetic-step} and \eqref{def:geometric-displacement} with $\alpha = 1/\|\nabla_{\!\eta} H\|_\eta$. With this choice, $\|\bm\xi_\eta\|_\eta = 1$ coincides with the unit normal vector to $\Sigma_E$. 

\begin{figure}[H]
    \centering
    \includegraphics[width=0.9\linewidth]{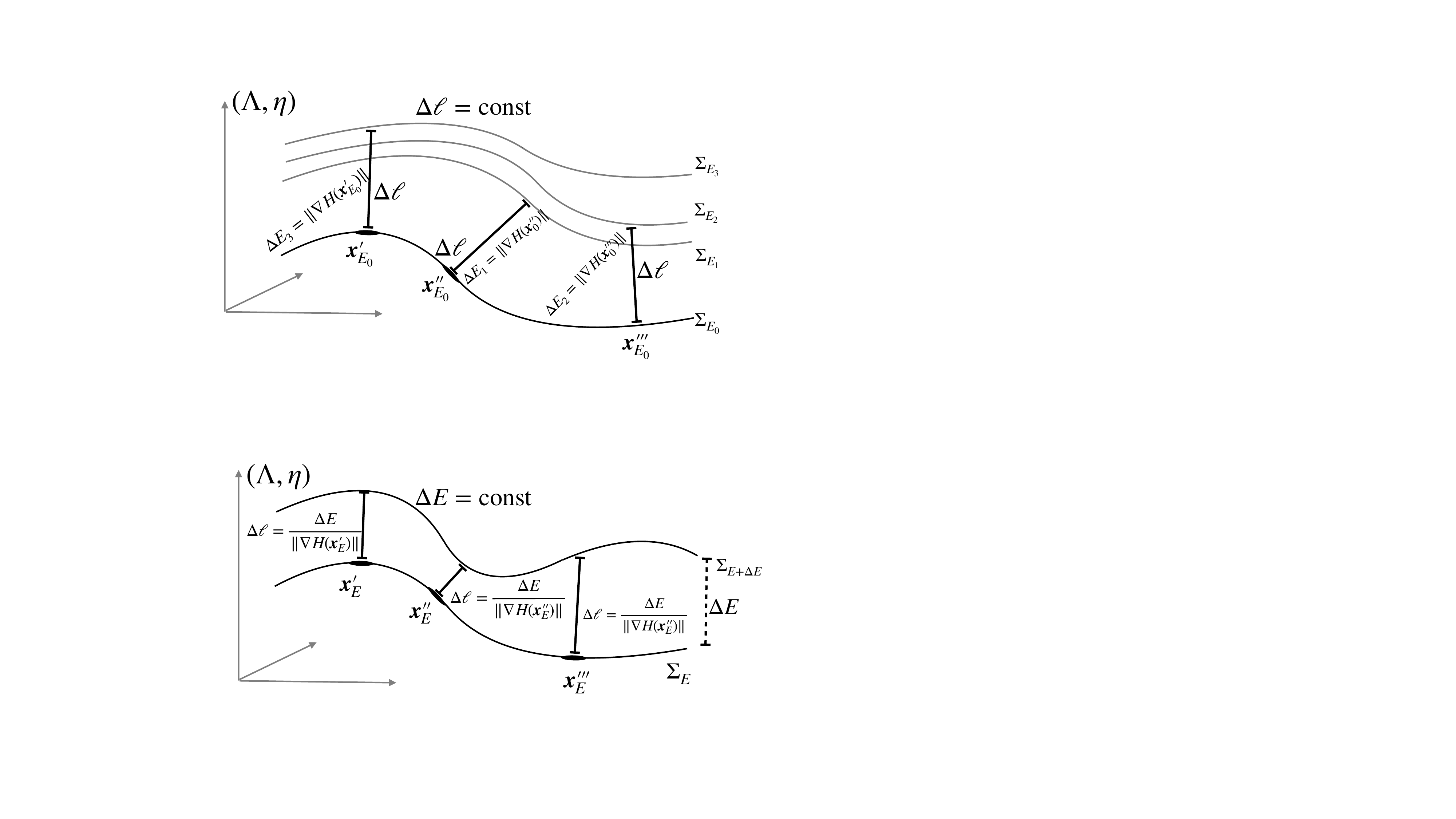}
    \caption{\textbf{Pictorial representation of the geometric clock}. The geometric step $\Delta\ell$ is fixed while the energetic step $\Delta E=\|\nabla_{\!\eta} H(\bm{x})\|_\eta \Delta\ell$ can vary. Different point on the same hypersurface are mapped on different hypersurfaces. We control the geometric distance but not the energetic step.}
    \label{fig:geometric-clock}
\end{figure}

The parameter $\epsilon$ measures geometric distance: 
\[
    d\ell = d\epsilon\,,
\]
and it is constant for every point on $\Sigma_E$. Therefore, the resulting flow is geometric, namely, a ``rigid'' $d\ell$-displacement of each point of $\Sigma_E$. In units of geometric steps, say, $d\ell=1$, the clock varies; indeed: 
\[
    dE=\|\nabla_\eta H(\bm{x})\|_\eta \;d\ell=\|\nabla_{\!\eta} H(\bm{x})\|_\eta\,,
\]
that is, the spacing of energy shells depends on the local gradient magnitude.
Since for any pairs $\bm{x}\neq\bm{x}'\in\Sigma_E$, we have, in general:
\[
    \|\nabla_{\!\eta} H(\bm{x})\|_{\eta}\neq\|\nabla_{\!\eta} H(\bm{x}')\|_\eta\,,
\]
then, two points on the same $\Sigma_E$ are mapped onto different hypersurfaces (see Fig.~\eqref{fig:geometric-clock}).

\section{The canonical ensemble as geometric deformation}

\subsection{Thermal deformation of the metric}

The microcanonical framework establishes a natural metric on phase space:
\begin{equation}
    \eta = \frac{dE \otimes dE}{\|\nabla_{\!\eta} H\|^2} + \sigma_E\,,
\end{equation}
with energy clock $\bm\xi = \nabla_{\!\eta} H / \|\nabla_{\!\eta} H\|^2_\eta$ satisfying $dH(\bm\xi_\eta) = 1$. The canonical ensemble emerges by exponentially rescaling the transverse direction:
\begin{equation}\label{eq:thermal-metric}
    \tilde{\eta}^{(\beta)} = \frac{e^{-2\beta E}}{\|\nabla_{\!\eta} H\|^2} dE \otimes dE + \sigma_E.
\end{equation}

The induced volume form becomes:
\begin{equation}
    d\tilde{\mu}_\Lambda = \sqrt{\det(\tilde{\eta})} \, dE \wedge dy^{2N-1} = \frac{e^{-\beta E}}{\|\nabla_{\!\eta} H\|} d\sigma_E \wedge dE,
\end{equation}
reproducing the Boltzmann factor. Normalizing gives the canonical distribution:
\begin{equation}
    d\mu^{(\beta)}_\Lambda =\frac{d\sigma_E}{\|\nabla_{\!\eta} H\|} \wedge e^{-\beta E}dE.
\end{equation}

\subsection{The thermal clock}

The gradient in the thermal metric satisfies $\tilde{\eta}(\tilde{\nabla} H, \bm v) = dH(\bm v)$. Since only the transverse part is rescaled:
\begin{equation}
    \tilde{\nabla} H = e^{2\beta E} \nabla_{\!\eta} H.
\end{equation}
The gradient grows exponentially with energy in the thermal geometry. However, its norm:
\begin{equation}
    \|\tilde{\nabla} H\|^2_{\tilde{\eta}} = e^{2\beta E} \|\nabla_{\!\eta} H\|_\eta^2
\end{equation}
also grows by the same factor, yielding:
\begin{equation}
    \bm\xi_{\tilde{\eta}} = \frac{\tilde{\nabla} H}{\|\tilde{\nabla} H\|^2_{\tilde{\eta}}} = \frac{e^{2\beta E} \nabla_{\!\eta} H}{e^{2\beta E} \|\nabla_{\!\eta} H\|_\eta^2} = \frac{\nabla_{\!\eta} H}{\|\nabla_{\!\eta} H\|_\eta^2} = \bm\xi_\eta.
\end{equation}

\textit{The clock remains invariant.} The thermal deformation stretches phase space distances exponentially with energy, but the generator of the energy foliation is unchanged: $dH(\bm\xi_{\tilde{\eta}}) = 1$ in both geometries.

\subsection{Origin from reparametrization}

The thermal metric \eqref{eq:thermal-metric} arises from a natural time reparametrization. Define the thermal time coordinate $s$ by:
\begin{equation}
    ds = e^{-\beta E} dE.
\end{equation}
Integrating:
\begin{equation}
    s(E) = -\frac{1}{\beta} e^{-\beta E} \quad \Rightarrow \quad E(s) = -\frac{1}{\beta} \ln(-\beta s).
\end{equation}
In the coordinate $s$, the metric takes the canonical form:
\begin{equation}
    \tilde{\eta}^{(\beta)} = \frac{ds \otimes ds}{\|\nabla_{\!\eta} H\|_\eta^2} + \sigma_E,
\end{equation}
identical in structure to the microcanonical metric but parametrized by $s$ instead of $E$. The transformation $E \mapsto s$ compresses high-energy regions: as $E \to \infty$, the interval $dE$ maps to an exponentially smaller $ds \sim e^{-\beta E} dE$. This geometric compression mirrors the probabilistic suppression $\rho \sim e^{-\beta E}$ of high-energy states.

\subsection{Geometric and physical interpretation}

The geometric framework is then able to incorporate the canonical ensemble and, in general, all other ensembles as alternative Riemannian geometries on the same phase space. The temperature $\beta$ controls the intensity of an exponential deformation in the energy direction.

In the thermal metric $\tilde{\eta}^{(\beta)}$, the ``distance'' between energy shells $\Sigma_E$ and $\Sigma_{E+dE}$ grows as $e^{\beta E}$. High-energy regions are geometrically remote. This encodes the physical intuition that accessing high-energy microstates requires overcoming an exponentially large entropic barrier.

The thermal geometry realizes the principle of maximum entropy subject to fixed average energy $\langle E \rangle$. The Boltzmann factor $e^{-\beta E}$ is the Lagrange multiplier enforcing this constraint. Geometrically, this translates into an exponential warping of phase space that concentrates volume near energies compatible with the thermal constraint.

The clock $\bm\xi$ is the same: it measures energy increments at a unit rate in both geometries. What changes is the metric distance required to traverse an energy interval $dE$. At low temperature ($\beta \to \infty$), the geometry becomes infinitely stretched at high energies, effectively confining the system to the ground state region. At high temperature ($\beta \to 0$), the deformation vanishes and the thermal geometry approaches the microcanonical one—the system explores phase space uniformly.

For macroscopic systems, fluctuations $\Delta E / \langle E \rangle \sim N^{-1/2}$ vanish in the thermodynamic limit. The canonical distribution $\rho \sim e^{-\beta E}$ becomes sharply peaked near $E^* = \langle E \rangle$, and the thermal geometry locally coincides with the microcanonical geometry restricted to $\Sigma_{E^*}$. The two geometries—and the two ensembles—become physically indistinguishable for macroscopic observables.

Thus, the geometric framework unifies the microcanonical and canonical descriptions: they correspond to different Riemannian structures on the same symplectic manifold, related by an energy-dependent conformal rescaling. Temperature is the parameter controlling this geometric deformation, with a clear physical interpretation as the inverse energy scale characterizing thermal fluctuations.

\section{Calculation of second-order GCF}
\label{app:calculation-2-GCF}

In this appendix, the subscript $\eta$ is intended for each object that requires it. We do not insert it in order to lighten the notation.

Given
\begin{align}
    \mathrm{Tr}[W_{\bm\xi}]
    =\frac{\Delta H}{\|\nabla H\|^{2}}
     -2\,\frac{\langle\nabla H,(\text{Hess}\,H)\,\nabla H\rangle}{\|\nabla H\|^{4}},
\end{align}
with
\begin{align}
    \bm\xi=\frac{\nabla H}{\|\nabla H\|^{2}},
\end{align}
we introduce the shorthand notation:
\[
\begin{split}
    u&:=\nabla H,\qquad 
    G:=\|\nabla H\|^{2},\qquad
    \\
    A&:=\Delta H,\qquad
    B:=\langle u,(\text{Hess}\,H)u\rangle.
\end{split}
\]
Then,
\begin{equation}
    \mathrm{Tr}[W_{\bm\xi}]=\frac{A}{G}-\frac{2B}{G^{2}}.
\end{equation}
Since $\partial_{E}=\mathscr{L}_{\bm\xi}=(\bm\xi\!\cdot\!\nabla)=(u/G)\!\cdot\!\nabla$, one finds
\begin{align}
    \partial_{E}\mathrm{Tr}[W_{\bm\xi}]
    &=\frac{\langle u,\nabla A\rangle}{G^{2}}
      -\frac{2A\langle u,\nabla G\rangle}{G^{3}}
      \nonumber\\
      &-\frac{2\langle u,\nabla B\rangle}{G^{3}}
      +\frac{4B\langle u,\nabla G\rangle}{G^{4}}.
\end{align}
Using the auxiliary relations
\[
\begin{split}
    &\langle u,\nabla G\rangle = 2B, \\ 
    &\langle u,\nabla B\rangle = 2\,\langle u,(\text{Hess}\,H)^{2}u\rangle
+\nabla^{3}H(u,u,u),
\end{split}
\]
we obtain
\begin{equation}
\begin{split}
    \partial_{E}\mathrm{Tr}[W_{\bm\xi}]=&\frac{\langle u,\nabla A\rangle}{G^{2}}
    -\frac{2AB}{G^{3}}+\frac{8B^{2}}{G^{4}}\\
    -&\frac{4\,\langle u,(\text{Hess}\,H)^{2}u\rangle+2\,\nabla^{3}H(u,u,u)}{G^{3}}.
\end{split}
\end{equation}
Furthermore,
\[
    \big(\mathrm{Tr}[W_{\bm\xi}]\big)^{2}
   =\frac{A^{2}}{G^{2}}
    -\frac{4AB}{G^{3}}
    +\frac{4B^{2}}{G^{4}}\,.
\]
Therefore:
\[
\begin{split}
    \partial_{E}\mathrm{Tr}[W_{\bm\xi}]&+\big(\mathrm{Tr}[W_{\bm\xi}]\big)^{2}\\
    &=
   \frac{A^{2}+\langle u,\nabla A\rangle}{G^{2}}
   +\frac{12B^{2}}{G^{4}}\\
   &-\frac{6AB+4\,\langle u,(\text{Hess}\,H)^{2}u\rangle
      +2\,\nabla^{3}H(u,u,u)}{G^{3}}.
\end{split}
\]
Collecting terms gives the final result:
\begin{widetext}
\begin{equation}
\begin{aligned}
\partial_{E}^{2}{\rm area}^{g}(E)
   &=\int_{\Sigma_E}\!\!
   \Bigg[
   \frac{A^{2}+\langle u,\nabla A\rangle}{G^{2}}
   -\frac{6AB+4\,\langle u,(\text{Hess}\,H)^{2}u\rangle
      +2\,\nabla^{3}H(u,u,u)}{G^{3}}
   +\frac{12B^{2}}{G^{4}}
   \Bigg]d\mu^g_E\\[4pt]
   &=\int_{\Sigma_E}\!\!
   \Bigg[
   \frac{(\Delta H)^{2}+\langle\nabla H,\nabla(\Delta H)\rangle}{\|\nabla H\|^{4}}
   -\frac{
      6\,(\Delta H)\langle\nabla H,(\text{Hess}\,H)\nabla H\rangle}{\|\nabla H\|^{6}}-\frac{
      4\,\langle\nabla H,(\text{Hess}\,H)^{2}\nabla H\rangle
      }{\|\nabla H\|^{6}}\\
      &\hspace{5cm}-\frac{2\,\nabla^{3}H(\nabla H,\nabla H,\nabla H)
      }{\|\nabla H\|^{6}}
   +\frac{
      12\,\langle\nabla H,(\text{Hess}\,H)\nabla H\rangle^{2}
      }{\|\nabla H\|^{8}}
   \Bigg]d\mu^g_E.
\end{aligned}
\end{equation}
Finally, the second-order GCF reads
\begin{equation}
    \begin{split}
\Upsilon^{(2)}_g(E)=\frac{\partial_{E}^{2}{\rm area}^{g}(E)}{\text{area}^g(E)}
   &=\int_{\Sigma_E}\!\!
   \Bigg[
   \frac{(\Delta H)^{2}+\langle\nabla H,\nabla(\Delta H)\rangle}{\|\nabla H\|^{4}}
   -\frac{
      6\,(\Delta H)\langle\nabla H,(\text{Hess}\,H)\nabla H\rangle}{\|\nabla H\|^{6}}-\frac{
      4\,\langle\nabla H,(\text{Hess}\,H)^{2}\nabla H\rangle
      }{\|\nabla H\|^{6}}\\
      &\hspace{5cm}-\frac{2\,\nabla^{3}H(\nabla H,\nabla H,\nabla H)
      }{\|\nabla H\|^{6}}
   +\frac{
      12\,\langle\nabla H,(\text{Hess}\,H)\nabla H\rangle^{2}
      }{\|\nabla H\|^{8}}
   \Bigg]d\rho^g_E.
    \end{split}
\end{equation}
with $d\rho^g_E=d\mu_E^g/\text{area}^g(E)$. In practice, in numerical simulations, one considers the microcanonical average 
\[
    \langle O\rangle_E=\int O(\pi,\phi)\delta(H(\pi,\phi)-E)~D\pi\,D\phi,
\]
and the GCF is estimated by computing
\begin{equation}\label{def:estimation-GCF}
    \begin{split}
\Upsilon^{(2)}_g(E)=\left\langle  \frac{(\Delta H)^{2}+\langle\nabla H,\nabla(\Delta H)\rangle}{\|\nabla H\|^{4}}\right\rangle_E&-\left\langle\frac{
      6\,(\Delta H)\langle\nabla H,(\text{Hess}\,H)\nabla H\rangle}{\|\nabla H\|^{6}}\right\rangle_E-\left\langle\frac{
      4\,\langle\nabla H,(\text{Hess}\,H)^{2}\nabla H\rangle
      }{\|\nabla H\|^{6}}\right\rangle_E\\
      &-\left\langle\frac{2\,\nabla^{3}H(\nabla H,\nabla H,\nabla H)
      }{\|\nabla H\|^{6}}\right\rangle_E
   +\left\langle\frac{
      12\,\langle\nabla H,(\text{Hess}\,H)\nabla H\rangle^{2}
      }{\|\nabla H\|^{8}}\right\rangle_E\,.
    \end{split}
\end{equation}
\end{widetext}

\section{Complex–exponential magnetization and angular derivatives for the 1D XY mean-field model}
\label{app:complex-mag}

This appendix gathers the identities and derivations used throughout Sec.~\ref{sec:geometric-change-XY-MF}.

\subsection{Complex magnetization and basic identities}
\label{app:complex-magn}
Let us introduce the complex magnetization
\begin{equation}
    \mathcal{M}:=\frac{1}{N}\sum_{j=1}^N e^{\,i\theta_j}
    = M\,e^{\,i\phi}\,,
\end{equation}
with $\phi\in[0,2\pi)$. Then, introducing the notation $M=\big|\mathcal{M}\big|$ and $\phi=\arg \mathcal{M}$, we obtain
\begin{equation}
    M\cos\phi=\frac{1}{N}\sum_j \cos\theta_j,\quad
    M\sin\phi=\frac{1}{N}\sum_j \sin\theta_j.
\end{equation}
For any angles $\theta_i,\theta_j$, we define $\Theta_i:=\theta_i-\phi$, the potential function (and any observable, in general) can be written in terms of the following basic sums:
\begin{align}
    \sum_{j=1}^N e^{-i\theta_j}&= N\,\overline{\mathcal{M}}=N\,M\,e^{-i\phi},\\
    \sum_{j=1}^N \cos(\theta_i-\theta_j)
    &= \Re\bigg(e^{i\theta_i}\sum_{j}e^{-i\theta_j}\bigg)\nonumber\\
    &= \Re\big(NM\,e^{i(\theta_i-\phi)}\big)=N\,M\cos\Theta_i,\\
    \sum_{j=1}^N \sin(\theta_i-\theta_j)
    &= \Im\bigg(e^{i\theta_i}\sum_{j}e^{-i\theta_j}\bigg)\nonumber\\
    &= \Im\big(NM\,e^{i(\theta_i-\phi)}\big)=N\,M\sin\Theta_i.
\end{align}
where $\Re$ and $\Im$ are, respectively, the real and imaginary parts of a complex number.
Double sums collapse similarly:
\begin{align}
    \sum_{i,j=1}^N \cos(\theta_i-\theta_j)
    &=\Re\bigg(\sum_i e^{i\theta_i}\sum_j e^{-i\theta_j}\bigg)\nonumber\\
    &= \Re\big(N^2 \mathcal{M}\overline{\mathcal{M}}\big)
    = N^2 M^2,\\
\sum_{i,j=1}^N \sin(\theta_i-\theta_j)
&=\Im\bigg(\sum_i e^{i\theta_i}\sum_j e^{-i\theta_j}\bigg)=0.
\end{align}

\subsection{1D XY mean-field Hamiltonian, potential, and its derivatives}

The 1D XY mean-field Hamiltonian reads
\begin{equation}
    H(\theta,p)=\sum_{i=1}^N \frac{p_i^2}{2} +\frac{J}{2N}\sum_{i,j=1}^N\bigl[1-\cos(\theta_i-\theta_j)\bigr].
\end{equation}
Using $\sum_{i,j} \cos(\theta_i-\theta_j)=N^2M^2$, one obtains the compact expression
\begin{equation}
V(\theta)=\frac{J}{2}N\,(1-M^2).
\end{equation}
The first derivatives of $V$ (namely, the components of $\nabla V$) are obtained by differentiating $V$ with respect to $\theta_i$:
\begin{align}
    \partial_{\theta_i}V
    =\frac{J}{N}\sum_{j}\sin(\theta_i-\theta_j).
\end{align}
By the complex–magnetization identity above, we obtain
\begin{equation}\label{eqn:partial-V}
    \partial_{\theta_i}V = J\,M\,\sin(\theta_i-\phi) = J\,M\,\sin\Theta_i.
\end{equation}

The second derivatives (namely, the components of $\text{Hess}V$) are obtained by differentiating Eq.~\eqref{eqn:partial-V} once more. For $i=j$, we have:
\begin{align}
\text{Hess}V_{ii}
&=\frac{J}{N}\sum_{j}\cos(\theta_i-\theta_j)
=\,J\,M\,\cos\Theta_i\,.
\end{align}
For $i\neq j$, we obtain:
\begin{align}
    \text{Hess}V_{ij}= -\frac{J}{N}\cos(\theta_i-\theta_j)\, .
\end{align}

\subsection{Laplacian, gradient norm, and contraction with the Hessian}

To compute geometric curvature observables such as the trace of the Weingarten operator, it is convenient to introduce a useful combination. 

We denote $A_{ij}:=\text{Hess}V_{ij}$, $u_i:=\partial_{\theta_i}V=J\,M\sin\Theta_i$ and $K:=\sum_{i=1}^N p_i^2$ (note that this is twice the kinetic energy). Then
\begin{align}
    \sum_{i=1}^N A_{ii}&=\frac{J}{N}\sum_{i,j}\cos(\theta_i-\theta_j)=J\,N\,M^2,
\end{align}
\begin{align}
    \sum_{i=1}^N u_i^2&=J^2 M^2 \sum_{i=1}^N \sin^2\Theta_i,
\end{align}
\begin{align}
    \sum_{i=1}^N u_i^2 A_{ii}&=\sum_{i=1}^N (J M \sin\Theta_i)^2\,(J M \cos\Theta_i)\nonumber\\
    &=J^3 M^3 \sum_{i=1}^N \sin^2\Theta_i\,\cos\Theta_i,
\end{align}
\begin{align}
    \sum_{i\neq j} u_i u_j A_{ij}&=\sum_{i\neq j} (J M \sin\Theta_i)(J M \sin\Theta_j)\nonumber\\
    &\qquad\qquad\qquad\qquad\times\Big(-\frac{J}{N}\cos(\theta_i-\theta_j)\Big)\nonumber\\
    &=-\frac{J^3 M^2}{N}\sum_{i\neq j}\sin\Theta_i\sin\Theta_j\cos(\theta_i-\theta_j)\nonumber.
\end{align}
For the kinetic part: $\partial_{p_i}H=p_i$, $\partial^2_{p_i}H=1$. Hence
\begin{align}
    \Delta H&:=\sum_{i=1}^N \partial^2_{p_i}H + \sum_{i=1}^N \partial^2_{\theta_i}V= N + J N M^2\,,
\end{align}
\begin{align}\label{def:nabla-Theta-dependence}
    \|\nabla H\|^2&=\sum_{i=1}^N (\partial_{p_i}H)^2 + \sum_{i=1}^N (\partial_{\theta_i}V)^2\nonumber\\
    &=\, K + J^2 M^2 \sum_{i=1}^N \sin^2\Theta_i\,,
\end{align}
\begin{align}
    \nabla H^{\!\top}&(\text{Hess}\,H)\nabla H=\nonumber\\
    &=\sum_{i=1}^N (\partial_{p_i}H)^2+ \sum_{i=1}^N u_i^2 A_{ii}+ \sum_{i\neq j} u_i u_j A_{ij}\nonumber\\
    &=K + J^3 M^3 \sum_{i}\sin^2\Theta_i\cos\Theta_i\nonumber\\
    &-\frac{J^3 M^2}{N}\sum_{i\neq j}\sin\Theta_i\sin\Theta_j\cos(\theta_i-\theta_j)\,.
\end{align}

\subsection{Weingarten trace and block decomposition}\label{ssec:decomposition-weingarten}

Exploiting the calculations above, the trace of the Weingarten operator
\begin{equation}
    \mathrm{Tr}\,W_{\bm\xi}=\frac{\Delta H}{\|\nabla H\|^2}-\frac{2\,\nabla H^{\!\top}(\text{Hess}\,H)\nabla H}{\|\nabla H\|^4},
\end{equation}
can be grouped into three terms:
\paragraph{Momentum block.}
\begin{equation}\label{eqn:kappa-p}
    \sum_i \kappa_{p_i}=\frac{N}{\|\nabla H\|^2}-\frac{2K}{\|\nabla H\|^4}\ .
\end{equation}

\paragraph{Diagonal angular block.}
\begin{equation}\label{eqn:kappa-d}
\sum_i \kappa^{\rm d}_{\theta_i}=\frac{J N M^2}{\|\nabla H\|^2}
-\frac{2J^3 M^3}{\|\nabla H\|^4}\sum_{i=1}^N \sin^2\Theta_i\,\cos\Theta_i\ .
\end{equation}

\paragraph{Off–diagonal (interaction) block.}
\begin{equation}\label{eqn:kappa-int}
    \sum_i \kappa^{\text{int}}_{ij}=\frac{2J^3 M^2}{N\,\|\nabla H\|^4}\sum_{i\ne j}\sin\Theta_i\,\sin\Theta_j\,\cos(\theta_i-\theta_j)\ .
\end{equation}

Thus,
\begin{equation}
    \begin{split}
        \mathrm{Tr}\,W_{\bm\xi}&=\sum_i \kappa_{p_i}+\sum_i \kappa^{\rm d}_{\theta_i}+\sum_i \kappa^{\text{int}}_{ij}.
    \end{split}
\end{equation}

\subsection{Small–$M$ angular sums}

In order to expand $\text{Tr}\,W_{\bm\xi}$ in power of $M$, we need to analyze the $\Theta_i$-dependent function within Eqs.~\eqref{eqn:kappa-p}-\eqref{eqn:kappa-int} and where the $\Theta_i$-dependence of $\|\nabla H\|^2$ is given in Eq.~\eqref{def:nabla-Theta-dependence}. We thus define:
\[
\begin{split}
    S_2&:=\sum_{i=1}^N \sin^2\Theta_i,\qquad
    S_{21}:=\sum_{i=1}^N \sin^2\Theta_i\,\cos\Theta_i,\\
    &\qquad S_{\mathrm{int}}:=\sum_{i\neq j}\sin\Theta_i\sin\Theta_j\cos(\theta_i-\theta_j).
\end{split}
\]

\subsection{Angular statistics and small-$M$ expansions}\label{app:angular-averages}

Let $\Theta_i=\theta_i-\phi$ and $M:=\big\|\frac{1}{N}\sum_i(\cos\theta_i,\sin\theta_i)\big\|\ll 1$. On the disordered branch ($M\simeq 0$), we model the angular fluctuations using the von Mises density
\[
f(\Delta)=\frac{e^{h\cos\Delta}}{2\pi I_0(h)},\qquad h\ll 1,
\]
whose moments satisfy (see standard Refs.~\cite{MardiaJupp,Fisher93,Jammalamadaka2001})
\[
    \langle \cos(k\Delta)\rangle = I_k(h)/I_0(h)\,,
\]
and for small $h$
\begin{align}
    I_0(h)&=1+\frac{h^2}{4}+\frac{h^4}{64}+O(h^6),\\
    I_1(h)&=\frac{h}{2}+\frac{h^3}{16}+O(h^5),\\
    I_2(h)&=\frac{h^2}{8}+\frac{h^4}{96}+O(h^6),\\
    I_3(h)&=\frac{h^3}{48}+O(h^5).
\end{align}
Another standard identity is $\langle \sin(k\Delta)\rangle=0$ by symmetry, whence
\[
\begin{split}
\langle \cos\Delta\rangle=\frac{I_1}{I_0}= \frac{h}{2}-\frac{h^3}{16}+O(h^5)=:M
\\\Rightarrow\quad h=2M+O(M^3).
\end{split}
\]
Small–$h$ expansions for the modified Bessel functions \(I_k\) follow from classical handbooks \cite{AbramowitzStegun}.
Using $\cos^2\Delta=\tfrac12(1+\cos2\Delta)$ and the small-$h$ expansions of $I_k$,
\[
\langle \cos^2\Delta\rangle=\frac12+\frac{1}{2}\frac{I_2}{I_0}
=\frac12+\frac{h^2}{16}+O(h^4)
=\frac12+\frac{M^2}{4}+O(M^4),
\]
hence
\begin{equation}
\langle \sin^2\Delta\rangle=\frac12-\frac{M^2}{4}+O(M^4).
\label{eq:sin2-avg}
\end{equation}
Moreover, using $\cos^3\Delta=\tfrac14(3\cos\Delta+\cos3\Delta)$,
\[
\begin{split}
\langle \cos^3\Delta\rangle=\frac{3}{4}M+\frac{1}{4}\frac{I_3}{I_0}
&=\frac{3}{4}M+\frac{h^3}{192}+O(h^5)
\\
&=\frac{3}{4}M+\frac{M^3}{24}+O(M^5),
\end{split}
\]
so that
\begin{equation}
\begin{split}
\langle \sin^2\Delta\,\cos\Delta\rangle
&=\langle \cos\Delta-\cos^3\Delta\rangle
\\
&=\frac{1}{4}M-\frac{1}{24}M^3+O(M^5).
\end{split}
\label{eq:sin2cos-avg}
\end{equation}

By the law of large numbers, sums over $i=1,\dots,N$ are replaced by $N$ times the expectations up to $O(\sqrt{N})$ fluctuations. Therefore,
\begin{equation}
\begin{split}
    S_2&:=\sum_{i=1}^N \sin^2\Theta_i= N\,\langle \sin^2\Delta\rangle\\
    &= N\Big(\frac12-\frac14 M^2\Big)+O(NM^4)+O(\sqrt{N}),
\end{split}
\label{eq:S2-exp}
\end{equation}
\begin{equation}
\begin{split}
    S_{21}:=\sum_{i=1}^N \sin^2\Theta_i\,\cos\Theta_i&= N\,\langle \sin^2\Delta\,\cos\Delta\rangle\\
    &= \frac{MN}{4}+O(NM^3).
\end{split}
\label{eq:S21-exp}
\end{equation}

For the scaling of $S_{\mathrm{int}}$
we use $\cos(\Theta_i-\Theta_j)=\cos\Theta_i\cos\Theta_j+\sin\Theta_i\sin\Theta_j$, and we get
\begin{align}
S_{\mathrm{int}}
&=\sum_{i\neq j}\sin\Theta_i\sin\Theta_j\cos(\Theta_i-\Theta_j)\nonumber\\
&=\sum_{i\neq j}\big(\sin\Theta_i\cos\Theta_i\big)\big(\sin\Theta_j\cos\Theta_j\big)
 \nonumber\\
 &\qquad\qquad\qquad\qquad+\sum_{i\neq j}\sin^2\Theta_i\,\sin^2\Theta_j \nonumber\\
&=\Big(\sum_i \sin\Theta_i\cos\Theta_i\Big)^2-\sum_i \sin^2\Theta_i\cos^2\Theta_i
 \nonumber\\
 &\qquad\qquad+\Big(\sum_i \sin^2\Theta_i\Big)^2-\sum_i \sin^4\Theta_i. \label{eq:Sint-decomp}
\end{align}
By symmetry $\sum_i \sin\Theta_i\cos\Theta_i=O(\sqrt{N})$, hence its square is $O(N)$ and subleading versus $N^2$ terms.
Set $A:=\sum_i\sin^2\Theta_i$; from \eqref{eq:S2-exp}, \[
    A=\frac{N}{2}-\frac{N}{4}M^2+O(NM^4)\,,
\]
Using 
\[
    \sin^4\Delta=\frac{3-4\cos 2\Delta+\cos 4\Delta}{8}\,,
\]
and $I_4/I_0=O(h^4)$, we get
\[
\begin{split}
\langle \sin^4\Delta\rangle=\frac{3}{8}-\frac{1}{2}\frac{I_2}{I_0}+\frac{1}{8}\frac{I_4}{I_0}
&=\frac{3}{8}-\frac{h^2}{16}+O(h^4)
\\
&=\frac{3}{8}-\frac{M^2}{4}+O(M^4),
\end{split}
\]
so that $\sum_i\sin^4\Theta_i=N\big(\tfrac{3}{8}-\tfrac{M^2}{4}\big)+O(NM^4)+O(\sqrt{N})$.
Plugging into \eqref{eq:Sint-decomp} and retaining $O(N^2)$ terms,
\begin{equation}
\begin{split}
    S_{\mathrm{int}}&=\bigg(\frac{N}{2}-\frac{N}{4}M^2\bigg)^2 - \sum_i \sin^4\Theta_i + O(N)\\
    &=\frac{N^2}{4}+O(N^2 M^2) - \frac{3N}{8}+O(N),
\end{split}
\label{eq:Sint-scaling}
\end{equation}
which yields, to leading order in $N$,
\begin{equation}
S_{\mathrm{int}}\simeq \frac{N^2}{4},\qquad (N\to\infty,\ M\to 0).
\end{equation}

\subsection{Denominator expansions and block sums}\label{app:denominators}

Recall $K:=\sum_i p_i^2$ and
\[
G:=\|\nabla H\|^2 = K + J^2 M^2 S_2.
\]
Write $c:=K/N$ (finite in the thermodynamic limit). Using \eqref{eq:S2-exp},
\[
G = N\Big(c+\frac{J^2}{2}M^2\Big)+O(NM^4).
\]
Therefore,
\begin{align}
    \frac{1}{G}&=\frac{1}{Nc}\Big(1+\frac{J^2}{2c}M^2\Big)^{-1}
    \nonumber\\
    &=\frac{1}{Nc}\Big(1-\frac{J^2}{2c}M^2+O(M^4)\Big), \label{eq:oneoverG}\\
    \frac{1}{G^2}&=\frac{1}{N^2c^2}\Big(1+\frac{J^2}{2c}M^2\Big)^{-2}
    \nonumber\\
    &=\frac{1}{N^2c^2}\Big(1-\frac{J^2}{c}M^2+O(M^4)\Big).\label{eq:oneoverG2}
\end{align}

Inserting \eqref{eq:S2-exp}, \eqref{eq:S21-exp}, \eqref{eq:Sint-scaling}, \eqref{eq:oneoverG}–\eqref{eq:oneoverG2} into the blocks
\[
\begin{split}
    \sum_i\kappa_{p_i}&=\frac{N}{G}-\frac{2K}{G^2},\\
    \sum_i\kappa^{\mathrm{d}}_{\theta_i}&=\frac{J N M^2}{G}-\frac{2J^3 M^3 S_{21}}{G^2},\\
    \sum_{i\neq j}\kappa^{\mathrm{int}}_{ij}&=\frac{2J^3 M^2}{N\,G^2}S_{\mathrm{int}},
\end{split}
\]
and keeping only $O(N^0)$ terms (the ones that survive as $N\to\infty$), reproduces the expressions for $\bm P$ and $\bm I$ of the main text.

\medskip
Then, in the small-$M$ limit and thermodynamic limit, we have
\[
S_2\simeq N\Big(\frac12-\frac14 M^2\Big),\qquad
S_{21}\simeq \frac{MN}{4},\qquad
S_{\mathrm{int}}\simeq \frac{N^2}{4},
\]
\[
\frac{1}{G}\simeq \frac{1}{Nc}\Big(1-\frac{J^2}{2c}M^2\Big),\qquad
\frac{1}{G^2}\simeq \frac{1}{N^2c^2}\Big(1-\frac{J^2}{c}M^2\Big),
\]

\section{Matrix determinant lemma for the Hessian}\label{app:hessianV-diagonalization}

In this appendix, we derive the secular equation for the spectrum of 
the angular Hessian $H_{qq}$ of the mean-field XY model. 

\subsection{Diagonal--rank-2 structure}
Recall that the Hessian can be written as
\begin{equation}
H_{qq}=D-\alpha UU^\top,\;\; 
\alpha=\frac{J}{N},\;\; 
U:=[x,y]\in\mathbb R^{N\times 2},
\end{equation}
where 
\begin{align}
    &\qquad \; D=\operatorname{diag}(JM\cos\Theta_1,\ldots,JM\cos\Theta_N),\\
    x&=(\cos\theta_1,\ldots,\cos\theta_N)^\top,\quad
    y=(\sin\theta_1,\ldots,\sin\theta_N)^\top.\nonumber
\end{align}
Thus $H_{qq}$ is a diagonal matrix perturbed by a rank-2 update.

\subsection{Sylvester determinant theorem}
The general identity (Sylvester theorem, also known as the matrix determinant lemma) reads
\begin{equation*}
\det(A+UV^\top)=\det(I_k+V^\top A^{-1}U)\,\det(A),
\end{equation*}
for any invertible $A\in\mathbb R^{n\times n}$ and $U,V\in\mathbb R^{n\times k}$.

\subsection{Application to $H_{qq}$}
We want the characteristic polynomial
\begin{equation*}
\det(H_{qq}-\lambda I)=\det\big((D-\lambda I)-\alpha UU^\top\big).
\end{equation*}
We identify
\begin{equation*}
A:=D-\lambda I,\;\;\; U':=\sqrt{-\alpha}\,U,\;\;\; V':=\sqrt{-\alpha}\,U,
\end{equation*}
so that
\begin{equation*}
(D-\lambda I)-\alpha UU^\top = A+U'V'^\top.
\end{equation*}
By Sylvester’s theorem,
\begin{equation*}
\det(H_{qq}-\lambda I)=\det(A)\,\det\!\left(I_2+V'^\top A^{-1}U'\right).
\end{equation*}
Since $U'=V'=\sqrt{-\alpha}\,U$, this becomes
\begin{equation*}
\det(H_{qq}-\lambda I)=\det(D-\lambda I)\,\det\!\Big(I_2-\alpha\,U^\top(D-\lambda I)^{-1}U\Big).
\label{A.1}
\end{equation*}

\subsection{Secular equation}
Equation (A.1) shows that the characteristic polynomial factorizes into
\begin{align*}
&\det(H_{qq}-\lambda I)=0
\;\;\Longleftrightarrow\;\;
\\
&\det(D-\lambda I)=0 \quad \text{or}\quad \det\Big(I_2-\alpha\,G(\lambda)\Big)=0,
\end{align*}
with
\begin{align*}
G(\lambda)&:=U^\top(D-\lambda I)^{-1}U
\\
&=\begin{pmatrix}
x^\top(D-\lambda I)^{-1}x & x^\top(D-\lambda I)^{-1}y\\
y^\top(D-\lambda I)^{-1}x & y^\top(D-\lambda I)^{-1}y
\end{pmatrix}.
\end{align*}
The first factor $\det(D-\lambda I)$ vanishes only at the poles of $G(\lambda)$, 
and does not correspond to actual eigenvalues of $H_{qq}$ (unless $x_i=y_i=0$ for some $i$). 
Hence the true spectrum is determined by the secular equation
\begin{equation*}
\det(I_2-\alpha\,G(\lambda))=0,
\end{equation*}
which is of dimension $2\times 2$, reflecting the rank-2 nature of the perturbation. 
Thus all $N$ eigenvalues of $H_{qq}$ are obtained as the roots of this secular equation, 
with at most two eigenvalues displaced far from the bulk.

\section{Numerical methods and integration schemes}\label{sec:numerical-sampling}

Consider $\phi$ as a generalized coordinates: this can be the real field configuration in the 2D $\phi^4$ model or the angle variable $\theta$ in the XY model.\\

\subsection{Configurational microcanonical Monte Carlo algorithm}
\label{app:micro_montecarlo}
The microcanonical ensemble has been reproduced by adopting the method proposed in Refs.~\cite{ray1996microcanonical,ray1991microcanonical} and consisting of a microcanonical Monte Carlo (MICROMC) algorithm in the configuration space. We have chosen a method based on a random proposal to enforce ergodicity. 

Note that the numerical procedure that we adopted and reported below is system-independent.

The microcanonical algorithm is based on the identification of a suitable density probability function that is used to produce a reliable sampling. To find such a probability function, we start with the microcanonical partition function:
\[
\Omega(E) = \int \delta(H[\pi, \phi] - E) D\pi\,D\phi\,,
\]
where
\[
    D\pi=\prod_{\bm{n}\in\mathbb{L}}d\pi_{\bm{n}},\qquad D\phi=\prod_{\bm{n}\in\mathbb{L}}d\phi_{\bm{n}}\,,
\]
and where the Hamiltonian is
\[
    H(\bm \pi, \bm\phi) = \sum_{\bm{n}\in\mathbb{L}} \frac{\pi_{\bm{n}}^2}{2}  + V(\phi)\,,
\]
for any potential function. Finally, $\delta$ represents the Dirac delta function. From now on, we introduce
$L=N^2$ to denote the total number of degrees of freedom. We focus on the momentum integral that separates as

\begin{equation}\label{def:integral_momentum_micro}
    I_p = \int \delta \left(\sum_{\bm{n}\in\mathbb{L}} \frac{\pi_{\bm{n}}^2}{2} + V(\phi) - E\right)D\pi\,.
\end{equation}
Thus, we notice that the kinetic energy variable,
\[
    K = \sum_{\bm{n}\in\mathbb{L}} \frac{p_{\bm{n}}^2}{2}\;,
\]
can be interpreted as the equation for a $L$-dimensional sphere of radius $r^2=2K$. Then, exploiting the spherical coordinates in $L$ dimensions, we have
\[
    \prod_{\bm{n}\in\mathbb{L}}d\pi_{\bm{n}} = \rho^{L-1}\,d\rho\,d\Omega_{L-1}\,,
\]
where \( d\Omega_{L-1} \) is the differential solid angle element on the unit sphere \( S^{L-1} \), expressed as:
\[
    d\Omega_{L-1} = \prod_{i=1}^{L-1} d\theta_i \sin^{L-1-i} \theta_i,\qquad
    \Omega_{L}=\int d\Omega_{L-1}\;.
\]
Then, we define 
\[
    \rho=\sqrt{2K},\quad d\rho=\frac{dK}{\sqrt{2K}}\implies \rho^{L-1}d\rho=(2K)^{L/2-1}\,dK\,.
\]
Finally, integral \eqref{def:integral_momentum_micro} rewrites
\begin{equation}\label{}
\begin{split}
    I_p &= \Omega_{L}\int \delta \left[K-(E- V(\phi))\right](2K)^{L/2-1}\;dK\\
    &= 2^{L-1}\Omega_{L}(E- V(\phi))^{L/2-1}\Theta(E - V(\phi))\,.
\end{split}
\end{equation}
Reintroducing above the integral over field configurations, we recover the \emph{configurational microcanonical partition function} that reads
\begin{equation}\label{def:configurational_entropy}
    \Omega(E) = \int  (E - V(\bm\phi))^{L/2 - 1} \Theta(E - V(\bm\phi))\;D^{N}\phi.
\end{equation}
Notice that we have dropped irrelevant constants that play no role in the calculation of expectation values.\\

The numerical algorithm then exploits the probability distribution function arising from Eq.~\eqref{def:configurational_entropy}
$$
    W_{E}[\bm\phi]= \left(E-V(\bm\phi)\right)^{L/2-1}\Theta[E-V(\bm\phi)]\,.
$$
The sampling is carried out by randomly picking a lattice site, $\bm{n}$, and proposing a new random configuration $\phi^{old}_{\bm{n}}\mapsto \phi^{new}_{\bm{n}}=\phi^{old}_{\bm{n}} + \eta\Delta\phi$ where $\eta$ is a random number uniformly sampled from the interval $[-1,1]$ and $\Delta\phi$ is adjusted to achieve a target acceptance rate of $50\%-60\%$. The new configuration is accepted according to the Monte Carlo acceptance probability
$$
    W(\bm\phi^{old}\to\bm\phi^{new}) = \min\left(1,\frac{W_{E}[\bm\phi^{new}]}{W_{E}[\bm\phi^{old}]}\right)\,.
$$
Notice that an efficient way to evaluate this acceptance probability is to rewrite the acceptance ratio as follows:
$$
    \frac{W_{E}[\bm\phi^{new}]}{W_{E}[\bm\phi^{old}]} = \exp\left[\bigg(\frac{L}{2}-1\bigg)\log\left(\frac{E-V(\bm\phi^{new})}{E-V(\bm\phi^{old})}\right)\right]\,.
$$
once we have checked that $E-V(\phi^{new})>0$.

\subsection{Mean-field 1D XY model: \texorpdfstring{$\mathcal{O}(1)$}{O(1)} microcanonical updates via magnetization}

For the mean-field (fully-connected) 1D XY model we parameterize the spins by angles
$\{\theta_n\}_{n=1}^{N}$ with conjugate momenta $\{p_n\}_{n=1}^{N}$ and Hamiltonian
\begin{equation}
H(\bm p,\bm\theta)=\sum_{n=1}^{N}\frac{p_n^{\,2}}{2}+V(\bm\theta)\,.
\end{equation}
A convenient form of the mean-field interaction is
\begin{equation}
V(\bm\theta)=\frac{J}{2N}\sum_{n,m=1}^{N}\Bigl[1-\cos(\theta_n-\theta_m)\Bigr]\,,
\end{equation}
where $J$ controls the coupling strength (the additive constant is irrelevant for the
microcanonical acceptance ratio, but we keep it to make $V\ge 0$ explicit).

\paragraph{Magnetization representation of the potential.}
Introduce the (unnormalized) magnetization components
\begin{equation}
M_x:=\sum_{n=1}^{N}\cos\theta_n,\;\;
M_y:=\sum_{n=1}^{N}\sin\theta_n,
\;\;
M^2:=\frac{M_x^2+M_y^2}{N^2}.
\end{equation}
Using
$\sum_{n,m}\cos(\theta_n-\theta_m)=\left(\sum_n\cos\theta_n\right)^2+\left(\sum_n\sin\theta_n\right)^2
= M_x^2+M_y^2$,
the interaction energy becomes the \emph{single-scalar} function
\begin{equation}\label{eq:V_MF_M2}
V(\bm\theta)=\frac{J N}{2}\Bigl(1-M^2\Bigr).
\end{equation}
This identity is the key numerical simplification: once $(M_x,M_y)$ are known, $V$ is obtained
without any $\mathcal{O}(N^2)$ pair sum.

\paragraph{Single-site proposal and \texorpdfstring{$\mathcal{O}(1)$}{O(1)} update.}
At each Monte Carlo step choose a site $n$ uniformly at random and propose
\begin{equation}
\theta_n^{\text{new}}=\theta_n^{\text{old}}+\eta\,\Delta\theta,
\qquad \eta\sim\mathcal{U}[-1,1],
\end{equation}
with $\Delta\theta$ tuned to reach the target acceptance rate.
Define the local trigonometric increments
\begin{equation}
\begin{split}
    \Delta c &:= \cos\theta_n^{\text{new}}-\cos\theta_n^{\text{old}},\\
    \Delta s &:= \sin\theta_n^{\text{new}}-\sin\theta_n^{\text{old}}.
\end{split}
\end{equation}
Then, the magnetization sums update exactly as
\begin{equation}\label{eq:MF_update_M}
M_x^{\text{new}} = M_x^{\text{old}} + \Delta c,\qquad
M_y^{\text{new}} = M_y^{\text{old}} + \Delta s,
\end{equation}
which is $\mathcal{O}(1)$ per the proposal. Consequently,
\begin{equation}
\begin{split}
    M_{\text{new}}^{2}&=\frac{\left(M_x^{\text{new}}\right)^2+\left(M_y^{\text{new}}\right)^2}{N^2},\\
    V_{\text{new}}&=\frac{J N}{2}\Bigl(1-M_{\text{new}}^{2}\Bigr),
\end{split}
\end{equation}
and similarly for the ``old'' configuration.\\

Now, the microcanonical acceptance rule becomes the same as the one introduced in the section above. For the mean-field XY model, the configurational microcanonical weight is
\begin{equation}
W_E[\bm\theta]=\bigl(E-V(\bm\theta)\bigr)^{\frac{N}{2}-1}\,
\Theta\!\bigl(E-V(\bm\theta)\bigr),
\end{equation}
so a proposed move is \emph{automatically rejected} if $E\le V_{\text{new}}$.
Otherwise, it is accepted with probability
\begin{equation}\label{eq:MF_acceptance}
\mathcal{A}(\text{old}\to\text{new})
=\min\!\left\{
1,\left(\frac{E-V_{\text{new}}}{E-V_{\text{old}}}\right)^{\frac{N}{2}-1}
\right\}.
\end{equation}
For numerical stability, one evaluates the ratio in logarithmic form:
\begin{equation}
\log\frac{W_E[\bm\theta^{\text{new}}]}{W_E[\bm\theta^{\text{old}}]}
=\left(\frac{N}{2}-1\right)
\Bigl[\log\!\bigl(E-V_{\text{new}}\bigr)-\log\!\bigl(E-V_{\text{old}}\bigr)\Bigr],
\end{equation}
and compares $\log u$ (with $u\sim\mathcal{U}[0,1]$) to the expression above.

\paragraph{Computational cost.}
Equations \eqref{eq:MF_update_M}--\eqref{eq:MF_acceptance} reduce each Metropolis proposal to a
constant-time update (two trigonometric evaluations plus a few floating-point operations),
in contrast to the $\mathcal{O}(N^2)$ evaluation of the fully-connected interaction by explicit
pair summation.

\subsection{Initial configuration}

Initial conditions are randomly proposed in order to start with a high-energy configuration, $E_{rand}$, larger than the desired input energy $E_{inp}$ and such that $E_{rand}-V(\phi_{ini})>0$. Then, we perform $10^{4}$ steps of equilibration with the MICROMC algorithm. At this stage, we search for the correct value for $\Delta\phi$. To do that, we start with a small value for $\Delta\phi$, say $0.001$ and run a few MICROMC steps using the desired input energy $E_{inp}$ and computing the acceptance rate, $N_{\text{acc}}$. If $N_{\text{acc}}\notin[0.5,0.6]$, then we repeat the procedure by replacing $\Delta\phi$ with $\Delta\phi+0.001$. It should be stressed that other strategies, such as the molecular dynamics algorithm, have also been used to select the initial conditions. Comparison of the thermodynamic observables obtained with both methods did not yield appreciable differences. Finally, once the suitable tune parameter's value has been obtained, the configuration is equilibrated for $10^4$ steps, and the trajectory is evolved with the MICROMC method and used for computing averages. For each energy value, we have produced $N_{trj}=32$ realizations for each system size. Each thermodynamic observable has been evaluated through $N_{\text{avg}}=10^{6}$ measurements performed in every $N_{\text{step}}=100$ MICROMC step. The microcanonical average of a given observable, $f$, is computed by
\begin{equation}\label{def:time_average}
    \langle f\rangle_{\varepsilon}=\frac{1}{ N_{\text{trj}}\cdot N_{\text{avg}}}\sum_{i=1}^{N_{trj}}\sum_{\alpha=1}^{N_{\text{avg}}}f(\phi^{(i)}_{\alpha})\,,
\end{equation}
where $f$ is the observable evaluated on the configuration $\phi^{(i)}_{\alpha}$ at the $\alpha$-th MC step for the $i$-th realization.

\subsection{Numerical calculation of the microcanonical entropy's derivatives}
\label{app:PHT_method}

The MIPA method requires the estimation of higher-order derivatives of the microcanonical entropy from a microcanonical sampling of the phase space. To this purpose, we adopt the PHT method~\cite{pearson1985laplace}, which allows a direct calculation of the first- and second-order derivatives of the microcanonical entropy as follows. Any energy derivative of the entropy function can be rewritten in terms of averages of the powers of the kinetic energy $k=(E-V(\phi)/L$. In this framework, the first- and second-order derivatives of the microcanonical entropy can be expressed as: 
\begin{align}
     \partial_{\varepsilon}S(\varepsilon)&= \left(\frac{1}{2}-\frac{1}{L}\right)\langle k^{-1}\rangle_{\varepsilon}\,, \label{eqn:derS-I}\\
    \partial^{2}_{\varepsilon}S(\varepsilon)
    &=L\bigg[\bigg(\dfrac{1}{2}-\dfrac{1}{L}\bigg) \bigg(\dfrac{1}{2}-\dfrac{2}{L}\bigg)\langle k^{-2} \rangle_{\varepsilon}
    \nonumber\\
    &\qquad\qquad\qquad- \bigg(\dfrac{1}{2}-\dfrac{1}{L}\bigg)^2 \langle k^{-1}\rangle^2_{\varepsilon} \bigg]\,, \label{eqn:derS-II}
\end{align}
where the microcanonical averages have been estimated by Eq.~\eqref{def:time_average} over the realizations.

\section{Numerical details for short-range and mean-field models}

The Hamiltonian for the 2D $\phi^4$ model with nearest neighbor interactions is 
\begin{equation}
\begin{split}\label{def:hamiltonian_phi4-app}
    H:=\sum_{\bm{i}}\Bigg[\frac{\pi^{2}_{\bm{i}}}{2}+\frac{\lambda}{4!}\phi^{4}_{\bm{i}}&-\frac{\mu^{2}}{2}\phi^{2}_{\bm{i}}+\frac{J}{4}\sum_{\bm{k}\in N(\bm{i})}(\phi_{\bm{i}}-\phi_{\bm{i}})^{2}\Bigg]
\end{split}
\end{equation}
and we choose values 
\[
    \lambda=3/5,\qquad \mu^2=2,\qquad J=1.
\]
The label $\bm{i}:=(i_1,i_2)$ is the two-dimensional array of integer numbers used for labeling the sites; finally, we denote $N(\bm{i})$ as the set of the nearest neighbors of the $\bm{i}$-th site.

The Hamiltonian of the 1D XY mean-field is given by
\begin{equation}
    H(\bm \theta,\bm p)=\sum_{i=1}^N \frac{p_i^2}{2}+\frac{J}{2N}\sum_{i,j}\bigl[1-\cos(\theta_i-\theta_j)\bigr],
\end{equation}
where $J=1$.

For the $\phi^4$-model, we chose the energy range $[9,16]$ with energy resolution $\Delta E=0.1$ and system sizes $N=64^4,\,128^2,\,256^2$.
For the XY model, the energy range is $[0.5,1]$ with energy resolution $\Delta E=0.01$, and we studied three different system sizes $N=8100,\,14400,40000$.

%==========================================================
\section{Microcanonical molecular-dynamics sampling for the 1D long-range XY chain}
\label{sec:md_fft_xy1d}

The 1D XY long-range model cannot be numerically investigated using the microcanonical Monte Carlo sampling method explained in the previous section due to the long-range interactions. Therefore, we use a molecular dynamics method combined with the Fast Fourier Transform (FFT) adapted from the one in Ref.~\cite{dagotto1988study}, which we explain in the upcoming section. 

\subsection{Model, periodic geometry, and Kac normalization}
We consider a 1D chain of $N$ rotors with angles $\theta_i\in[0,2\pi)$ and conjugate momenta $p_i\in\mathbb{R}$, $i=1,\dots,N$, with periodic boundary
conditions $\theta_{i+N}\equiv \theta_i$.
Long-range couplings depend only on the lattice displacement $n$ (nearest-image
convention). We define the interaction kernel
\begin{equation}
\begin{split}
  f_n \equiv f(|n|_N),\qquad |n|_N := \min(n,N-n),\\
  f_0=0,\qquad f_n = |n|_N^{-\alpha}\ \ (1\le n\le N-1),
\end{split}
\end{equation}
with $\alpha = 1+\sigma$ in the notation used in our long-range XY work. The Kac factor
(normalization ensuring extensivity) is
\begin{equation}
  N_\alpha \equiv \sum_{n=1}^{N-1} f_n
  = 2\sum_{n=1}^{N_{\rm f}}\frac{1}{n^{\alpha}}
  +\delta_{N\,\mathrm{even}}\frac{1}{(N/2)^{\alpha}} ,
\end{equation}
where $N_{\rm f }:=\lfloor (N-1)/2\rfloor$.
The Hamiltonian is written as
\begin{equation}
  H
  = \sum_{i=1}^N \frac{p_i^2}{2}
  + \frac{J}{2N_\alpha}\sum_{i=1}^N\sum_{n=1}^{N-1} f_n\,
  \Bigl[1-\cos\bigl(\theta_i-\theta_{i+n}\bigr)\Bigr],
\end{equation}
where indices are understood modulo $N$ (e.g.\ $i+n\equiv i+n-N$ if $i+n>N$).
This ``$1-\cos$'' form is convenient because it makes the fully aligned state
have $V=0$. The FFT implementation below uses an equivalent but faster-to-evaluate
representation of the same potential.

\subsection{FFT representation of the long-range force}
A direct evaluation of the long-range force costs $\mathcal{O}(N^2)$ per step.
To reduce this to $\mathcal{O}(N\log N)$ we follow the standard Dagotto-Moreo
strategy~\cite{dagotto1988study}: rewrite the interaction as a discrete convolution computable by FFT.

Introduce complex ``spins''
\begin{equation*}
  z_i := e^{-\,\mathrm{i}\theta_i},\qquad z_i^\ast = e^{\mathrm{i}\theta_i}.
\end{equation*}
Define the complex field (a convolution on the ring)
\begin{equation}
  A_i := (f * z)_i \equiv \sum_{n=0}^{N-1} f_n\, z_{i-n},
  \qquad (i-n\ \text{mod}\ N),
  \label{eq:Ai_convolution}
\end{equation}
where $f_n$ is the real kernel above, symmetrized by periodicity ($f_{N-n}=f_n$).
Then one has the identity
\begin{equation}
  \sum_{i=1}^N \Re\bigl(z_i^\ast A_i\bigr)
  = \sum_{i=1}^N\sum_{n=1}^{N-1} f_n \cos(\theta_i-\theta_{i+n}),
\end{equation}
so the potential can be evaluated as
\begin{equation}
  V(\bm \theta)
  = \frac{J}{2N_\alpha}\Bigl(N N_\alpha - \sum_{i=1}^N \Re\bigl(z_i^\ast A_i\bigr)\Bigr).
  \label{eq:V_fft_form}
\end{equation}
From \eqref{eq:V_fft_form} the force can be expressed compactly in terms of the
imaginary part of the same local quantity:
\begin{equation}
  \dot{\theta}_i = p_i,\qquad
  \dot{p}_i = -\frac{\partial V}{\partial \theta_i}
           = -\frac{J}{N_\alpha}\,\Im\bigl(z_i^\ast A_i\bigr).
  \label{eq:eom_fft_force}
\end{equation}
This is exactly the object computed in the code as $\Im\left(e^{\mathrm{i}\theta_i}A_i\right)$.

\paragraph{FFT evaluation.}
Let $\mathcal{F}$ denote the discrete Fourier transform (DFT) on $N$ points and
$\mathcal{F}^{-1}$ its inverse. Precompute once the DFT of the kernel:
\begin{equation*}
  \widehat{f}_k := \mathcal{F}[f]_k,\qquad k=0,\dots,N-1,
\end{equation*}
and, at each MD step, compute
\begin{equation}
  \widehat{z}_k := \mathcal{F}[z]_k,\qquad
  \widehat{A}_k := \widehat{f}_k\,\widehat{z}_k,\qquad
  A_i := \mathcal{F}^{-1}[\widehat{A}]_i.
  \label{eq:fft_pipeline}
\end{equation}
The convolution theorem makes \eqref{eq:fft_pipeline} equivalent to \eqref{eq:Ai_convolution},
but with $\mathcal{O}(N\log N)$ cost. We call FFTW via its official Fortran 2003 interface (\texttt{fftw3.f03}) provided by the FFTW distribution~\cite{frigo2005fftw}.

\subsection{Symplectic time integration (microcanonical MD)}
To sample the microcanonical measure at fixed energy $E$, we evolve the Hamiltonian
flow using a second-order symplectic splitting (velocity-Verlet/leapfrog), which controls the long-time energy drift and provides robust microcanonical trajectories.

Given $(\theta^m,p^m)$ at time $t_m=m\Delta t$, compute the force
$F_i(\theta) := -\partial_{\theta_i}V(\theta)$ via \eqref{eq:eom_fft_force}--\eqref{eq:fft_pipeline},
then update:
\begin{align*}
  p_i^{m+\frac12} &= p_i^{m} + \frac{\Delta t}{2}\,F_i(\theta^m), \\
  \theta_i^{m+1}  &= \theta_i^{m} + \Delta t\, p_i^{m+\frac12}, \\
  p_i^{m+1}       &= p_i^{m+\frac12} + \frac{\Delta t}{2}\,F_i(\theta^{m+1}).
\end{align*}
The step size $\Delta t$ is chosen
so that the relative energy drift $\Delta H/H$ remains negligible during the production time window: we used $\Delta t=0.01$.

\subsection{Microcanonical sampling protocol (time averages)}
The microcanonical expectation of an observable $\mathcal{O}(\theta,p)$ is estimated as a time average along the MD trajectory (after equilibration):
\begin{equation*}
\begin{split}
  \langle \mathcal{O}\rangle_E \;&\approx\; \frac{1}{M}\sum_{m=1}^{M}
  \mathcal{O}\!\left(\theta(t_m),p(t_m)\right),
  \\
  t_m &= m\,\Delta t_\mathrm{samp},
\end{split}
\end{equation*}
with a sampling stride $\Delta t_\mathrm{samp}$ (typically an integer multiple of $\Delta t$)
chosen to reduce autocorrelation; we set $\Delta t_{\rm samp}=100$.

\paragraph{Initialization at fixed energy.}
A practical initialization at target total energy $E$ proceeds as follows:
(i) draw $\theta_i$ independently uniformly in $[0,2\pi)$;
(ii) draw trial momenta $\tilde p_i$ (e.g.\ Gaussian), enforcing zero total momentum
$p_i^{(0)} := \tilde p_i - \frac{1}{N}\sum_j \tilde p_j$;
(iii) compute $V(\theta^{(0)})$ using \eqref{eq:V_fft_form};
(iv) rescale momenta by a single factor to match the desired kinetic energy
$K^{(0)}=E-V(\theta^{(0)})$:
\begin{equation}
  p_i^{(0)} \leftarrow \lambda\, p_i^{(0)},\qquad
  \lambda := \sqrt{\frac{2\,(E-V(\theta^{(0)}))}{\sum_{j=1}^N (p_j^{(0)})^2}}.
\end{equation}
This guaranties $H(\theta^{(0)},p^{(0)})=E$ up to floating-point roundoff.

\paragraph{Typical measured quantities.}
During production, we record, for instance, the magnetization
\begin{equation}
  M := \frac{1}{N}\left|\sum_{i=1}^N e^{\mathrm{i}\theta_i}\right|,
\end{equation}

This method produces MD trajectories $\{p_n,\theta_n\}_n$ for which we can compute the kinetic energy relevant for computing the energy derivatives of entropy according to the PHT method introduced in Sec.~\ref{app:PHT_method}. Similarly, all geometric quantities can be computed once the Laplacian, gradient, and Hessian of the Hamiltonian function have been calculated. The second-order GCF $\Upsilon_{(g)}^2$ has been computed according to Eq.~\eqref{def:estimation-GCF}.
%==========================================================

\section{Numerical solution of the Entropy Flow Equation (EFE)}\label{sec:efe-solution-procedure}

\paragraph{Equation and unknowns.}
We solve the EFE
\begin{equation}\label{eq:EFE}
    S_g''(E)+\big(S_g'(E)\big)^2 \;=\; \Upsilon^{(2)}_g(E),
\end{equation}
on an interval \(E\in[E_{\min},E_{\max}]\).
Set
\[
    g(E):=S_g'(E),\qquad S_g'(E)=g(E),\qquad S_g''(E)=g'(E),
\]
so that \eqref{eq:EFE} reduces to the Riccati ODE
\begin{equation}\label{eq:Riccati}
    g'(E)+g(E)^2 \;=\; \Upsilon^{(2)}_g(E).
\end{equation}
Once \(g\) is known, the entropy follows by a single quadrature,
\begin{equation}\label{eq:S-from-g}
    S_g(E)=S_g(E_0)+\int_{E_0}^{E} g(\tilde E)\,d\tilde E.
\end{equation}

\paragraph{Input data: geometric source.}
The source \(\Upsilon^{(2)}_g\) is computed numerically at a discrete set of energies
\(\{(E_i,\Upsilon_i)\}_{i=1}^M\), with \(E_{\min}\le E_1<\dots<E_M\le E_{\max}\).
To obtain a smooth right–hand side in \eqref{eq:Riccati}, we construct a high–order
interpolant
\begin{equation*}
    \widehat\Upsilon(E)\;:=\;\mathcal{I}_{\!p}\big[\{(E_i,\Upsilon_i)\}\big](E),
\end{equation*}
using a polynomial spline of order \(p=9\) (Mathematica’s \texttt{Interpolation} with \texttt{InterpolationOrder\(\to\)9}).
This yields a \(C^{p-1}\) function \(\widehat\Upsilon\) on \([E_{\min},E_{\max}]\) that we use in place of \(\Upsilon^{(2)}_g\) in \eqref{eq:Riccati}.

\paragraph{Initial data.}
We pick a left endpoint \(E_0\in[E_{\min},E_{\max})\) and prescribe:
\begin{equation*}
    g(E_0)=\beta_0,\qquad S_g(E_0)=0,
\end{equation*}
where \(\beta_0\) is the microcanonical inverse temperature at \(E_0\) obtained from numerical simulation through Eq.~\eqref{eqn:derS-I}, and $S_g(E_0)$ sets the additive entropy. since it cancels in derivatives, it can be fixed to zero.

\paragraph{Time–integration (energy stepping).}
We integrate the first–order system
\begin{equation*}
\begin{cases}
S'(E)=g(E),\\[2pt]
g'(E)=\widehat\Upsilon(E)-g(E)^2,
\end{cases}
\qquad E\in[E_0,E_{\max}],
\end{equation*}
with an explicit high–order Runge–Kutta method. In practice, we used Mathematica’s
\texttt{NDSolve} with
\begin{widetext}
\[
\begin{split}
\texttt{Method}\;=\;\{\texttt{"TimeIntegration"}\to
\{\texttt{"ExplicitRungeKutta"},\ \texttt{"DifferenceOrder"}\to 9\}\},
\end{split}
\]
\end{widetext}
and uniform step \(\Delta E\) on \([E_0,E_{\max}]\).
The numerical solution returns \(S(E)\), \(g(E)\), and, where needed, \(g'(E)\).

\paragraph{Discretization and sampling.}
For visualization and post-processing, we evaluate \((g,g')\) on the uniform grid
\[
E_j \;=\; E_0 + j\,\Delta E,\qquad j=0,1,\dots,J,\quad E_J\le E_{\max}.
\]
This produces the arrays \(\{(E_j,g(E_j))\}\) and \(\{(E_j,g'(E_j))\}\) used in the figures.

\paragraph{Stability and cross–checks.}
The Riccati form \eqref{eq:Riccati} can be validated against the equivalent second–order
\emph{linear} problem via the substitution \(g=\psi'/\psi\), namely
\begin{equation}\label{eq:linear-check}
\begin{split}
    \psi''(E)\;=\;\widehat\Upsilon(E)\,&\psi(E),\qquad
g(E)=\frac{\psi'(E)}{\psi(E)},\\
S_g(E)&=S_0+\ln\frac{\psi(E)}{\psi(E_0)}.
\end{split}
\end{equation}
We solved \eqref{eq:linear-check} with the same RK scheme and verified that the reconstructed \(g\) matches the direct Riccati integration within the stated tolerances. Convergence was monitored by halving \(\Delta E\) and by lowering the interpolant order \(p\) (from 9 to 7) to ensure insensitivity to the smoothing.

\paragraph{Reproducibility notes.}
(i) \(\widehat\Upsilon(E)\) is evaluated only within the range \(\{E_i\}\) used in numerical simulations (no extrapolation).
(ii) All ODE solvers use a fixed step and order (RK9); decreasing \(\Delta E\) by a factor of 2 changes \(g\) and \(g'\) by less than the line width.
(iii) The additive constant \(S_0\) is fixed by matching \(S_g\) to the reference at \(E_0\) (or by setting \(S_0=0\)); derivatives are independent of \(S_0\).

\paragraph{Output.}
The procedure returns smooth functions \(g(E)=S_g'(E)\) and \(g'(E)=S_g''(E)\) on \([E_0,E_{\max}]\), together with the entropy \(S_g(E)\) from \eqref{eq:S-from-g}. These are the curves reported in the figures and used to compare with microcanonical observables.

\section{Exact microcanonical curvature for the mean-field XY (HMF) model}
\label{app:HMF_entropy_curvature}

In this appendix, we derive an explicit analytic expression for the microcanonical entropy and its curvature in the mean-field XY model, which also represents the canonical limit of the one-dimensional $\alpha$-XY chain with Kac-type normalization
for $\alpha<1$.

\subsection{Model and microcanonical density of states}

We consider the Hamiltonian (with $k_B=1$ and the coupling constant set to unity)
\begin{equation*}
\begin{split}
  H_N(\bm p,\bm \theta)
  = \sum_{i=1}^N \frac{p_i^2}{2}
    + V_N(\{\theta_i\}),
  \\
  V_N(\bm \theta) = \frac{1}{2N}\sum_{i,j=1}^N \bigl[1-\cos(\theta_i-\theta_j)\bigr].
\end{split}
\end{equation*}
The magnetization vector and its modulus are defined as follows:
\begin{equation*}
  \bm{M}
  = \frac{1}{N}\sum_{i=1}^N (\cos\theta_i,\sin\theta_i),
  \qquad M = |\bm{M}|,
\end{equation*}
so that the potential energy per particle takes the simple mean-field form
\begin{equation*}
  u(M) := \frac{V_N}{N} = \frac{1-M^2}{2},
\end{equation*}
and the energy per particle is
\begin{equation*}
  \varepsilon = \frac{E}{N} = \frac{K}{N} + u(M).
\end{equation*}

The microcanonical density of states at fixed energy $E$ is
\begin{equation*}
  \Omega_N(E)
  = \int d^N\theta\,d^N p\;
    \delta\!\bigl(E - H_N(\theta,p)\bigr).
\end{equation*}
Integrating out the momenta yields
\begin{equation}
  \Omega_N(E)
  = C_N \int_{[0,2\pi]^N} d^N\theta\;
    \bigl[E - V_N(\theta)\bigr]_+^{\frac{N}{2}-1},
  \label{eq:OmegaN_theta_integral_app}
\end{equation}
where $C_N$ is an explicit $N$-dependent prefactor (the volume of an
$N$-dimensional sphere) and $[x]_+$ denotes the positive part.

For mean-field Hamiltonians of the form
\begin{equation*}
  H_N = \sum_{i=1}^N \frac{p_i^2}{2} + N\,u(M),
\end{equation*}
the configurational part of \eqref{eq:OmegaN_theta_integral_app}
admits a large-deviation description in terms of the order parameter $M$:
the number of angle configurations with fixed magnetization $M$ scales as
$\exp\{N\,s_{\mathrm{conf}}(M)\}$, where $s_{\mathrm{conf}}(M)$ is the
{\it configurational entropy density} at fixed $M$.
In the thermodynamic limit, the entropy density
$s(\varepsilon) = \lim_{N\to\infty} N^{-1}\ln\Omega_N(N\varepsilon)$
can then be written as a variational problem over $M$~\cite{Campa2000Canonical,campa2009statistical}.

\subsection{Configurational entropy at fixed magnetization}

For the XY model in an external field $h$, the single-spin partition function is
\begin{equation*}
  Z_1(h) = \int_0^{2\pi} d\theta\, e^{h\cos\theta}
         = 2\pi I_0(h),
\end{equation*}
where $I_n$ is the modified Bessel function of the first kind, and
the single-spin magnetization is
\begin{equation}
  M(h) = \langle \cos\theta \rangle_h
       = \frac{I_1(h)}{I_0(h)}.
  \label{eq:m_of_h}
\end{equation}
The conjugate field $h$ is determined implicitly by \eqref{eq:m_of_h}
for each given $M\in[0,1]$.

The configurational entropy density at fixed $M$ is the Legendre transform of
$\ln Z_1(h)$:
\begin{equation*}
  s_{\mathrm{conf}}(M)
  = \ln\bigl[2\pi I_0(h(M))\bigr] - h(M)\,M.
\end{equation*}
It is convenient to view $s_{\mathrm{conf}}$ as a function of $h$,
\begin{equation*}
  \tilde{s}_{\mathrm{conf}}(h)
  := \ln\bigl[2\pi I_0(h)\bigr] - h\,M(h),
\end{equation*}
so that $s_{\mathrm{conf}}(M) = \tilde{s}_{\mathrm{conf}}(h(M))$.

Taking the derivative with respect to $h$ and using $I_0'(h)=I_1(h)$, we obtain
\begin{equation*}
  \frac{d\tilde{s}_{\mathrm{conf}}}{dh}
  = \frac{I_0'(h)}{I_0(h)} - M(h) - h\,\frac{dM}{dh}
  = -\,h\,\frac{dM}{dh}.
\end{equation*}
By the chain rule,
\begin{equation*}
  \frac{d s_{\mathrm{conf}}}{dM}
  = \frac{d\tilde{s}_{\mathrm{conf}}}{dh}\,\frac{dh}{dM}
  = -\,h(M),
\end{equation*}
so that the first derivative is simply $s_{\mathrm{conf}}'(M)=-h(M)$,
as expected from Legendre duality.

The second derivative is
\begin{equation*}
  s_{\mathrm{conf}}''(M)
  = -\,\frac{dh}{dM}
  = -\,\frac{1}{\dfrac{dM}{dh}}.
\end{equation*}
From \eqref{eq:m_of_h} and the Bessel identity
$I_\nu'(h) = \tfrac12[I_{\nu-1}(h)+I_{\nu+1}(h)]$, one finds
\begin{equation*}
\begin{split}
  \frac{dM}{dh}
  &= \frac{I_1'(h)\,I_0(h) - I_1(h)\,I_0'(h)}{I_0(h)^2}
  \\
  &= \frac{1}{2}\bigg(1 + \frac{I_2(h)}{I_0(h)}\bigg)- \left(\frac{I_1(h)}{I_0(h)}\right)^2.
\end{split}
\end{equation*}
Defining
\begin{equation}
  A(h) := \frac{dM}{dh}
        = \frac{1}{2}\left(1 + \frac{I_2(h)}{I_0(h)}\right) - M(h)^2,
  \label{eq:Adef_app}
\end{equation}
the configurational curvature can be written compactly as
\begin{equation*}
  s_{\mathrm{conf}}''(M)
  = -\,\frac{1}{A\bigl(h(M)\bigr)}.
\end{equation*}

\subsection{Variational representation of $s(\varepsilon)$}

Combining the kinetic and configurational contributions, the entropy density
in the thermodynamic limit can be expressed as~\cite{Campa2000Canonical,campa2009statistical}
\begin{equation*}
\begin{split}
  s(\varepsilon)
  &= \sup_{M\in[0,1]}\;
    \Phi(\varepsilon,M) + \text{const},
  \\
  \Phi(\varepsilon,M)
  &:= s_{\mathrm{conf}}(M)
     + \frac{1}{2}\ln\bigl[\varepsilon - u(M)\bigr],
\end{split}
\end{equation*}
with $u(M)=(1-M^2)/2$.
The maximizer $M^*(\varepsilon)$ is determined by
\begin{equation}
  \frac{\partial \Phi}{\partial M}(\varepsilon,M^*) = 0.
  \label{eq:stationarity_m_app}
\end{equation}
Along the physical branch $M^*(\varepsilon)$, the entropy is
\begin{equation*}
  s(\varepsilon)
  = \Phi\bigl(\varepsilon,M^*(\varepsilon)\bigr).
\end{equation*}

Differentiating with respect to $\varepsilon$ and using
\eqref{eq:stationarity_m_app}, we obtain
\begin{equation*}
  \frac{ds}{d\varepsilon}
  = \frac{\partial \Phi}{\partial \varepsilon}
    + \frac{\partial \Phi}{\partial M}\,\frac{d M^*}{d\varepsilon}
  = \frac{\partial \Phi}{\partial \varepsilon}
  = \frac{1}{2\bigl[\varepsilon - u(M^*)\bigr]},
\end{equation*}
so that $s'(\varepsilon)=\beta(\varepsilon)$, the inverse temperature.

\subsection{Canonical solution and specific heat}

In canonical ensemble, the HMF model is described by the inverse temperature
$\beta=1/T$ and the equilibrium magnetization $M(\beta)$, which satisfy
\begin{equation*}
  M(\beta) = \frac{I_1(h)}{I_0(h)},
  \qquad
  h = \beta\,M(\beta).
\end{equation*}
The energy per particle is
\begin{equation}
  \varepsilon(\beta)
  = \frac{T}{2} + \frac{1-M(\beta)^2}{2}
  = \frac{1}{2\beta} + \frac{1-M(\beta)^2}{2}.
  \label{eq:eps_of_beta_app}
\end{equation}

Let us derive the derivative $dM/d\beta$.
Writing $M(\beta)=F(h)$ with $F(h)=I_1(h)/I_0(h)$ and $h(\beta)=\beta M(\beta)$,
we have
\begin{equation*}
  \frac{dM}{d\beta}
  = F'(h)\,\frac{dh}{d\beta}
  = A(h)\,\bigg(M + \beta\,\frac{dM}{d\beta}\bigg),
\end{equation*}
where $A(h)=dM/dh$ is given by \eqref{eq:Adef_app}.
Solving for $dM/d\beta$ yields
\begin{equation}
  \frac{dM}{d\beta}
  = \frac{A(h)\,M}{1 - \beta A(h)}.
  \label{eq:dmdBeta_app}
\end{equation}

Using \eqref{eq:eps_of_beta_app}, the derivative of the energy with respect
to $\beta$ is
\begin{equation*}
  \frac{d\varepsilon}{d\beta}
  = -\,\frac{1}{2\beta^2} - M(\beta)\,\frac{dM}{d\beta}.
\end{equation*}
Substituting \eqref{eq:dmdBeta_app} gives
\begin{equation*}
  \frac{d\varepsilon}{d\beta}
  = -\,\frac{1}{2\beta^2}
    - \frac{A(h)\,M^2}{1 - \beta A(h)}.
\end{equation*}
The canonical specific heat per particle is
\begin{equation*}
  c(\beta)
  = \frac{d\varepsilon}{dT}
  = \frac{d\varepsilon}{d\beta}\,\frac{d\beta}{dT}
  = -\,\beta^2\,\frac{d\varepsilon}{d\beta},
\end{equation*}
which yields
\begin{equation}
  c(\beta)
  = \frac{1}{2}
    + \frac{\beta^2 A(h)\,M^2}{1 - \beta A(h)}.
  \label{eq:c_beta_final_app}
\end{equation}

\subsection{Microcanonical curvature $s''(\varepsilon)$}\label{ssec:mathematica-code}

Under ensemble equivalence (which holds for the HMF model), the second
derivative of the microcanonical entropy can be expressed in terms of the
canonical specific heat as
\begin{equation*}
  s'(\varepsilon) = \beta(\varepsilon),
  \qquad
  s''(\varepsilon) = \frac{d\beta}{d\varepsilon}
                   = -\,\frac{\beta^2}{c(\varepsilon)}.
\end{equation*}
Using \eqref{eq:c_beta_final_app} and viewing $\varepsilon$ as a function
of $\beta$ via \eqref{eq:eps_of_beta_app}, we obtain the parametric
representation
\begin{equation}
  s''\bigl(\varepsilon(\beta)\bigr)
  = -\,\frac{\beta^2}{
        \dfrac{1}{2}
      + \dfrac{\beta^2 A(h)\,M^2}{1 - \beta A(h)}
      },
  \label{eq:spp_parametric_app}
\end{equation}
with
\begin{equation*}
  M = \frac{I_1(h)}{I_0(h)},
  \;\;\;
  A(h) = \frac{1}{2}\left(1 + \frac{I_2(h)}{I_0(h)}\right) - M^2,
  \;\;\;
  h = \beta M.
\end{equation*}
Equation~\eqref{eq:spp_parametric_app} provides a fully analytic expression for the microcanonical curvature $s''(\varepsilon)$ of the mean-field XY model, valid on both sides of the second-order phase transition at $\beta_c=2$ (or $\varepsilon_c=3/4$). As a consistency check, in the paramagnetic phase $M=0$, one has $c(\beta)=1/2$ and therefore $s''(\varepsilon)=-2\beta^2$; using $\varepsilon = T/2 + 1/2$, this reduces to
$s''(\varepsilon) = -1/[2(\varepsilon - 1/2)^2]$, in agreement with the
explicit paramagnetic branch.

Below is the Mathematica code to plot this function.

\begin{lstlisting}[language=Mathematica,
caption={Computation of $m(\beta)$, $\varepsilon(\beta)$ and $s''(\varepsilon)$ across the para/ferro branches.},
label={app:code_s2}]


(* Spontaneous magnetization m[beta]:
   - paramagnetic branch: m = 0 for beta <= 2
   - ferromagnetic branch: solve self-consistency equation for beta > 2
*)

m[\[Beta]_?NumericQ] /; \[Beta] <= 2 := 0;

m[\[Beta]_?NumericQ] :=
  m[\[Beta]] =
    m /. FindRoot[
      BesselI[1, \[Beta] m]/BesselI[0, \[Beta] m] - m,
      {m, 0.9}
    ];

(* Auxiliary field *)
h[\[Beta]_?NumericQ] := \[Beta]*m[\[Beta]];

(* Susceptibility-like combination appearing in fluctuations *)
A[\[Beta]_?NumericQ] :=
  1/2 (1 + BesselI[2, h[\[Beta]]]/BesselI[0, h[\[Beta]]]) -
  m[\[Beta]]^2;

(* Energy density epsilon(beta) *)
eps[\[Beta]_?NumericQ] := 1/(2 \[Beta]) + (1 - m[\[Beta]]^2)/2;

(* Specific heat c(beta) *)
c[\[Beta]_?NumericQ] :=
  1/2 + \[Beta]^2 A[\[Beta]] m[\[Beta]]^2/(1 - \[Beta] A[\[Beta]]);

(* Second derivative of the entropy with respect to epsilon:
   s''(epsilon) = -beta^2 / c(beta)
*)
s2[\[Beta]_?NumericQ] := -\[Beta]^2/c[\[Beta]];

(* Ferromagnetic branch: use a very small step close to beta = 2 *)
dataF = Table[{eps[\[Beta]], s2[\[Beta]]}, {\[Beta], 2.0001, 4, 0.0005}];

(* Paramagnetic branch *)
dataP = Table[{eps[\[Beta]], s2[\[Beta]]}, {\[Beta], 0.2, 2, 0.01}];

(* Merge and sort points by energy *)
dataAll    = SortBy[Join[dataF, dataP], First];
dataSorted = SortBy[dataAll, First];

ListLinePlot[
  dataSorted,
  PlotRange -> All,
  AxesLabel -> {"\[Epsilon]", "s''(\[Epsilon])"},
  PlotTheme -> "Scientific"
];

\end{lstlisting}

\end{document}